\documentclass[12pt]{article}
\usepackage[utf8]{inputenc}
\usepackage{comment}
\usepackage{cite}
\input epsf.sty
\usepackage{cite}

\usepackage[margin=2cm]{geometry}
\usepackage{epsfig,amsfonts,amssymb,framed,amsmath,xcolor}
\usepackage{slashed,upgreek}
\usepackage{hyperref}
\input epsf.sty
\usepackage{hhline}
\usepackage{bm}
\usepackage{array}
\usepackage{float}
\usepackage{mathtools}
\usepackage[makeroom]{cancel}
\usepackage{bm}

\newcommand{\be}{\begin{eqnarray}\displaystyle}
\newcommand{\ee}{\end{eqnarray}}

\newcommand{\bi}{\begin{itemize}}
\newcommand{\ei}{\end{itemize}}

\newcommand{\bse}{\begin{subequations}}
\newcommand{\ese}{\end{subequations}}

\newcommand{\f}{\frac}

\newcommand{\p}{\partial}

\newcommand{\non}{\nonumber}

\newcolumntype{P}[1]{>{\centering\arraybackslash}p{#1}}

\begin{document}
\numberwithin{equation}{section}

\baselineskip 24pt

\begin{center}

{\Large \textbf{ Subleading Soft Theorem for arbitrary number of external soft photons and gravitons}}

\end{center}

\vskip .6cm
\medskip

\vspace*{4.0ex}

\baselineskip=18pt

\centerline{\large \textbf{  Sayali Atul Bhatkar$^{\epsilon}$\  and \ Biswajit Sahoo$^{\varepsilon}$ }}

\vspace*{4.0ex}

\centerline{\large \it ~$^{\epsilon}$Indian Institute of Science Education and Research,}
\centerline{\large \it  Homi Bhabha Rd, Pashan, Pune 411 008, India.}
\vspace{5mm}
\centerline{\large \it $^{\varepsilon}$Harish-Chandra Research Institute, HBNI}
\centerline{\large \it  Chhatnag Road, Jhusi,
Allahabad 211019, India.}

\vspace*{1.0ex}
\centerline{\small E-mail: sayali.bhatkar@students.iiserpune.ac.in ,\  biswajitsahoo@hri.res.in}

\vspace*{5.0ex}

\centerline{\bf Abstract} \bigskip

We obtain the subleading soft theorem for a generic theory of quantum gravity, for arbitrary number of soft photons and gravitons and for arbitrary number of finite energy particles with arbitrary mass and spin when all the soft particles are soft in the same rate. This  result is valid at tree level for spacetime dimensions equal to four and five and to all loop orders in spacetime dimensions greater than five. We verify that in classical limit, low energy photon and graviton radiation decouple from each other.



\vfill \eject

\baselineskip 18pt

\tableofcontents

\begin{section}{Introduction and Result}

Soft theorem relates an amplitude with arbitrary number of finite energy(hard) particles and low energy(soft) massless particles (photons/gravitons) to an amplitude without the soft particles\cite{Gell-Mann,Low1,Low2,saito,burnett,bell,duca,weinberg1,weinberg2,jackiw1,jackiw2}.
Recently soft theorem has been explored mostly for one external soft photon or graviton in specific theories \cite{1103.2981,1404.4091,1404.7749,1405.1015,1405.1410,1405.2346,1405.3413,1405.3533,1406.6574,1406.6987,1406.7184,1407.5936,
1407.5982,1408.4179,1410.6406,1412.3699,1504.01364,1507.08882,1509.07840,1604.00650,
1604.03893,1611.02172,1611.07534,1611.03137}. One of the current interests of pursuing work on soft theorem is to understand the symmetry of quantum gravity S-matrix. For one external soft graviton the soft theorem is interpreted as the Ward identity of generalised asymptotically flat spacetime symmetry group\cite{1312.2229,1401.7026,1411.5745,1506.05789,1509.01406,1605.09094,1608.00685,
1612.08294,1701.00496,1612.05886,1703.05448}. Similarly for one external soft photon the soft theorem turns out to be Ward identity of large gauge transformation symmetry for electromagnetic theory\cite{1703.05448,1407.3789,1506.02906,1412.2763,1407.3814,
1709.03850,1605.09677,1505.05346,1605.09731,1805.05651}. Single soft theorem was studied for a generic effective field theory using BCFW formalism in \cite{1611.07534} and corrections to the subleading soft factor due to all possible non-minimal couplings were derived.
Soft theorem is studied also in string theory for arbitrary number of external hard particles\cite{ademollo,shapiro,1406.4172,1406.5155,1411.6661,1502.05258,1505.05854,
1507.08829,1511.04921,1512.00803,1601.03457,1604.03355,1808.04845,1610.03481,1702.03934,1703.00024}.
For specific theories subleading soft graviton theorem with two or arbitrary number of external soft gravitons has been derived\cite{1503.04816,1504.05558,1504.05559,1507.00938,1604.02834,1705.06175,1607.02700,
1702.02350,1709.07883}.
Very recently, soft graviton theorem has been derived for a generic theory of quantum gravity with arbitrary number of external particles with arbitrary mass and spin, starting from 1PI effective action and covariantizing with respect to soft graviton background \cite{1702.03934,1703.00024,1706.00759,1707.06803}. For a generic theory of quantum gravity multiple soft graviton theorem was derived in \cite{1707.06803} when all the gravitons are soft in the same rate(simultaneous limit). In \cite{1803.03023,1808.09965} the double soft graviton theorem was analysed from asymptotic symmetry point of view and symmetry interpretation was found for consecutive limit i.e. when one of the soft graviton is softer than the other one. A new direction was taken up in seminal papers\cite{1801.07719,1804.09193,1806.01872} where classical limit of multiple soft graviton theorem was studied to obtain the long wavelength gravitational radiation emitted in classical scattering processes. Recently in four dimensions soft theorem was explored beyond tree level in \cite{1808.03288} using extension of the infrared treatment developed in\cite{yennie,grammer} and it was shown that new non-analytic terms appear in soft momentum expansion in the form of logarithmic functions of soft energy.  \\

Our goal in this paper is to derive subleading soft theorem for a generic theory of quantum gravity, for arbitrary number of soft photons and gravitons and for arbitrary number of finite energy particles with arbitrary mass and spin when all the soft particles are soft in the same rate. This result can be applied in three possible directions which need to be  pursued. One is to find the Ward identity interpretation of this soft theorem and understand the asymptotic symmetry. Another is to consider classical limit analogues to \cite{1801.07719,1804.09193,1806.01872} and derive the power spectrum for long wavelength photon and graviton radiation in classical scattering in an example where both electromagnetic and gravitational interactions are important. The other is to test the CHY formula\cite{1306.6575,1307.2199,1309.0885,1409.8256,1412.3479} for Einstein-Maxwell theory taking soft limit analogously as it was done for perturbative Einstein gravity in \cite{1709.07883}.

As it is well known there are IR divergences in D=4 and notion of S-matrix is itself ill-defined. As discussed in \cite{1707.06803}, this is  seen in 1PI effective vertices containing soft particles, which have soft momentum factors in denominator. Thus, our result will hold only at tree level in $D=4,5$. In $D>5$, there are no additional divergences even if loop momentum goes soft. So, our result holds to all loops. We hope that we can extend our analysis in $D=4$ to extract logarithmic terms even for multiple external soft photon-graviton case generalising the strategy developed in \cite{1808.03288}.\\

We will combine soft graviton and photon into a single soft field and write down the soft theorem for this composite soft particle. We denote hard particle polarisations, momenta and charges by $\lbrace \epsilon_{i},p_{i},Q_{i}\rbrace$ for $i=1,2,\cdots,N$. Soft particle polarisations and momenta are denoted by $\lbrace \xi_{r},k_{r}\rbrace$ with $\xi_r(k_r)\equiv \ \lbrace e_r(k_r),\varepsilon_r(k_r)\rbrace$ where $e_r, \varepsilon_r$ respectively being the polarisation of soft photon and soft graviton for $r=1,2,\cdots,M$. Our final result for soft theorem with $N$ number of finite energy particles (hard particles) and $M$ number of soft composite particles up to subleading order in soft momenta expansion  is as follows:

\be \label{FINAL}
&&\ \Gamma^{(N+M)}\ \big(\lbrace\epsilon_{i},p_{i}\rbrace ;\ \lbrace e_{r},k_{r}\rbrace ;\ \lbrace \varepsilon_{s},k_{s}\rbrace\big)\non\\[15pt]
&=&\ \Bigg{\lbrace}\prod_{j=1}^{N} \epsilon_{j,\alpha_{j}}(p_{j})\Bigg{\rbrace}\ \Bigg[\ \Big[\Bigg{\lbrace}\prod_{r=1}^{M}\ \Bigg(  S^{(0)}_{r}(\gamma) +\  S^{(0)}_{r}(g)\Bigg)\Bigg{\rbrace}\ \Gamma\Big]^{\alpha_{1}\cdots \alpha_{N}}\non\\
&&\ +\ \sum_{s=1}^{M}\ \Big[\ \Bigg{\lbrace}\prod_{\substack{r=1\\r\neq s}}^{M}\   \Bigg(S^{(0)}_{r}(\gamma)\ +\ S^{(0)}_{r}(g)\Bigg)\Bigg{\rbrace}\  \Big(S^{(1)}_{s}(\gamma)\ +\ \mathcal{N}_{s}(\gamma)\ +\ S^{(1)}_{s}(g)\  \Big)\ \Gamma\ \Big]^{\alpha_{1}\cdots\alpha_{N}} \non\\
&&\ +\ \sum_{\substack{r,u=1\\r< u}}^{M}\Big[ \ \Bigg{\lbrace}\prod_{\substack{s=1\\s\neq r,u}}^{M}\ \Bigg( S^{(0)}_{s}(\gamma)\ +\ S^{(0)}_{s}(g)\Bigg)\Bigg{\rbrace}\sum_{i=1}^{N}\ \lbrace p_{i}\cdot(k_{r}+k_{u})\rbrace ^{-1}\non\\
&&\ \Bigg( \mathcal{M}_{pp}\big(p_{i};\ e_{r},k_{r};\ e_{u},k_{u}\big)\ +\ \mathcal{M}_{gg}\big(p_{i};\ \varepsilon_{r},k_{r};\ \varepsilon_{u},k_{u}\big)\ +\ \mathcal{M}_{pg}\big(p_{i};\ e_{r},k_{r};\ \varepsilon_{u},k_{u}\big)\ \non\\
&&\ +\ \mathcal{M}_{pg}\big(p_{i};\ e_{u},k_{u};\ \varepsilon_{r},k_{r}\big)\Bigg)\ \Gamma \Big]^{\alpha_{1}\cdots \alpha_{N}}\ \Bigg] \ .
\ee

where

\be
\Big[S^{(0)}_{r}(\gamma)\Gamma\Big]^{\alpha_{1}\alpha_{2}\cdots\alpha_{N}}\ &=&\ \sum_{i=1}^{N}\ \f{Q_{\beta_{i}}^{\ \alpha_{i}}\ e_{r,\mu} p_{i}^{\mu}}{p_{i}\cdot k_{r}}\ \Gamma^{\alpha_{1}\cdots\alpha_{i-1}\beta_{i}\alpha_{i+1}\cdots\alpha_{N}}\label{leadingsoftphoton} , \\[12pt]
\Big[S^{(0)}_{r}(g)\Gamma\Big]^{\alpha_{1}\cdots\alpha_{N}}\ &=&\ \sum_{i=1}^{N}\ \f{\varepsilon_{r,\mu\nu}\  p_{i}^{\mu}p_{i}^{\nu}}{p_{i}\cdot k_{r}} \ \Gamma^{\alpha_{1}\cdots\alpha_{N}}\ ,
\ee
\be
\Big[ S^{(1)}_{r}(\gamma)\ \Gamma\Big]^{\alpha_{1}\cdots\alpha_{N}}\ &=&\ \sum_{i=1}^{N}\ \f{Q_{\beta_{i}}^{\ \alpha_{i}}\ e_{r,\mu}\ k_{r,\nu}}{p_{i}\cdot k_{r}}\ \Bigg(p_{i}^{\mu}\f{\p \Gamma^{\alpha_{1}\cdots\alpha_{i-1}\beta_{i}\alpha_{i+1}\cdots\alpha_{N}}}{\p p_{i\nu}}\ -\ p_{i}^{\nu}\f{\p \Gamma^{\alpha_{1}\cdots\alpha_{i-1}\beta_{i}\alpha_{i+1}\cdots\alpha_{N}}}{\p p_{i\mu}}\Bigg)\ , \\[12pt]
\Big[S^{(1)}_{r}(g)\ \Gamma\Big]^{\alpha_{1}\cdots\alpha_{N}}\ &=&\ \sum_{i=1}^{N}\ \f{\varepsilon_{r,b\mu}\ p_{i}^{\mu}\ k_{ra}}{p_{i}\cdot k_{r}}\ \ \Bigg(p_{i}^{b}\f{\p \Gamma^{\alpha_{1}\cdots \alpha_{N}}}{\p p_{ia}} - p_{i}^{a}\f{\p \Gamma^{\alpha_{1}\cdots\alpha_{N}}}{\p p_{ib}}\ +(J^{ab})_{\beta_{i}}\ ^{\alpha_{i}}\ \Gamma^{\alpha_{1}\cdots \alpha_{i-1}\beta_{i}\alpha_{i+1}\cdots\alpha_{N}}\Bigg),\non\\
\ee
\be
\Big[\mathcal{N}_{s}(\gamma)\ \Gamma\Big]^{\alpha_{1}\cdots\alpha_{N}} &=&\ \sum_{i=1}^{N}\ (p_{i}\cdot k_{s})^{-1}\ \big(e_{s,\mu}k_{s\nu}-e_{s,\nu}k_{s\mu}\big)\ \big[\mathcal{N}_{(i)}^{\mu\nu}(-p_{i})\big]^{\alpha_{i}}\ _{\beta_{i}}\ \Gamma^{\alpha_{1}\cdots\alpha_{i-1}\beta_{i}\alpha_{i+1}
\cdots\alpha_{N}} \ , \non\\
\ee

\be
&&\ \mathcal{M}_{pp}\big(p_{i};\ e_{1},k_{1};\ e_{2},k_{2}\big)\non\\ [15pt]
&&=\  Q_{i}^{T}Q_{i}^{T}\ (p_{i}\cdot k_{1})^{-1}\ (p_{i}\cdot k_{2})^{-1}\  \Big[-\ (k_{1}\cdot k_{2})\ (e_{1}\cdot p_{i})\ (e_{2}\cdot p_{i})\ +\ (p_{i}\cdot k_{2})\ (e_{1}\cdot p_{i})\ (e_{2}\cdot k_{1})\non\\
&&\ +\ (p_{i}\cdot k_{1})\ (e_{2}\cdot p_{i})\ (e_{1}\cdot k_{2})\ -\ (e_{1}\cdot e_{2})\ (p_{i}\cdot k_{1})\ (p_{i}\cdot k_{2})\Big]\non\\
&&\ -\ (k_{1}\cdot k_{2})^{-1} \Bigg[\ -\f{1}{D-2}\ p_{i}^{2}\ (k_{1}\cdot k_{2})\ (e_{1}\cdot e_{2})\ +\ \f{1}{D-2}\ p_{i}^{2}\ (e_{1}\cdot k_{2})\ (e_{2}\cdot k_{1})
\ +\ (e_{1}\cdot e_{2})\ (p_{i}\cdot k_{1})\non\\
&&\ (p_{i}\cdot k_{2})\ +\ (k_{1}\cdot k_{2})\ (e_{1}\cdot p_{i})(e_{2}\cdot p_{i})\ -\ (e_{1}\cdot k_{2})\ (e_{2}\cdot p_{i})\ (p_{i}\cdot k_{1})\ -\ (e_{1}\cdot p_{i})\ (e_{2}\cdot k_{1})\ (p_{i}\cdot k_{2})\Bigg],\non\\ 
\ee

\be
&&\mathcal{M}_{pg}\big(p_{i};\ e_{1},k_{1}; \varepsilon_{2},k_{2}\big)\ \non\\[15pt]
&&=\  Q_{i}^{T}\ \Bigg[(p_{i}\cdot k_{1})^{-1}\ (p_{i}\cdot k_{2})^{-1}\  \Bigg(2\ (e_{1}\cdot p_{i})\ (p_{i}\cdot k_{2})\ (p_{i}\cdot \varepsilon_{2}\cdot k_{1})\ +\ (e_{1}\cdot k_{2})\non\\
&&\ (p_{i}\cdot k_{1})\ (p_{i}\cdot \varepsilon_{2}\cdot p_{i})  -\ (e_{1}\cdot p_{i})\ (p_{i}\cdot \varepsilon_{2}\cdot p_{i})\ (k_{1}\cdot k_{2})\ -\ 2(p_{i}\cdot \varepsilon_{2}\cdot e_{1})\ (p_{i}\cdot k_{1})\ (p_{i}\cdot k_{2})\Bigg)\non\\
&&\  +\ (k_{1}\cdot k_{2})^{-1}\ \Bigg((e_{1}\cdot p_{i})\ (k_{1}\cdot \varepsilon_{2}\cdot k_{1})\ +\ (k_{1}\cdot k_{2})\ (p_{i}\cdot \varepsilon_{2}\cdot e_{1})\ -\ (p_{i}\cdot k_{1})(e_{1}\cdot \varepsilon_{2}\cdot k_{1})\non\\
&&\ -\ (e_{1}\cdot k_{2})\ (p_{i}\cdot \varepsilon_{2}\cdot k_{1})\Bigg)\Bigg]\non\\
\ee
and
\be
&&\mathcal{M}_{gg}\big(p_{i}; \ \varepsilon_{1},k_{1};\ \varepsilon_{2},k_{2}\ \big)\non\\
&&=\  (p_{i}\cdot k_{1})^{-1}\ (p_{i}\cdot k_{2})^{-1}\ \Big[ -(k_{1}\cdot k_{2})\ (p_{i}\cdot \varepsilon_{1}\cdot p_{i})\ (p_{i}\cdot \varepsilon_{2}\cdot p_{i}) \ +\ 2(p_{i}\cdot k_{2})\ (p_{i}\cdot \varepsilon_{1}\cdot p_{i})\ (p_{i}\cdot \varepsilon_{2}\cdot k_{1})\non\\
&&\ \ +\ 2(p_{i}\cdot k_{1})\ (p_{i}\cdot \varepsilon_{2}\cdot p_{i})\ (p_{i}\cdot \varepsilon_{1}\cdot k_{2})\ -\ 2(p_{i}\cdot k_{1})\ (p_{i}\cdot k_{2})\ (p_{i}\cdot \varepsilon_{1}\cdot \varepsilon_{2}\cdot p_{i})\Big]\non\\[10pt]
&&\ +\ (k_{1}\cdot k_{2})^{-1}\ \Big[ -(k_{2}\cdot \varepsilon_{1}\cdot \varepsilon_{2}\cdot p_{i})\ (k_{2}\cdot p_{i})\ -\ (k_{1}\cdot \varepsilon_{2}\cdot \varepsilon_{1}\cdot p_{i})\ (k_{1}\cdot p_{i})\non\\
&&\ +\ (k_{2}\cdot \varepsilon_{1}\cdot \varepsilon_{2}\cdot p_{i})\ (k_{1}\cdot p_{i})+(k_{1}\cdot \varepsilon_{2}\cdot \varepsilon_{1}\cdot p_{i})\ (k_{2}\cdot p_{i})\ -\ (\varepsilon_{1,\rho\sigma}\ \varepsilon_{2}^{\rho\sigma})(k_{1}\cdot p_{i})(k_{2}\cdot p_{i})\non\\
&&\ -\ 2(p_{i}\cdot \varepsilon_{1}\cdot k_{2})(p_{i}\cdot \varepsilon_{2}\cdot k_{1})\ +\ (p_{i}\cdot \varepsilon_{2}\cdot p_{i})(k_{2}\cdot \varepsilon_{1}\cdot k_{2})\ +\ (p_{i}\cdot \varepsilon_{1}\cdot p_{i})\ (k_{1}\cdot \varepsilon_{2}\cdot k_{1})\Big].\non\\
\ee
The amplitude without soft particles $\Gamma(\lbrace \epsilon_{i},p_{i}\rbrace)$ contains the momentum conserving delta function and defined only in terms of hard particle degrees of freedom as
\be
\Gamma(\lbrace \epsilon_{i},p_{i}\rbrace)\ &=&\ \Bigg{\lbrace}\prod_{j=1}^{N} \epsilon_{j,\alpha_{j}}(p_{j})\Bigg{\rbrace}\ \Gamma^{\alpha_{1}\alpha_{2}\cdots \alpha_{N}}\ .
\ee
The indices $\alpha,\beta,\cdots$ run over all kind of fields and also takes care of the particle polarisation information. $J^{ab}$ represents the spin angular momentum generator in some reducible representation of $D$-dimensional Lorentz group $SO(D-1,1)$ on the fields and $Q_{\alpha}^{\ \beta}$ represents the global $U(1)$ generator on the fields. At this stage the indices $a,b,\cdots$ and $\mu,\nu,\cdots$ are the space time coordinate or momentum indices and we assumed Einstein summation convention for these indices. We used indices $i,j,\cdots$ for hard particles and $r,s,\cdots$ for soft photons/gravitons and there no Einstein summation convention is assumed. The Minkowski metric we used is with mostly $+$ sign convention.
 
Here we would like clarify an important point to make connection with the existing results. In \eqref{leadingsoftphoton}, the leading soft factor for photon looks very different from the usual expression as here it involves the charge operator. But since the external particle is an eigenstate of charge operator, we have for i'th particle with charge $q_i$ : $ Q_{\beta_{i}}^{\ \alpha_{i}}\epsilon_{i,\alpha_{i}}=q_i\epsilon_{i,\beta_{i}}$. Then the leading single soft photon theorem with soft photon momentum $k$ (M=1 and $\varepsilon_1=0$)becomes: \be
\Gamma^{N+1}=\Big[\prod_{j=1}^{N} \epsilon_{j,\alpha_{j}}(p_{j})\Big]\Big[ S^{(0)}(\gamma) \Gamma\ \Big]^{\alpha_{1}\cdots\alpha_{N}}=\sum_{i=1}^{N} \f{e_{\mu} p_{i}^{\mu}}{p_{i}\cdot k} \epsilon_{i,\alpha_{i}}(p_{i})Q_{\beta_{i}}^{\ \alpha_{i}}\ \Gamma^{\beta_i} =\sum_{i=1}^{N} \f{e_{\mu} p_{i}^{\mu}}{p_{i}\cdot k} q_i\ \Gamma^{N}. \ee
Here, we see that we get back the conventional leading soft factor appeared in literature. Similar logic works for multiple case as well, the charge operator in $S^{(0)},S^{(1)}$ or the contact terms acts on the polarisation tensors and gives corresponding U(1) charges as multiplicative factors and we can recover the expected form of soft factors. 

Since photon can couple to hard particles via its field strength in a non-universal way, the subleading soft part of the theorem depends on the theory and the theory dependent part is the $\mathcal{N}_{(i)}^{\mu\nu}(-p_{i})$ factor which contains the effect of  the non-minimal coupling term of the soft photon with two finite energy fields via field strength. Another point we note is that our multiple soft photon-graviton theorem is dependent on spacetime dimension $D$, which appears in the expression of $\mathcal{M}_{pp}$.\\

As mentioned earlier, \eqref{FINAL} is written in a combined way considering soft graviton and photon as two components of  a single soft field. One can extract multiple soft photon theorem from \eqref{FINAL} by setting all the graviton polarisations $\lbrace \varepsilon_{r}\rbrace$ to be zero. Similarly to get multiple soft graviton theorem from \eqref{FINAL} one have to set photon polarisations $\lbrace e_{r}\rbrace$ to zero and the result agrees with\cite{1707.06803}. To get multiple soft photon-graviton theorem with $M_{1}$ number of soft photons with polarisations and momenta $\lbrace e_{r},\ell_{r}\rbrace$ for $r=1,2,\cdots, M_{1}$ and $M_{2}$ number of soft gravitons with polarisations and momenta $\lbrace \varepsilon_{s},k_{s} \rbrace$ for $s=1,2,\cdots,M_{2}$ from \eqref{FINAL} we need to set $\varepsilon_{r}=0$ for $r=1,2,\cdots,M_{1}$ and $e_{r}=0$ for $r=M_{1}+1,\cdots,M_{1}+M_{2}\equiv M$ and then replace $k_{s}$ by $\ell_{s}$ for $s=1,2,\cdots,M_{1}$ and $(\varepsilon_{M_{1}+s},k_{M_{1}+s})$ by $(\varepsilon_{s},k_{s})$ for $s=1,2,\cdots,M_{2}$. The explicit form of the multiple soft photon-graviton result is given in \eqref{final}.\\

The paper is organised as follows : we derive vertices and propagators needed to prove soft theorem  in section \ref{ver-prop} . In section \ref{strategy} we write down the general strategy that will be used to manipulate the amplitudes to write them as soft theorems. Then, we derive the single, double and multiple soft theorems in successive sections \ref{single} , \ref{2soft} , \ref{Msoft} . Finally, we extract the soft theorem for arbitrary number of soft photons and gravitons and analyse it's classical limit to get long wavelength power spectrum of photon and graviton radiation  in section \ref{special} .

\end{section}

\begin{section}{Evaluation of vertices and propagators}\label{ver-prop}

We will consider a generic theory of quantum gravity which is UV complete and background independent and is given in the form of 1PI effective action. For this generic theory to derive the relevant vertices and propagators we will follow the strategy developed in \cite{1702.03934}\cite{1703.00024}\cite{1706.00759}\cite{1707.06803}. First we will expand the 1PI effective action in terms of all the fields of the theory (including photon and graviton) around their vacuum values. Then we will gauge fix the action using Lorentz covariant gauge fixing condition to get well defined propagators having simple poles. Now the vertices having coupling with soft gravitons and/or photons are derived by covariantizing the gauge fixed action with respect soft photon and graviton background. We assume that all the fields carry tangent space indices and when covariantizing, we have to replace all the ordinary derivatives by covariant derivatives multiplied with inverse vielbeins in the soft photon-graviton background. We parametrize the metric in the following way :

\be
g_{\mu\nu}\ =\ \big(e^{2S\eta}\eta\big)_{\mu\nu}\ =\ \eta_{\mu\nu}\ +\ 2S_{\mu\nu}\ +\ 2\ S_{\mu\rho}S^{\rho}_{\nu}+\cdots,\ \hspace{8mm}\ for\hspace{4mm} S_{\mu\nu}=S_{\nu\mu}\ ,\ S^{\mu}_{\mu}=0.
\ee
The vielbeins are similarly expanded as 
\be
e_{\mu}^{a}\ =\ \delta_{\mu}^{a}\ +\ S_{\mu}^{a}\ +\ \f{1}{2}S_{\mu b}\ S^{ba}\ +\ \cdots \ ,\ E_{a}^{\mu}\ =\ \delta_{a}^{\mu}\ -\ S_{a}^{\mu}\ +\ \f{1}{2}\ S_{ab}\ S^{b\mu}\ +\ \cdots\ .
\ee
Here we use $a,b,c,\cdots$ as tangent space indices, $\mu,\nu,\rho,\cdots $ as curved space indices and
all the indices are raised and lowered by the flat metric $\eta$. If $\lbrace\Phi_{\alpha}\rbrace$ represents all sets of real component fields present in the theory which belongs to some reducible representation of the local Lorentz group SO(1,D-1), then in the covariantization procedure a set of ordinary derivatives operating on $\Phi_{\alpha}$  transforms as

\be
\p_{a_{1}}\p_{a_{2}}...\p_{a_{n}}\Phi_{\alpha}\ \rightarrow \ E_{a_{1}}^{\mu_{1}}E_{a_{2}}^{\mu_{2}}\cdot\cdot\cdot E_{a_{n}}^{\mu_{n}}\ D_{\mu_{1}}D_{\mu_{2}}\cdot\cdot\cdot D_{\mu_{n}}\ \Phi_{\alpha},\label{covariant}
\ee

where

\be
D_{\mu}\Phi_{\alpha}\ =\ \p_{\mu}\Phi_{\alpha}\ +\ \f{1}{2}\ \omega_{\mu}^{ab}\ (J_{ab})_{\alpha}\ ^{\gamma}\ \Phi_{\gamma}\ -\ iQ_{\alpha}\ ^{\gamma}A_{\mu}\ \Phi_{\gamma}. \label{firstderivative}
\ee
Here $A_{\mu}$ represents the background $U(1)$ gauge field and $Q_{\alpha} ^{\ \gamma}$ is the $U(1)$ charge matrix operating on field $\Phi_{\gamma}$ in this big reducible representation.  $J^{ab}$ is the spin angular momenta normalised in a way that when it operates on a covariant vector field $\phi_{c}$, it takes the form
\be
(J^{ab})_{c}\ ^{d}\ =\ \delta^{a}\ _{c}\eta^{bd}\ -\ \delta^{b}\ _{c}\eta^{ad}.
\ee
In the analysis of subleading soft theorem we need the spin connection only upto first order in $S_{\mu\nu}$,
\be
\omega_{\mu}^{ab}\ =\ \p^{b}S_{\mu}^{a}\ -\ \p^{a}S_{\mu}^{b}.
\ee
Then we covariantize two derivative terms:
\be
&D_{\mu}D_{\nu}\Phi_{\alpha}&=\ \p_{\mu}[\p_{\nu}\Phi_{\alpha}\ +\ \f{1}{2}\ \omega_{\nu}^{ab}\ (J_{ab})_{\alpha} ^{\ \gamma}\ \Phi_{\gamma}]\ -\ iQ_{\alpha}^{\ \gamma}A_{\nu}\ \p_\mu\Phi_{\gamma}\non\\
&&-\ \Gamma^{\rho}_{\mu\nu} [\f{1}{2}\ \omega_{\rho}^{ab}\ (J_{ab})_{\alpha}\ ^{\gamma}\ \Phi_{\gamma}-\ iQ_{\alpha}\ ^{\gamma}A_{\rho}\ \Phi_{\gamma}]  -\ iQ_{\alpha} ^{\ \gamma}\ A_{\mu}\ [\p_{\nu}\Phi_{\alpha}\ +\ \f{1}{2}\ \omega_{\nu}^{ab}\ (J_{ab})_{\gamma}\ ^{\beta}\ \Phi_{\beta}]\non\\ 
&&+\f{1}{2}\ \omega_\mu^{ab}\ (J_{ab})_{\alpha}\ ^{\gamma} [\p_{\nu}\Phi_\gamma+\f{1}{2}\ \omega_{\nu}^{cd}\ (J_{cd})_{\gamma}\ ^{\beta}\ \Phi_{\beta}-\ iQ_{\gamma}\ ^{\beta}A_{\nu}\ \Phi_{\beta}] \non\\ 
&&-\ \Gamma^{\rho}_{\mu\nu} \ \p_{\rho}\Phi_{\alpha}\ - \ iQ_{\alpha} ^{\ \gamma}\ \p_{\mu}A_{\nu}\ \Phi_{\gamma} \ -\ Q_{\alpha}^{\ \beta}\ Q_{\beta} ^{\ \gamma}\ A_{\mu}A_{\nu}\ \Phi_{\gamma} \, \label{2der}
\ee
where
\be
\Gamma^{\rho}_{\mu\nu}\ =\  \Big(\ \p_{\mu}S_{\nu}^{\rho}+\p_{\nu}S_{\mu}^{\rho}-\p^{\rho}S_{\mu\nu}\Big)\ +\  \hbox{terms quadratic in $S_{\mu\nu}$}\ .
\ee
In our analysis for multiple soft theorem upto subleading order in soft momenta, while covariantizing the terms having single soft field we need to keep upto  first derivative of  the soft field and for the terms having two soft fields we need to keep only the terms which do not have any derivatives on the soft fields. So, we ignore such terms present in first three lines of \eqref{2der}. Other terms in these lines involve soft couplings similar to the couplings obtained from \eqref{firstderivative}. The last line contains three new kind of couplings between hard and soft particles : the first term containing Christoffel connection appears from the definition of left covariant derivative operation on a vector. The second term comes from the operation of left derivative on the gauge field contained in the right covariant derivative and the third term contains two gauge fields come from two covariant derivatives. Also we can drop the terms containing Christoffel connection following the argument given in the appendix of \cite{1703.00024}. \\

The convention we follow is that all the particles are ingoing and in any diagram the thick lines represent hard particles and thin lines represent combined soft particles (photons and gravitons). The mode expansion of external soft photon and graviton fields with momentum $k^{\mu}$ are

\be
A_{\mu}(x)\ &=&\ e_{\mu}(k)\ e^{ik.x}\ ,\hspace{10mm}\ k^{\mu}e_{\mu}=0 \ , \label{phgauge}\\
 S_{\mu\nu}(x)\ &=&\ \varepsilon_{\mu\nu}(k)\ e^{ik.x}\ ,\hspace{10mm}\ \varepsilon_{\mu\nu}\ \eta^{\mu\nu}\ =\ 0\ ,\ k^{\mu}\varepsilon_{\mu\nu}\ =\ 0\ =\ k^{\nu}\varepsilon_{\mu\nu}.\label{grgauge}
\ee
Above we choose graviton polarisation to be symmetric traceless for simplicity, so that  $\sqrt{-det(g)}$ is always one in the covariantized action. In diagram computations we treat soft photon and graviton fields are two components of a single soft field having polarisation $\xi(k)=\ \big( e(k),\varepsilon(k) \big)$. \\

To determine the vertices containing two finite energy fields and one or two soft particles we start by covariantising the quadratic part of 1PI effective action

\be \label{action}
S^{(2)}\ &=&\ \f{1}{2}\ \int\frac{d^{D}q_{1}}{(2\pi)^{D}}\ \f{d^{D}q_{2}}{(2\pi)^{D}}\ \Phi_{\alpha}(q_{1})\mathcal{K}^{\alpha\beta}(q_{2})\ \Phi_{\beta}(q_{2})\ (2\pi)^{D}\delta^{(D)}(q_{1}+q_{2})\ ,
\ee
with the kinetic term satisfying :
\be
\mathcal{K}^{\alpha\beta}(q)\ =\ \mathcal{K}^{\beta\alpha}(-q)\ .\label{ant-symm}
\ee
If $\Phi_{\alpha}$ is a grassmann odd field then the r.h.s of \eqref{ant-symm} contains an extra -ve sign but final result does not depend on this. The full renormalised propagator of the internal finite energy particle with momentum $q^{\mu}$ is given by,
\be
i\mathcal{K}^{-1}_{\alpha\beta}(q)\equiv (q^{2}+M^{2})^{-1}\ \Xi_{\alpha\beta}(q) \ , \label{Xi}
\ee
where $M$ is the renormalised mass of the particle. 

For covariantization with respect to photon background we need the action \eqref{action} to be invariant under global $U(1)$ transformation $\Phi_{\alpha}\rightarrow [exp\ (iQ\theta)]_{\alpha}\ ^{\beta}\ \Phi_{\beta}$ , where $\theta$ is the parameter of $U(1)$ global transformation and $Q$ is the $U(1)$ transformation generator on real field.\footnote{Since we choose $\lbrace\Phi_{\alpha}\rbrace$ to be real component fields $Q$ needs to be completely imaginary hermitian matrix i.e. $Q^{T}=-Q$, though throughout we have not used this property.} This requires the following property of the kinetic operator,
\be
Q_{\gamma} ^{\ \alpha}\ \mathcal{K}^{\gamma\beta}\ +\ \mathcal{K}^{\alpha\gamma}\ Q_{\gamma} ^{\ \beta} =\ 0.
\ee
In matrix notation this translates to,
\be
Q^{T}\mathcal{K}\ +\ \mathcal{K}Q\ =0 .\label{u1K}
\ee
This also impose the following condition on the numerator of the propagator:
\be
Q\ \Xi \ +\ \Xi\  Q^{T}\ =\ 0.\label{u1Xi}
\ee

Now the vertex associated with two finite energy fields and one soft field with polarisation and momentum $(\xi,\ k)$ upto linear order in soft momentum, will be obtained from the following covariantized part of \eqref{action}
\be
S^{(3)}\ &=&\ \f{1}{2}\ \int \frac{d^{D}q_{1}}{(2\pi)^{D}}\ \f{d^{D}q_{2}}{(2\pi)^{D}} \ (2\pi)^{D}\delta^{(D)}(q_{1}+q_{2}+k)\non\\
&&\ \Phi_{\alpha}(q_{1})\ \Bigg[\ - \ e_{\mu}\f{\p\mathcal{K}^{\alpha\gamma}(q_{2})}{\p q_{2\mu}}\ Q_{\gamma} ^{\ \beta} -\ \f{1}{4}\ (k_{\mu}e_{\nu}\ +\ k_{\nu}e_{\mu})\ \f{\p^{2}\mathcal{K}^{\alpha
\gamma}(q_{2})}{\p q_{2\mu}\p q_{2\nu}}\ Q_{\gamma} ^{\ \beta}\non\\
&&\ -\ \varepsilon_{\mu\nu}q_{2}^{\nu}\ \f{\p \mathcal{K}^{\alpha\beta}(q_{2})}{\p q_{2\mu}}\ +\ \f{1}{2}\ (k_{b}\varepsilon_{a\mu}-k_{a}\varepsilon_{b\mu})\ \f{\p \mathcal{K}^{\alpha\gamma}(q_{2})}{\p q_{2\mu}}\ (J^{ab})_{\gamma}\ ^{\beta} \Bigg]\ \Phi_{\beta}(q_{2})\ . \label{3v}
\ee

In the above expression the first term inside the square bracket comes from the presence of gauge field within covariant derivative. Second term comes from the symmetric part of derivative operating on the gauge field when two covariant derivatives operate on the field $\Phi_{\beta}$. Source of third term is multiplication by inverse vielbein to transform flat space index to curved space index. The fourth term comes from the spin connection part in covariant derivative. There is another source of three point  vertex associated with soft photon coming from the non-minimally coupled field strength with two finite energy fields having most general form :
\be
\bar{S}^{(3)}\ \equiv \f{1}{2}\ \int\f{d^{D}q_{1}}{(2\pi)^{D}}\f{d^{D}q_{2}}{(2\pi)^{D}}\ (2\pi)^{D}\delta^{(D)}(q_{1}+q_{2}+k)\ F_{\mu\nu}(k)\ \Phi_{\alpha}(q_{1})\ \mathcal{B}^{\alpha\beta,\mu\nu}(q_{2})\ \Phi_{\beta}(q_{2})\label{non-minimal}
\ee
where,

\be
F_{\mu\nu}(k)=i\big[k_{\mu}e_{\nu}(k)-k_{\nu}e_{\mu}(k)\big]\non
\ee
with,
\be
&&Q_{ \gamma}^{\ \alpha}\ \mathcal{B}^{\gamma\beta,\mu\nu}(q_{2})\  +\ \mathcal{B}^{\alpha\gamma,\mu\nu}(q_{2})\ Q_{\gamma} ^{\ \beta} =\ 0\\ \label{u1B}
&&\mathcal{B}^{\alpha\beta,\mu\nu}(q_{2})\ =\ -\mathcal{B}^{\alpha\beta,\nu\mu}(q_{2})\\
&&\mathcal{B}^{\alpha\beta,\mu\nu}(q_{2})\ =\ \mathcal{B}^{\beta\alpha,\mu\nu}(-q_{1}-k).\label{B}
\ee
If $\Phi_{\alpha}$ be a grassman odd field then  RHS of the above equation will carry an extra -ve sign. 
\begin{figure}[H]
\begin{center}
\includegraphics[scale=0.8]{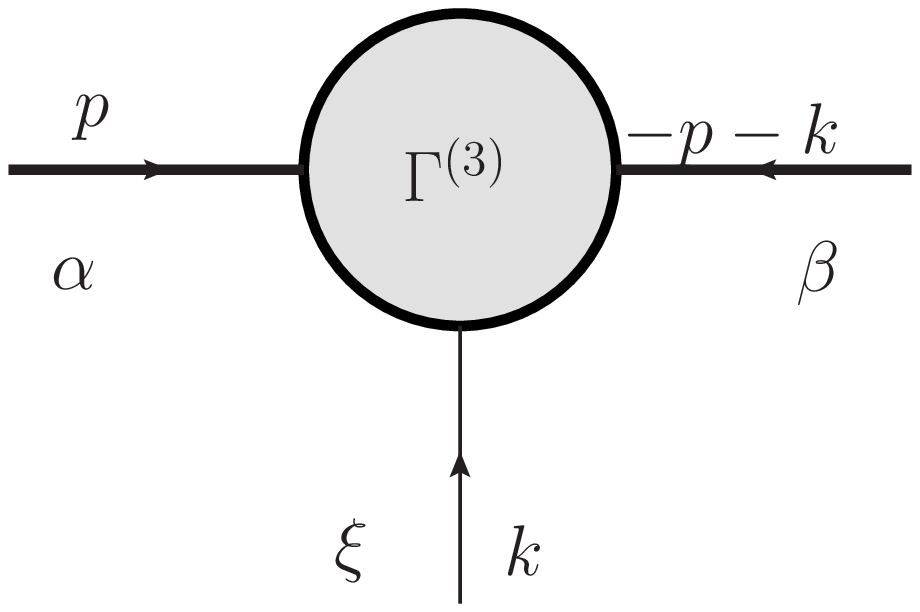}
\caption{ 1PI vertex involving two finite energy particles and one soft particle.}\label{phi-phi-xi}
\end{center}
\end{figure}

 The expression of 1PI vertex corresponding to Fig.\eqref{phi-phi-xi}, derived from \eqref{3v} and \eqref{non-minimal} is :
\be
&&\ \Gamma^{(3)\alpha\beta}(\xi , k;\ p,-p-k)\non\\
&&\ =\ +\f{i}{2}\ \Bigg[\ e_{\mu}(k)\f{\p \mathcal{K}^{\alpha\gamma}(-p-k)}{\p p_{\mu}}\ Q_{\gamma} ^{\ \beta} -\ \ e_{\mu}(k)\f{\p \mathcal{K}^{\beta\gamma}(p)}{\p p_{\mu}}\ Q_{\gamma} ^{\ \alpha}\ -\f{1}{4}\ (k_{\mu}e_{\nu}+k_{\nu}e_{\mu})\f{\p^{2}\mathcal{K}^{\alpha\gamma
}(-p-k)}{\p p_{\mu}\p p_{\nu}}Q_{\gamma} ^{\ \beta}\non\\
&&\ -\f{1}{4}\ (k_{\mu}e_{\nu}+k_{\nu}e_{\mu})\f{\p^{2}\mathcal{K}^{\beta\gamma
}(p)}{\p p_{\mu}\p p_{\nu}}\ Q_{\gamma} ^{\ \alpha} +\ i\ (k_{\mu}e_{\nu}-k_{\nu}e_{\mu})\ \mathcal{B}^{\alpha\beta,\mu\nu}(-p-k)\ +\ i\ (k_{\mu}e_{\nu}-k_{\nu}e_{\mu})\mathcal{B}^{\beta\alpha,\mu\nu}(p)\non\\[10pt]
&&\ -\varepsilon_{\mu\nu}(k)(p+k)^{\nu}\f{\p \mathcal{K}^{\alpha\beta}(-p-k)}{\p p_{\mu}}\ -\ \varepsilon_{\mu\nu}(k)p^{\nu}\f{\p \mathcal{K}^{\beta\alpha}(p)}{\p p_{\mu}}\non\\
&& -\ \f{1}{2}(k_{b}\varepsilon_{a\mu}-k_{a}\varepsilon_{b\mu})\f{\p \mathcal{K}^{\alpha\gamma}(-p-k)}{\p p_{\mu}}(J^{ab})_{\gamma}\ ^{\beta}
\ +\ \f{1}{2}\ (k_{b}\varepsilon_{a\mu}-k_{a}\varepsilon_{b\mu})\f{\p \mathcal{K}^{\beta\gamma}(p)}{\p p_{\mu}}(J^{ab})_{\gamma}\ ^{\alpha}\Bigg]\ .
\ee

Using \eqref{ant-symm},\eqref{phgauge},\eqref{grgauge},
\eqref{B},\eqref{u1K} and making Taylor series expansion in $k^{\mu}$ we write down the matrix form of $\Gamma^{(3)}$ up to linear order in $k^{\mu}$ :

\be
&&\ \Gamma^{(3)}(\xi ,k;\ p,-p-k)\non\\[10pt]
&&=\ i\Bigg[ +\ e_{\mu}\f{\p \mathcal{K}(-p)}{\p p_{\mu}}\ Q\ +\ \f{1}{2}\ e_{\mu}k_{\nu}\ \f{\p^{2}\mathcal{K}(-p)}{\p p_{\mu}\p p_{\nu}}\ Q\ +\ i\ (k_{\mu}e_{\nu}-k_{\nu}e_{\mu})\ \mathcal{B}^{\mu\nu}(-p)\non\\
&&\ -\ \varepsilon_{\mu\nu}p^{\nu}\ \f{\p \mathcal{K}(-p)}{\p p_{\mu}}\ -\f{1}{2}\ \varepsilon_{\mu\nu}p^{\nu}\ k_{\rho}\ \f{\p^{2}\mathcal{K}(-p)}{\p p_{\mu}\p p_{\rho}}\ +\f{1}{2}\ \varepsilon_{b\mu}k_{a}\ \f{\p \mathcal{K}(-p)}{\p p_{\mu}}\ (J^{ab})\ -\f{1}{2}\ \varepsilon_{b\mu}k_{a}\ (J^{ab})^{T}\ \f{\p \mathcal{K}(-p)}{\p p_{\mu}}\Bigg]\ . \non\\\label{Gamma3}
\ee

Next we want to derive the four point vertex involving two finite energy particles and two soft particles. The diagrams which contain this kind of vertex are already in subleading order relative to the diagrams having only $\Gamma^{(3)}$ vertices. So we need to evaluate this vertex only up to leading order in soft momenta. Hence we will drop terms having spin connection, Christoffel connection, derivative over gauge fields as well as the non-minimal coupling piece in the covariantized action. We need to pick terms quadratic in soft fields, i.e. photon-photon, photon-graviton and graviton-graviton fields. The part of the covariantized action from where we will get the required four point vertex  involving two soft fields with polarisations and momenta $(\xi(k_{1}),k_{1})$ and $(\xi(k_{2}),k_{2})$ is:

\be
 S^{(4)}\ &=&\ \f{1}{2}\ \int \frac{d^{D}q_{1}}{(2\pi)^{D}}\ \f{d^{D}q_{2}}{(2\pi)^{D}} \ (2\pi)^{D}\delta^{(D)}(q_{1}+q_{2}+k_{1}+k_{2})\non\\
&&\ \Phi_{\alpha}(q_{1})\ \Bigg[ \ e_{\mu}(k_{1})e_{\nu}(k_{2})\ \f{\p^{2}\mathcal{K}^{\alpha\delta}(q_{2})}{\p q_{2\mu}\p q_{2\nu}}\ Q_{\delta} ^{\ \gamma}\ Q_{\gamma} ^{\ \beta}\ +\  \varepsilon_{\mu}^{\nu}(k_{1})\ e_{\nu}(k_{2})\ \f{\p \mathcal{K}^{\alpha\gamma}(q_{2})}{\p q_{2\mu}}Q_{\gamma} ^{\ \beta}\non\\
&&\ +\ \varepsilon_{\mu}^{\nu}(k_{1})\ q_{2\nu}e_{\rho}(k_{2})\ \f{\p^{2}\mathcal{K}^{\alpha\gamma}(q_{2})}{\p q_{2\mu}\p q_{2\rho}}\ Q_{\gamma} ^{\ \beta} +\ \varepsilon_{\mu}^{\nu}(k_{2})\ e_{\nu}(k_{1})\ \f{\p \mathcal{K}^{\alpha\gamma}(q_{2})}{\p q_{2\mu}}Q_{\gamma} ^{\ \beta}\non\\
&&\ +\ \varepsilon_{\mu}^{\nu}(k_{2})\ q_{2\nu}e_{\rho}(k_{1})\ \f{\p^{2}\mathcal{K}^{\alpha\gamma}(q_{2})}{\p q_{2\mu}\p q_{2\rho}}\ Q_{\gamma} ^{\ \beta}\ +\  \varepsilon_{\mu\nu}(k_{1})\varepsilon_{\rho\sigma}(k_{2})\ q_{2}^{\nu}q_{2}^{\sigma}\ \f{\p^{2}\mathcal{K}^{\alpha\beta}(q_{2})}{\p q_{2\mu}\p q_{2\rho}}\non\\
&&\ +\ \f{1}{2}\ \varepsilon_{\mu}^{b}(k_{1})\varepsilon_{b\nu}(k_{2})\ q_{2}^{\nu}\ \f{\p \mathcal{K}^{\alpha\beta}(q_{2})}{\p q_{2\mu}}\  +\ \f{1}{2}\ \varepsilon_{\mu}^{b}(k_{2})\varepsilon_{b\nu}(k_{1})\ q_{2}^{\nu}\ \f{\p \mathcal{K}^{\alpha\beta}(q_{2})}{\p q_{2\mu}}\ \Bigg]\ \Phi_{\beta}(q_{2}) \ . \label{4v}
\ee

\begin{figure}[H]
\begin{center}
\includegraphics[scale=0.8]{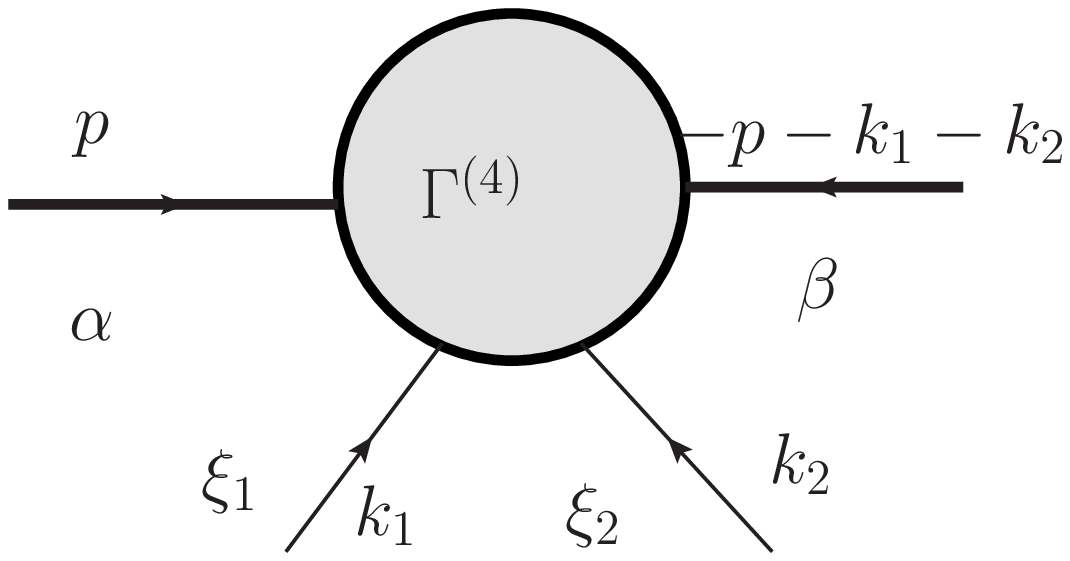}
\caption{ 1PI vertex involving two finite energy particles and two soft particles}\label{phi-phi-xi-xi}
\end{center}
\end{figure}
Symmetrising \eqref{4v} with respect to two hard fields and using \eqref{ant-symm} and \eqref{u1K} the expression for the vertex in Fig.\eqref{phi-phi-xi-xi} in matrix form with polarisations of soft particles  relabelled by $(\xi(k_{1}),\xi(k_{2}))\equiv (\xi_{1},\xi_{2})$ takes the form:

\be
&&\Gamma^{(4)}(\xi_{1},k_{1};\ \xi_{2},k_{2};\ p,-p-k_{1}-k_{2})\non\\
&&=\ i\ \Bigg[  e_{1,\mu}e_{2,\nu}\ \f{\p^{2}\mathcal{K}(-p)}{\p p_{\mu}\p p_{\nu}}\ QQ\ -\  \varepsilon_{1,\mu\nu}e_{2}^{\nu}\ \f{\p \mathcal{K}(-p)}{\p p_{\mu}}\ Q\ -\  \varepsilon_{1,\mu\nu}p^{\nu}\ e_{2,\rho}\ \f{\p^{2}\mathcal{K}(-p)}{\p p_{\mu}\p p_{\rho}}\ Q\non\\
&&\ -\ \varepsilon_{2,\mu\nu}e_{1}^{\nu}\ \f{\p \mathcal{K}(-p)}{\p p_{\mu}}\ Q\ -\ \ \varepsilon_{2,\mu\nu}p^{\nu}\ e_{1,\rho}\ \f{\p^{2}\mathcal{K}(-p)}{\p p_{\mu}\p p_{\rho}}\ Q\ +\ \varepsilon_{1,\mu\nu}\varepsilon_{2,\rho\sigma}\ p^{\nu}p^{\sigma}\ \f{\p^{2}\mathcal{K}(-p)}{\p p_{\mu}\p p_{\rho}}\non\\
&&\ +\ \f{1}{2}\ \big(\varepsilon_{1,\mu}^{\ b}\varepsilon_{2,b\nu}\ +\ \varepsilon_{2,\mu}^{\ b}\varepsilon_{1,b\nu}\big)p^{\nu}\ \f{\p \mathcal{K}(-p)}{\p p_{\mu}}\Bigg]\ . \label{Gamma4}
\ee

\begin{figure}[H]
\begin{center}
\includegraphics[scale=0.7]{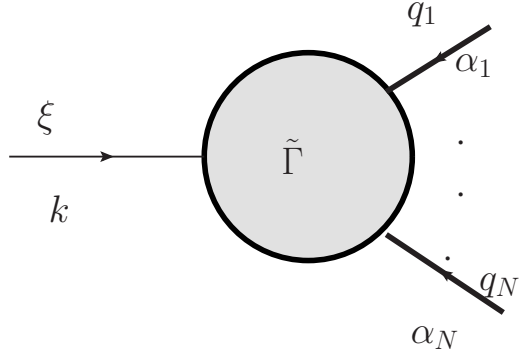}
\caption{ An amputated Greens function with N-number of external finite energy particles and one soft particle. This diagram represents sum over the diagrams where the soft particle is not attached to any external leg. }\label{Gammatilde}
\end{center}
\end{figure}

The amputated Greens function  of Fig.\eqref{Gammatilde} can be evaluated by covariantizing the N particle amputated Greens function $\Gamma(q_{1},q_{2},\cdots,q_{N})$ with respected to soft background field (photon and graviton). Since the diagrams that contain $\tilde{\Gamma}$ are already in subleading order, we only need its leading contribution.
\be
\widetilde{\Gamma}(\xi,k;\ q_{1},q_{2},\cdots,q_{N})\ &=&-\sum_{i=1}^{N} \Big{\lbrace}\prod_{\substack{j=1\\ j\neq i}}^{N}\epsilon_{\alpha_{j}}\Big{\rbrace}\ \epsilon_{\alpha_{i}}\ \Bigg[Q _{\beta_{i}} ^{\ \alpha_{i}}\ e_{\mu}(k)\ \f{\p}{\p q_{i\mu}}\ +\ \varepsilon_{\mu\nu}(k) \delta_{\beta_{i}} ^{\alpha_{i}}\ q_{i}^{\nu}\ \f{\p}{\p q_{i\mu}}\Bigg] \Gamma^{\alpha_{1}\cdots\beta_{i}\cdots\alpha_{N}}\ .\non\\ \label{tildeGamma}
\ee

\begin{figure}[H]
\begin{center}
\includegraphics[scale=0.7]{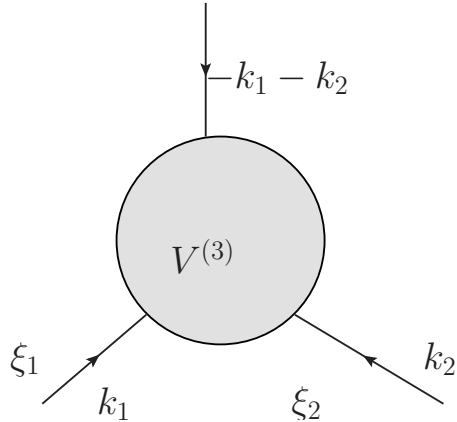}
\caption{ 1PI vertex involving two external soft particles and one internal soft particle.}\label{xi-xi-xi}
\end{center}
\end{figure}

Now we want to evaluate the vertex given in Fig.\eqref{xi-xi-xi} associated with three soft fields. This kind of vertex cannot be obtained by covariantizing the 1PI effective action but one needs to explicitly add most general gauge and general coordinate invariant terms involving only soft fields in the action. Since the diagrams containing this kind of vertex are already in subleading order,  we need to evaluate the vertex contribution of Fig.\eqref{xi-xi-xi} only up to leading order. The relevant action to determine this vertex is :

\be
S^{\ soft}\ &=&\ -\f{1}{2}\ \int d^{D}x\ \sqrt{-det(g)}\ R\ -\f{1}{4}\ \int d^{D}x\ \sqrt{-det(g)}\ g^{\mu\rho}g^{\nu\sigma}\ F_{\mu\nu}F_{\rho\sigma}\ . \label{Ssoft}
\ee

The contribution of $V^{(3)}$ have three sources depending on the choice of two external soft particles $(\xi_{1},\xi_{2})$. When both the external soft particles are photons then the internal soft line has to be graviton and the contribution to $V^{(3)}$ with internal soft graviton carrying indices $(\rho , \sigma)$ comes from the second part of \eqref{Ssoft}, it is \cite{9411092} :
\be
&&\ V^{(3)}_{PP\ \rho\sigma}\ (e_{1},k_{1};e_{2},k_{2};\ -k_{1}-k_{2})\non\\
 &=&\ 
(-i)\ \Bigg[\eta_{\rho\sigma}\big(-(k_{1}\cdot k_{2})\ (e_{1}.e_{2})\ +\ (e_{1}\cdot k_{2})\ (e_{2}\cdot k_{1})\ \big)\ +\ (e_{1}.e_{2})\ (k_{1\rho}k_{2\sigma}+k_{2\rho}k_{1\sigma})\non\\
&& +\ (k
_{1}\cdot k_{2})\ (e_{1,\rho}e_{2,\sigma}+e_{1,\sigma}e_{2,\rho})\ -(e_{1}\cdot k_{2})\ k_{1\sigma}e_{2,\rho}\  -\ (e_{2}\cdot k_{1})\ e_{1,\rho}k_{2\sigma}\non\\
&&\ -(e_{1}\cdot k_{2})\ k_{1\rho}e_{2,\sigma}\ -\ (e_{2}\cdot k_{1})\ e_{1,\sigma}k_{2\rho}\Bigg]\ . \label{V3pp}
\ee
When one of the external particle is photon and the other one is graviton then the internal soft particle has to be photon and the contribution to $V^{(3)}$ with internal soft photon carrying index $\mu$ from the second part of \eqref{Ssoft} turns out to be \cite{9411092} :

\be
&&V_{PG\ \mu}^{(3)}(e,k_{1};\ \varepsilon,k_{2}\ ;\ -k_{1}-k_{2})\non\\
&=&\ (-2i)\Big[-e_{\mu}(k_{1})\varepsilon_{\rho\sigma}(k_{2})k_{1}^{\rho}k_{1}^{\sigma}\ -\ (k_{1}\cdot k_{2})\varepsilon_{\mu\nu}(k_{2})e^{\nu}(k_{1})\ +\ e_{\rho}(k_{1})\varepsilon^{\rho\sigma}(k_{2})k_{1\sigma}k_{1\mu}\ +\ e_{\nu}(k_{1})k_{2}^{\nu}\ \varepsilon_{\mu\rho}(k_{2})k_{1}^{\rho}\Big].\non\\ \label{V3pg}
\ee
When both the external soft particles are gravitons then the internal soft particle has to be graviton and the contribution to $V^{(3)}$ with internal soft graviton carrying indices $(\mu,\nu)$ from the first part of \eqref{Ssoft} in de Donder gauge turns out to be  \cite{1707.06803,9411092} :

\be
&&\ V^{(3)}_{GG\ \mu\nu}\ \big(\varepsilon_{1},k_{1},\varepsilon_{2},k_{2}; -k_{1}-k_{2}\big)\non\\[10pt]
&& = \frac{i}{2}\varepsilon_{1,ab}\ \varepsilon_{2,cd}\Big[\ \Big{\lbrace} \eta_{\mu\nu}\eta^{ac}
\eta^{bd}k_{1}^{\rho}k_{2\rho}\ -\ 2\eta^{ad}\delta^{c}_{\nu}k_{2}^{b}
k_{2\mu}\ -\ 2\eta^{cb}\delta^{a}_{\nu}k_{a}^{d}k_{1\mu}\ +\ 2\eta^{ad}\delta^{c}_{\nu}
k_{1\mu}k_{2}^{b} \non\\
 &&\ +\ 2\eta^{cb}\delta^{a}_{\mu}k_{1}^{d}k_{2\nu}\ -\ 2\eta^{ac}\eta^{bd}k_{1\mu}
k_{2\nu}\ -\ 4\delta^{a}_{\nu}\delta^{c}_{\mu}k_{1}^{d}k_{2}^{b}\ +\ 2\delta^{c}_{\mu}
\delta^{d}_{\nu}k_{2}^{b}k_{2}^{a}\ +\ 2\delta^{a}_{\mu}\delta^{b}_{\nu}k_{1}^{d}
k_{1}^{c}\Big{\rbrace} +\ \Big{\lbrace}\mu\leftrightarrow\nu\Big{\rbrace}\Big].\non\\ \label{V3gg}
\ee
Hence the total contribution of the vertex in Fig.\eqref{xi-xi-xi} contracted with some arbitrary off-shell polarisation tensor $\hat{\xi}_{\alpha}\equiv (\hat{e}_{\mu},\hat{\varepsilon}_{\mu\nu}) $, takes the form:

\be
&&\ \hat{\xi}_{\alpha} V^{(3)\alpha}(\xi_{1},k_{1};\ \xi_{2},k_{2};\ -k_{1}-k_{2})\non\\[10pt]
&&=\ \hat{\varepsilon}_{\mu\nu}\ V^{(3)\mu\nu}_{PP}\ (e_{1},k_{1};e_{2},k_{2};\ -k_{1}-k_{2})\ +\ \hat{e}_{\mu}\ V_{PG}^{(3)\mu}(e_{1},k_{1};\ \varepsilon_{2},k_{2}\ ;\ -k_{1}-k_{2})\non\\
&&\ +\ \hat{e}_{\mu} V_{PG}^{(3)\mu}(e_{2},k_{2};\ \varepsilon_{1},k_{1}\ ;\ -k_{1}-k_{2})\ +\ \hat{\varepsilon}_{\mu\nu}\ V^{(3),\mu\nu}_{GG}\ \big(\varepsilon_{1},k_{1},\varepsilon_{2},k_{2}; -k_{1}-k_{2}\big)\ . \label{V3}
\ee

Let us now focus on deriving propagators. The hard particle propagator is already given in \eqref{Xi}. The soft graviton propagator with momentum $k$ in de Donder gauge with endpoints $(\mu,\nu)$ and $(\rho,\sigma)$ takes the  form:

\be
G_{\mu\nu,\rho\sigma}(k)\ &=&\ -\f{i}{2}\ \f{1}{k^{2}}\ \Bigg(\eta_{\mu\rho}\eta_{\nu\sigma}\ +\ \eta_{\mu\sigma}\eta_{\nu\rho}\ -\ \f{2}{D-2}\ \eta_{\mu\nu}\eta_{\rho\sigma}\Bigg)\ . \label{grprop}
\ee

The soft photon propagator with momentum $k$ in Feynman gauge with the endpoints $\mu$ and $\nu$ takes the form:

\be
D_{\mu\nu}(k)\ =\ \f{-i\ \eta_{\mu\nu}}{k^{2}}\ . \label{phprop}
\ee

\end{section}

\begin{section}{Strategy for evaluation of diagrams}\label{strategy}

Before describing the strategy we first write down the identities which we have to use while computing diagrams. If the polarisation of $i$'th hard particle be $\epsilon_{i}(p_{i})$ then it satisfies the on-shell condition

\be
\epsilon_{i}^{T}(p_{i})\ \mathcal{K}_{i}(-p_{i})\ =\ 0\ . \label{st1}
\ee
  We  can write down the hard particle propagator definition given in \eqref{Xi} for $i$'th particle, in matrix notation (removing the polarisation indices) as :
\be
\mathcal{K}_{i}(-p_{i})\Xi_{i}(-p_{i})\ =\ i\ (p_{i}^{2}+M_{i}^{2}) \ ,\\
\Xi_{i}(-p_{i})\mathcal{K}_{i}(-p_{i})\ =\ i\ (p_{i}^{2}+M_{i}^{2})\ . 
\ee
Now operating one momentum derivative on above relations we get

\be
\f{\p \mathcal{K}_{i}(-p_{i})}{\p p_{i\mu}}\Xi_{i}(-p_{i})\ =\ 2ip_{i}^{\mu}\ -\ \mathcal{K}_{i}(-p_{i})\f{\p \Xi_{i}(-p_{i})}{\p p_{i\mu}}\ , \label{st2}
\ee
\be
\f{\p \Xi_{i}(-p_{i})}{\p p_{i\mu}}\mathcal{K}_{i}(-p_{i})\ =\ -\Xi_{i}(-p_{i})\f{\p \mathcal{K}_{i}(-p_{i})}{\p p_{i\mu}}\ + \ 2ip_{i}^{\mu}\ . \label{st6}
\ee
Operating another momentum derivative on \eqref{st2} we get
\be
\f{\p^{2}\mathcal{K}_{i}(-p_{i})}{\p p_{i\mu}\p p_{i\nu}}\Xi_{i}(-p_{i})\ =\ 2i\eta^{\mu\nu}\ -\ \f{\p \mathcal{K}_{i}(-p_{i})}{\p p_{i\mu}}\f{\p \Xi_{i}(-p_{i})}{\p p_{i\nu}}\ -\ \f{\p \mathcal{K}_{i}(-p_{i})}{\p p_{i\nu}}\f{\p \Xi_{i}(-p_{i})}{\p p_{i\mu}}\ -\ \mathcal{K}_{i}(-p_{i})\f{\p^{2}\Xi_{i}(-p_{i})}{\p p_{i\mu}\p p_{i\nu}}.\non\\ \label{st3}
\ee
Lorentz covariance of $\mathcal{K}$ and $\Xi$ implies:

\be
(J^{ab})^{T}\ \mathcal{K}_{i}(-p_{i})\ =\ -\mathcal{K}_{i}(-p_{i})\ (J^{ab})\ +\ p_{i}^{a}\f{\p \mathcal{K}_{i}(-p_{i})}{\p p_{ib}}\ -\ p_{i}^{b}\f{\p \mathcal{K}_{i}(-p_{i})}{\p p_{ia}}\ , \label{st4}
\ee

\be
(J^{ab})\ \Xi_{i}(-p_{i})\ =\ -\ \Xi_{i}(-p_{i})\ (J^{ab})^{T}\ -\ p_{i}^{a}\f{\p \Xi_{i}(-p_{i})}{\p p_{ib}}\ +\ p_{i}^{b}\f{\p \Xi_{i}(-p_{i})}{\p p_{ia}}\ . \label{st5}
\ee

To evaluate the diagrams we follow the steps :
\begin{enumerate}
\item First we write down the contribution of the diagram in terms of $\mathcal{K}$ and $\Xi$ using the vertices and propagators derived in sec.\eqref{ver-prop} with the hard particle polarisation tensors multiplied from the left.
\item Move all the $J^{ab}$ and $(J^{ab})^{T}$ factors to the extreme right using \eqref{st4}, \eqref{st5} and their derivatives so that the polarisation indices of these directly contract with $\Gamma^{\alpha_{1},\alpha_{2},\cdots,\alpha_{N}}$\ . Similarly move the $U(1)$ charge matrix $Q$ to the extreme right using \eqref{u1K}, \eqref{u1Xi}, \eqref{u1B} and their momentum derivative forms.
\item Expand all the $\mathcal{K}$, $\Xi$ and $\Gamma^{\alpha_{1},\alpha_{2},\cdots,\alpha_{N}}$ in Taylor series in power of soft momenta up to the required order. 

\item Now transfer the derivatives on $\mathcal{K}$ to $\Xi$ to the maximal possible extent using \eqref{st2}, \eqref{st3}. For some special cases we need to move derivative from $\Xi$ to $\mathcal{K}$, we have to use \eqref{st6}.

\item At the end of the manipulation we need to use the on-shell condition \eqref{st1}.

\end{enumerate}

\end{section}

\begin{section}{Subleading soft theorem for one external soft particle}\label{single}

In this section, we will write down soft theorem till subleading order in soft momentum for one external composite particle ( graviton and photon) going soft.  The aim is to obtain the theory-dependent terms in single soft theorem, as the universal pieces are well known in the literature.

Let us start by defining the N-particle amplitude without soft insertion after stripping out the $i$'th particle polarisation tensor 
\be
\Gamma^{(i),\alpha_{i}}(p_{i})\ &=&\ \Bigg{\lbrace}\prod_{\substack{j=1\\j\neq i}}^{N}\ \epsilon_{j,\alpha_{j}}\Bigg{\rbrace}\ \Gamma^{\alpha_{1}\alpha_{2}\cdots \alpha_{N}}(p_{1},\cdots,p_{N})\ . 
\ee
\begin{figure}[H]
\begin{center}
\includegraphics[scale=0.7]{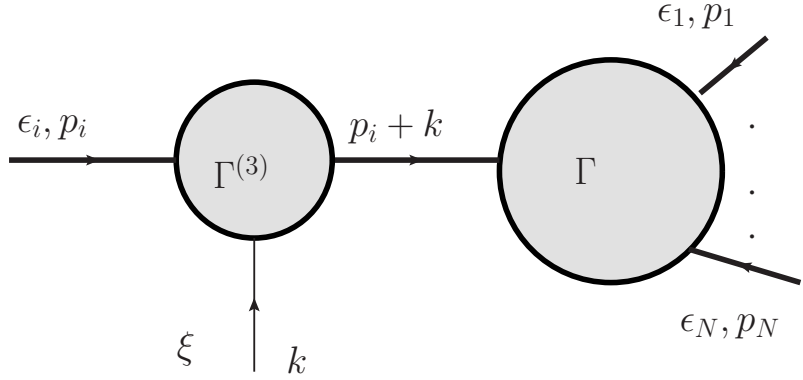}
\caption{ Diagram having soft particle attached to external finite energy leg via 1PI three point vertex which starts contributing at leading order in soft momentum .}\label{soft1a}
\end{center}
\end{figure}
The contribution of Fig.\eqref{soft1a} in terms of vertices and propagator is given by

\be
A_{1}\ &=&\ \lbrace 2p_{i}\cdot k\rbrace^{-1}\ \epsilon_{i}^{T}(p_{i})\ \Gamma^{(3)}(\xi,k;\ p_{i},-p_{i}-k)\ \Xi_{i}(-p_{i}-k)\ \Gamma^{(i)}(p_{i}+k)\ . \label{A1}
\ee
Using the expression for $\Gamma^{(3)}$ given in \eqref{Gamma3} and following the strategy described in sec.\eqref{strategy} up to subleading order we get

\be
&&\Gamma^{(3)}(\xi,k;\ p_{i},-p_{i}-k)\ \Xi_{i}(-p_{i}-k)\non\\
&&=\ \Bigg[2\ e_{\mu}p_{i}^{\mu}\ \ Q_{i}^{T}+\ i\ e_{\mu}\ \mathcal{K}_{i}(-p_{i})\f{\p \Xi_{i}(-p_{i})}{\p p_{i\mu}}\ \ Q_{i}^{T}+\ 2\big(e_{\mu}k_{\nu}-e_{\nu}k_{\mu}\big)\ \mathcal{N}_{(i)}^{\mu\nu}(-p_{i})\ +\ \mathcal{K}_{i}(-p_{i})\mathcal{Q}^{P}_{(i)}(p_{i},k) \non\\
&&\hspace{7mm} +\ 2\varepsilon_{\mu\nu}p_{i}^{\mu}p_{i}^{\nu}\ +\ i\varepsilon_{\mu\nu}p_{i}^{\nu}\ \mathcal{K}_{i}(-p_{i})\ \f{\p \Xi_{i}(-p_{i})}{\p p_{i\mu}}\  +\ 2\ \varepsilon_{b\mu}p_{i}^{\mu}k_{a}\ (J^{ab})^{T}\ +\ \mathcal{K}_{i}(-p_{i})\ \mathcal{Q}_{(i)}^{G}(p_{i},k)\Bigg]\label{Gamma3Xi}
\ee
where,
\be
\mathcal{N}^{\mu\nu}_{(i)}(-p_{i})\ &=&\ \f{i}{4}\ \f{\p \mathcal{K}_{i}(-p_{i})}{\p p_{i\nu}}\ \f{\p \Xi_{i}(-p_{i})}{\p p_{i\mu}}\ Q_{i}^{T}\ +\ \f{1}{2}\ \mathcal{B}^{\mu\nu}_{(i)}(-p_{i})\Xi_{i}(-p_{i})\ ,\label{N}
\ee

\be
\mathcal{Q}^{P}_{(i)}(p_{i},k)\ &=&\ \f{i}{2}\ e_{\mu}k_{\nu}\ \f{\p ^{2}\Xi_{i}(-p_{i})}{\p p_{i\mu}\p p_{i\nu}}\ Q_{i}^{T}, 
\ee

\be
\mathcal{Q}_{(i)}^{G}(p_{i},k)\ &=&\ \f{i}{2}\ (p_{i}.k)\ \varepsilon_{b\mu}\ \f{\p ^{2}\Xi_{i}(-p_{i})}{\p p_{i\mu}\p p_{ib}}\ +\ i\varepsilon_{b\mu}k_{a}\ \f{\p \Xi_{i}(-p_{i})}{\p p_{i\mu}}\ (J^{ab})^{T}\ .
\ee
Substituting \eqref{Gamma3Xi} in \eqref{A1} and using \eqref{st1} after expanding $\Gamma^{(i)}(p_{i}+k)$ up to linear order in $k$ we get
\be
 A_{1}\ 
&=&\ \lbrace p_{i}\cdot k\rbrace^{-1}\ \Bigg[ \ e_{\mu}p_{i}^{\mu}\epsilon_{i}^{T} Q_{i}^{T}\ \Gamma^{(i)}(p_{i})\ +\ e_{\mu}p_{i}^{\mu}\ k_{\rho}\ \epsilon_{i}^{T} Q_{i}^{T}\ \f{\p \Gamma^{(i)}(p_{i})}{\p p_{i\rho}}\ +\ (e_{\mu}k_{\nu}-e_{\nu}k_{\mu})\ \epsilon_{i}^{T}\ \mathcal{N}_{(i)}^{\mu\nu}(-p_{i})\ \Gamma^{(i)}(p_{i})\non\\
&&\ +\ \varepsilon_{\mu\nu}p_{i}^{\mu}p_{i}^{\nu}\ \epsilon_{i}^{T}\Gamma^{(i)}(p_{i})\ +\ \varepsilon_{\mu\nu}p_{i}^{\mu}p_{i}^{\nu}\ k_{\rho}\ \epsilon_{i}^{T}\f{\p \Gamma^{(i)}(p_{i})}{\p p_{i\rho}}\ +\ \varepsilon_{b\mu}p_{i}^{\mu}k_{a}\ \epsilon_{i}^{T}(J^{ab})^{T}\ \Gamma^{(i)}(p_{i})\Bigg]\ .  \label{FinalA1}
\ee

\begin{figure}[H]
\begin{center}
\includegraphics[scale=0.7]{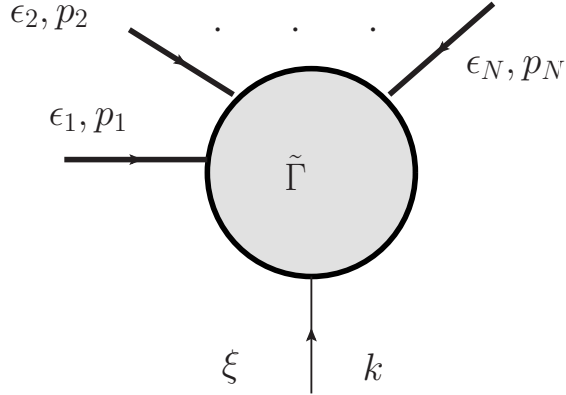}
\caption{ Diagram start contributing at subleading order being the soft particle is not attached to any external line.}\label{soft1b}
\end{center}
\end{figure}

Contribution of Fig.\eqref{soft1b} one can easily read out from \eqref{tildeGamma} to be

\be
A_{2}\ &=&\ -\sum_{j=1}^{N}\ \Bigg[ e_{\mu}(k)\ \epsilon_{j}^{T}\ Q_{j}^{T}\ \f{\p \Gamma^{(j)}(p_{j})}{\p p_{j\mu}}\ +\ \varepsilon_{\mu\nu}(k)p_{j}^{\nu}\ \epsilon_{j}^{T}\ \f{\p \Gamma^{(j)}(p_{j})}{\p p_{j\mu}}\Bigg]\ . \label{A2}
\ee

Now summing over contributions for soft insertion in different legs  for Fig.\eqref{soft1a} and adding the contribution \eqref{A2} we get the subleading soft theorem for one external soft particle

\be
&& \Gamma^{(N+1)}(e,\varepsilon,k;\ \lbrace \epsilon_{i},p_{i}\ \rbrace)\non\\
&&=\ \sum_{i=1}^{N}\ \lbrace p_{i}\cdot k\rbrace^{-1}\ \epsilon_{i}^{T}\ \Big{\lbrace} \ e_{\mu}p_{i}^{\mu}\ Q_{i}^{T}\ +\ \varepsilon_{\mu\nu}p_{i}^{\mu}p_{i}^{\nu}\Big{\rbrace}\ \Gamma^{(i)}(p_{i})\non\\
&&\ +\ \sum_{i=1}^{N}\ \lbrace p_{i}\cdot k\rbrace^{-1}\  e_{\mu}k_{\rho}\ \epsilon_{i}^{T}\ Q_{i}^{T}\ \Bigg{\lbrace} p_{i}^{\mu}\f{\p \Gamma^{(i)}(p_{i})}{\p p_{i\rho}}\ -\ p_{i}^{\rho}\f{\p \Gamma^{(i)}(p_{i})}{\p p_{i\mu}} \Bigg{\rbrace}\non\\
&&\ +\sum_{i=1}^{N}\ \lbrace p_{i}\cdot k\rbrace^{-1}\ \varepsilon_{b\mu}p_{i}^{\mu}\ k_{a}\ \epsilon_{i}^{T}\Bigg{\lbrace} p_{i}^{b}\f{\p \Gamma^{(i)}(p_{i})}{\p p_{ia}}\ -\ p_{i}^{a}\f{\p \Gamma^{(i)}(p_{i})}{\p p_{ib}}\ +\ (J^{ab})^{T}\Gamma^{(i)}(p_{i})\Bigg{\rbrace}\non\\
&&\ +\ \sum_{i=1}^{N}\ \lbrace p_{i}\cdot k\rbrace^{-1}\  (e_{\mu}k_{\nu}-e_{\nu}k_{\mu})\ \epsilon_{i}^{T}\ \mathcal{N}_{(i)}^{\mu\nu}(-p_{i})\ \Gamma^{(i)}(p_{i})\ . \label{singlesoft}
\ee
From here, we can recover the single soft graviton and photon theorems respectively by setting photon and graviton polarisation to be zero. We note that the theory dependent term $\mathcal{N}_{(i)}^{\mu\nu}(-p_{i})$ come from the soft photon coupling via its field strength. We compare this non-universal piece with the result derived in \cite{1611.07534} in Appendix \ref{appB} for two kind of non-minimal couplings. We will eventually see that the same non-universal piece appears in the multiple soft theorem result.

The above result is invariant under the following gauge transformations after using charge, momentum and angular momentum conservation\footnote{Exact statement of momentum, angular momentum and charge conservation relations are given in eq.\eqref{momentum},\eqref{angularmomentum} and \eqref{charge}.} 
\be
&&e_{\mu}(k)\ \rightarrow e_{\mu}(k)\ +\ \lambda(k)\ k_{\mu}\ , \\
&& \varepsilon_{\mu\nu}(k)\ \rightarrow \varepsilon_{\mu\nu}(k)\ +\ \zeta_{\mu}(k)k_{\nu}\ +\ \zeta_{\nu}(k)k_{\mu}
\ee
for arbitrary scalar function $\lambda(k)$ and arbitrary vector function $\zeta_{\mu}(k)$ satisfying $\zeta_{\mu}(k)k^{\mu}=0$.

\end{section}

\begin{section}{Subleading soft theorem for two external soft particles}\label{2soft}

We consider two external soft particles with polarisations and and momenta $(\xi_{1},k_{1})\equiv (e_{1},\varepsilon_{1},k_{1})$ and $(\xi_{2},k_{2})\equiv (e_{2},\varepsilon_{2},k_{2})$. The leading contribution comes from the kind of diagrams having both the soft particles attached to external leg via $\Gamma^{(3)}$ vertices as shown in Fig.\eqref{soft2a} and Fig.\eqref{soft2c}. The other diagrams shown in Fig.\eqref{soft2d},\eqref{soft2f} and \eqref{soft2g} start contributing from subleading order in soft momenta.

\begin{figure}[H]
\begin{center}
\includegraphics[scale=0.7]{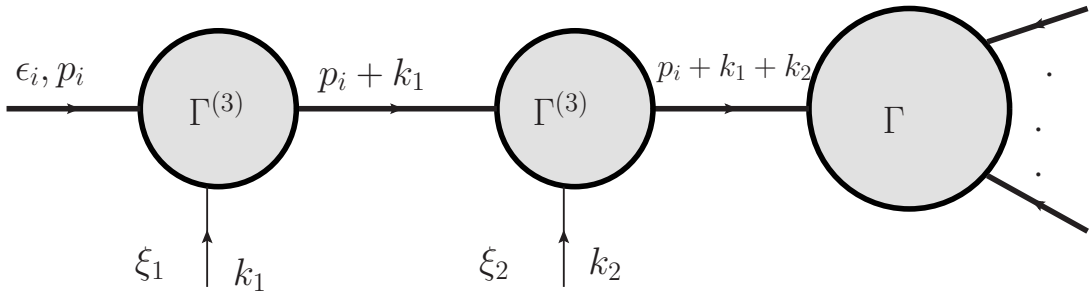}
\caption{ Diagram with two soft particles attached to same leg start contributing at leading order in soft momenta.}\label{soft2a}
\end{center}
\end{figure}

We begin with the evaluation of the contribution of Fig.\eqref{soft2a}, which in terms of vertices and propagators has the form :

\be
B_{1}\ &=&\ \lbrace 2p_{i}\cdot k_{1}\rbrace^{-1}\lbrace 2p_{i}\cdot (k_{1}+k_{2})+2k_{1}\cdot k_{2}\rbrace^{-1}\ \epsilon_{i}^{T}\ \Gamma^{(3)}(\xi_{1},k_{1};\ p_{i},-p_{i}-k_{1} )\ \Xi_{i}(-p_{i}-k_{1})\non\\[7pt]
&& \times\ \Gamma^{(3)}(\xi_{2},k_{2};\ p_{i}+k_{1},-p_{i}-k_{1}-k_{2} )\ \Xi_{i}(-p_{i}-k_{1}-k_{2})\ \Gamma^{(i)}(p_{i}+k_{1}+k_{2})\ . 
\ee
Using eq.\eqref{Gamma3Xi}, we get

\be
B_{1}\ &=&\ \lbrace 2p_{i}\cdot k_{1}\rbrace^{-1}\lbrace 2p_{i}\cdot (k_{1}+k_{2})+2k_{1}\cdot k_{2}\rbrace^{-1}\ \epsilon_{i}^{T}\non\\
&&\ \Bigg[2\ e_{1,\mu}p_{i}^{\mu}\ Q_{i}^{T}\ +\ i\ e_{1,\mu}\ \mathcal{K}_{i}(-p_{i})\f{\p \Xi_{i}(-p_{i})}{\p p_{i\mu}}\ Q_{i}^{T}\ +\ 2\big(e_{1,\mu}k_{1\nu}-e_{1,\nu}k_{1\mu}\big)\ \mathcal{N}_{(i)}^{\mu\nu}(-p_{i})\non\\
&&\ +\ \mathcal{K}_{i}(-p_{i})\mathcal{Q}^{P}_{(i)}(p_{i},k_{1}) \ +\ 2\varepsilon_{1,\mu\nu}p_{i}^{\mu}p_{i}^{\nu}\ +\ i\varepsilon_{1,\mu\nu}p_{i}^{\nu}\ \mathcal{K}_{i}(-p_{i})\ \f{\p \Xi_{i}(-p_{i})}{\p p_{i\mu}}\  +\ 2\ \varepsilon_{1,b\mu}p_{i}^{\mu}k_{1a}\ (J^{ab})^{T}\non\\
&&\ +\ \mathcal{K}_{i}(-p_{i})\ \mathcal{Q}_{(i)}^{G}(p_{i},k_{1})\Bigg]\ \times \ \Bigg[2\ e_{2,\mu}(p_{i}+k_{1})^{\mu}\ Q_{i}^{T}\ +\ i\ e_{2,\mu}\ \mathcal{K}_{i}(-p_{i}-k_{1})\f{\p \Xi_{i}(-p_{i}-k_{1})}{\p p_{i\mu}}\ Q_{i}^{T}\non\\
&&\ +\ 2\big(e_{2,\mu}k_{2\nu}-e_{2,\nu}k_{2\mu}\big)\ \mathcal{N}_{(i)}^{\mu\nu}(-p_{i})
\ +\ \mathcal{K}_{i}(-p_{i})\mathcal{Q}^{P}_{(i)}(p_{i},k_{2}) \ +\ 2\varepsilon_{2,\mu\nu}(p_{i}+k_{1})^{\mu}(p_{i}+k_{1})^{\nu}\non\\
&&\ +\ i\varepsilon_{2,\mu\nu}(p_{i}+k_{1})^{\nu}\ \mathcal{K}_{i}(-p_{i}-k_{1})\ \f{\p \Xi_{i}(-p_{i}-k_{1})}{\p p_{i\mu}}\  +\ 2\ \varepsilon_{2,b\mu}p_{i}^{\mu}k_{2a}\ (J^{ab})^{T}\ +\ \mathcal{K}_{i}(-p_{i})\ \mathcal{Q}_{(i)}^{G}(p_{i},k_{2})\Bigg]\non\\
&&\ \Bigg{\lbrace} \Gamma^{(i)}(p_{i})\ +\ (k_{1}+k_{2})_{\tau}\f{\p \Gamma^{(i)}(p_{i})}{\p p_{i\tau}}\Bigg{\rbrace}\ . 
\ee
Now first substituting explicit form of $\mathcal{N}_{(i)}^{\mu\nu}(-p_{i})$ from \eqref{N} in the above expression and then following the strategy described in Sec.\eqref{strategy} we get:

\be
B_{1}\  &=&\ \lbrace p_{i}\cdot k_{1}\rbrace^{-1}\lbrace p_{i}\cdot (k_{1}+k_{2})+k_{1}\cdot k_{2}\rbrace^{-1}\ \epsilon_{i}^{T}\ \non\\
&&\  \Bigg[\ e_{1,\mu}p_{i}^{\mu}Q_{i}^{T}\ e_{2,\nu}p_{i}^{\nu}Q_{i}^{T} \Gamma^{(i)}(p_{i})\ +\  e_{1,\mu}p_{i}^{\mu}Q_{i}^{T}\ e_{2,\nu}k_{1}^{\nu}Q_{i}^{T}\Gamma^{(i)}(p_{i})\ +\  e_{1,\mu}p_{i}^{\mu}Q_{i}^{T}\ e_{2,\nu}p_{i}^{\nu}Q_{i}^{T}(k_{1}+k_{2})_{\tau}\f{\p \Gamma^{(i)}(p_{i})}{\p p_{i\tau}}\non\\
&& +\ \f{i}{2}\ (p_{i}\cdot k_{1})\ e_{1,\mu}e_{2,\nu}\ \f{\p \mathcal{K}_{i}(-p_{i})}{\p p_{i\mu}}\f{\p \Xi_{i}(-p_{i})}{\p p_{i\nu}}\ Q_{i}^{T}Q_{i}^{T}\Gamma^{(i)}(p_{i})\ +\ (e_{1}\cdot p_{i})\ \big(e_{2,\mu}k_{2\nu}-e_{2,\nu}k_{2\mu}\big)\non\\
&&\ \mathcal{N}_{(i)}^{\mu\nu}(-p_{i})\ Q_{i}^{T}\Gamma^{(i)}(p_{i}) \ +\ (e_{2}\cdot p_{i})\ \big(e_{1,\mu}k_{1\nu}-e_{1,\nu}k_{1\mu}\big)\ \mathcal{N}_{(i)}^{\mu\nu}(-p_{i})\ Q_{i}^{T}\Gamma^{(i)}(p_{i})\non\\[10pt]
&&\ +\ e_{1,\mu}p_{i}^{\mu}\ \varepsilon_{2,\rho\sigma}p_{i}^{\rho}p_{i}^{\sigma}\ Q_{i}^{T}\Gamma^{(i)}(p_{i})\ +\ e_{1,\mu}p_{i}^{\mu}Q_{i}^{T}\  \varepsilon_{2,\rho\sigma}p_{i}^{\rho}p_{i}^{\sigma}\ (k_{1}+k_{2})_{\tau}\f{\p \Gamma^{(i)}(p_{i})}{\p p_{i\tau}}\non\\
&&\ +\ 2\ e_{1,\mu}p_{i}^{\mu}Q_{i}^{T}\ \varepsilon_{2,\rho\sigma}p_{i}^{\rho}k_{1}^{\sigma}\ \Gamma^{(i)}(p_{i})\ +\ e_{1,\mu}p_{i}^{\mu}\  \varepsilon_{2,b\rho}p_{i}^{\rho}k_{2a}\ Q_{i}^{T}(J^{ab})^{T}\Gamma^{(i)}(p_{i})\non\\
&&\ +\ \f{i}{2}\ (p_{i}\cdot k_{1})\ \varepsilon_{2,\rho\sigma}p_{i}^{\rho}e_{1,\mu} \ \f{\p \mathcal{K}_{i}(-p_{i})}{\p p_{i\mu}}\f{\p \Xi_{i}(-p_{i})}{\p p_{i\sigma}}\ Q_{i}^{T}\Gamma^{(i)}(p_{i})\ +\ \varepsilon_{2,\rho\sigma}p_{i}^{\rho}p_{i}^{\sigma}\ \big(e_{1,\mu}k_{1\nu}-e_{1,\nu}k_{1\mu}\big)\non\\
&&\ \mathcal{N}_{(i)}^{\mu\nu}(-p_{i})\ \Gamma^{(i)}(p_{i})\non\\[10pt]
&&\ +\ e_{2,\mu}p_{i}^{\mu}Q_{i}^{T}\ \varepsilon_{1,\rho\sigma}p_{i}^{\rho}p_{i}^{\sigma}\ \Gamma^{(i)}(p_{i})\ +\ e_{2,\mu}p_{i}^{\mu}Q_{i}^{T}\ \varepsilon_{1,\rho\sigma}p_{i}^{\rho}p_{i}^{\sigma}\ (k_{1}+k_{2})_{\tau}\f{\p \Gamma^{(i)}(p_{i})}{\p p_{i\tau}}\non\\
&&\ +\ \varepsilon_{1,\mu\nu}p_{i}^{\mu}p_{i}^{\nu}\ e_{2,\rho}k_{1}^{\rho}\ Q_{i}^{T}\Gamma^{(i)}(p_{i})\ +\ e_{2,\mu}p_{i}^{\mu}\ \varepsilon_{1,b\rho}p_{i}^{\rho}k_{1a}\ (J^{ab})^{T}Q_{i}^{T}\Gamma^{(i)}(p_{i})\non\\
&&\ +\ \f{i}{2}\ (p_{i}\cdot k_{2})\ \varepsilon_{1,\rho\sigma}p_{i}^{\rho}e_{2,\mu} \ \f{\p \mathcal{K}_{i}(-p_{i})}{\p p_{i\mu}}\f{\p \Xi_{i}(-p_{i})}{\p p_{i\sigma}}\ Q_{i}^{T}\Gamma^{(i)}(p_{i})\ +\ \varepsilon_{1,\rho\sigma}p_{i}^{\rho}p_{i}^{\sigma}\ \big(e_{2,\mu}k_{2\nu}-e_{2,\nu}k_{2\mu}\big)\non\\
&&\ \mathcal{N}_{(i)}^{\mu\nu}(-p_{i})\ \Gamma^{(i)}(p_{i})\non\\[10pt]
&&\ +\ \varepsilon_{1,\rho\sigma}p_{i}^{\rho}p_{i}^{\sigma}\ \varepsilon_{2,\mu\nu}p_{i}^{\mu}p_{i}^{\nu}\ \Gamma^{(i)}(p_{i})\ +\ 2\varepsilon_{1,\rho\sigma}p_{i}^{\rho}p_{i}^{\sigma}\ \varepsilon_{2,\mu\nu}p_{i}^{\mu}k_{1}^{\nu}\ \Gamma^{(i)}(p_{i})\ +\ \varepsilon_{1,\rho\sigma}p_{i}^{\rho}p_{i}^{\sigma}\ \varepsilon_{2,b\mu}p_{i}^{\mu}k_{2a}\non\\
&&\ (J^{ab})^{T}\Gamma^{(i)}(p_{i})\ +\ \varepsilon_{2,\rho\sigma}p_{i}^{\rho}p_{i}^{\sigma}\ \varepsilon_{1,b\mu}p_{i}^{\mu}k_{1a}\ (J^{ab})^{T}\Gamma^{(i)}(p_{i})\ +\ \varepsilon_{1,\rho\sigma}p_{i}^{\rho}p_{i}^{\sigma}\ \varepsilon_{2,\mu\nu}p_{i}^{\mu}p_{i}^{\nu}\ (k_{1}+k_{2})_{\tau}\f{\p \Gamma^{(i)}(p_{i})}{\p p_{i\tau}}\non\\
&&\ +\f{i}{2}\ (p_{i}\cdot k_{1})\ \varepsilon_{1,\mu\sigma}\varepsilon_{2,\rho\nu}\ p_{i}^{\sigma}p_{i}^{\nu}\ \f{\p \mathcal{K}_{i}(-p_{i})}{\p p_{i\mu}}\f{\p \Xi_{i}(-p_{i})}{\p p_{i\rho}}\ \Gamma^{(i)}(p_{i})\Bigg]\ . \label{B1}
\ee

To this, we need to add the contribution of the diagram analogous to Fig.\eqref{soft2a} with $(\xi_{1},k_{1})\leftrightarrow (\xi_{2},k_{2})$ exchange,
\be
B_{1}^{\prime}\ &=&\ B_{1}\Big{|}_{(\xi_{1},k_{1})\leftrightarrow (\xi_{2},k_{2})}\label{B1'}\ . 
\ee

\begin{figure}[H]
\begin{center}
\includegraphics[scale=0.6]{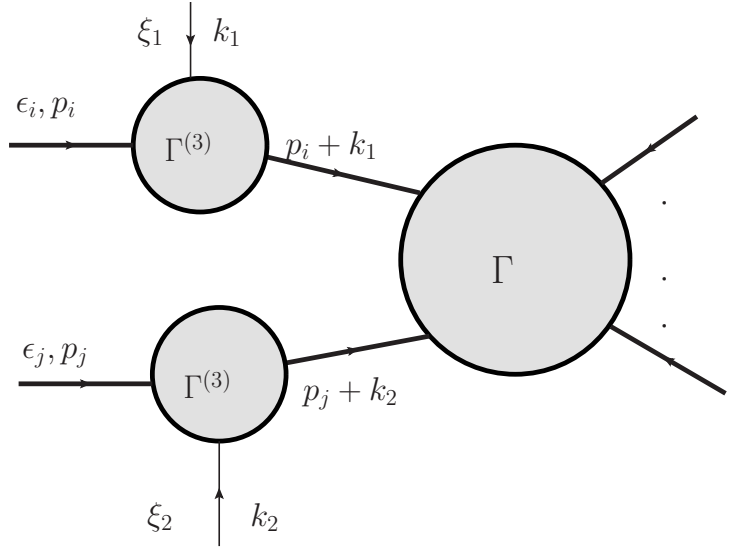}
\caption{  Diagram with two soft particles attached to two different legs start contributing at leading order in soft momenta.}\label{soft2c}
\end{center}
\end{figure}

The diagram where two of the soft particles are attached to two different legs as shown in Fig.\eqref{soft2c}, one can easily compute following the procedure used for computation of $A_{1}$. So using the result of \eqref{FinalA1} for $i$'th and $j$'th leg, the contribution of Fig.\eqref{soft2c} becomes

\be
B_{2}\ &=&\ \lbrace p_{i}\cdot k_{1}\rbrace^{-1}\ \lbrace p_{j}\cdot k_{2}\rbrace^{-1}\ \epsilon_{i}^{T}\epsilon_{j}^{T}\non\\
&&\Bigg[ e_{1,\mu}p_{i}^{\mu}Q_{i}^{T}\ e_{2,\nu}p_{j}^{\nu}Q_{j}^{T}\ \Gamma^{(ij)}(p_{i},p_{j}) \ +\  e_{1,\mu}p_{i}^{\mu}Q_{i}^{T}\ e_{2,\nu}p_{j}^{\nu}Q_{j}^{T}\ \Bigg(k_{1\tau}\ \f{\p \Gamma^{(ij)}(p_{i},p_{j})}{\p p_{i\tau}}\non\\
&&\ +\ k_{2\tau}\ \f{\p \Gamma^{(ij)}(p_{i},p_{j})}{\p p_{j\tau}}\Bigg)\ +\ e_{1,\rho}p_{i}^{\rho}\ \big(e_{2,\mu}k_{2\nu}-e_{2,\nu}k_{2\mu}\big)\ \mathcal{N}_{(j)}^{\mu\nu}(-p_{j})\ Q_{i}^{T}\Gamma^{(ij)}(p_{i},p_{j})\non\\
&&\ +\ e_{2,\rho}p_{j}^{\rho}\ \big(e_{1,\mu}k_{1\nu}-e_{1,\nu}k_{1\mu}\big)\ \mathcal{N}_{(i)}^{\mu\nu}(-p_{i})\ Q_{j}^{T}\Gamma^{(ij)}(p_{i},p_{j})\non\\[10pt]
&&\ +\ e_{1,\mu}p_{i}^{\mu}Q_{i}^{T}\ \varepsilon_{2,\rho\sigma}p_{j}^{\rho}p_{j}^{\sigma}\ \Gamma^{(ij)}(p_{i},p_{j}) \ +\ e_{1,\mu}p_{i}^{\mu}Q_{i}^{T}\ \varepsilon_{2,\rho\sigma}p_{j}^{\rho}p_{j}^{\sigma}\ \Bigg(k_{1\tau}\ \f{\p \Gamma^{(ij)}(p_{i},p_{j})}{\p p_{i\tau}}\ +\ k_{2\tau}\  \f{\p \Gamma^{(ij)}(p_{i},p_{j})}{\p p_{j\tau}}\Bigg)\non\\
&&\ +\ e_{1,\mu}p_{i}^{\mu} Q_{i}^{T}\ \varepsilon_{2,b\sigma}p_{j}^{\sigma}k_{2a}\ (J^{ab})^{T}\ \Gamma^{(ij)}(p_{i},p_{j})\ +\ \varepsilon_{2,\rho\sigma}p_{j}^{\rho}p_{j}^{\sigma}\ \big(e_{1,\mu}k_{1\nu}-e_{1,\nu}k_{1\mu}\big)\ \mathcal{N}_{(i)}^{\mu\nu}(-p_{i})\ \Gamma^{(ij)}(p_{i},p_{j})\non\\[10pt]
&&\ +\ e_{2,\mu}p_{j}^{\mu}Q_{j}^{T}\ \varepsilon_{1,\rho\sigma}p_{i}^{\rho}p_{i}^{\sigma}\ \Gamma^{(ij)}(p_{i},p_{j}) \ +\ e_{2,\mu}p_{j}^{\mu}Q_{j}^{T}\ \varepsilon_{1,\rho\sigma}p_{i}^{\rho}p_{i}^{\sigma}\ \Bigg(k_{1\tau} \f{\p \Gamma^{(ij)}(p_{i},p_{j})}{\p p_{i\tau}}\ +\ k_{2\tau}\ \f{\p \Gamma^{(ij)}(p_{i},p_{j})}{\p p_{j\tau}}\Bigg)\non\\
&&\ +\  e_{2,\mu}p_{j}^{\mu}Q_{j}^{T}\ \varepsilon_{1,b\sigma}p_{i}^{\sigma}k_{1a}\ (J^{ab})^{T}\ \Gamma^{(ij)}(p_{i},p_{j})\ +\ \varepsilon_{1,\rho\sigma}p_{i}^{\rho}p_{i}^{\sigma}\ \big(e_{2,\mu}k_{2\nu}-e_{2,\nu}k_{2\mu}\big)\ \mathcal{N}_{(j)}^{\mu\nu}(-p_{j})\ \Gamma^{(ij)}(p_{i},p_{j})\non\\[10pt]
&&\ +\ \varepsilon_{1,\mu\nu}p_{i}^{\mu}p_{i}^{\nu}\ \varepsilon_{2,\rho\sigma}p_{j}^{\rho}p_{j}^{\sigma}\ \Gamma^{(ij)}(p_{i},p_{j}) \ +\ \varepsilon_{1,\mu\nu}p_{i}^{\mu}p_{i}^{\nu}\ \varepsilon_{2,\rho\sigma}p_{j}^{\rho}p_{j}^{\sigma}\ \Bigg( k_{1\tau} \f{\p \Gamma^{(ij)}(p_{i},p_{j})}{\p p_{i\tau}}\ +\ k_{2\tau} \f{\p \Gamma^{(ij)}(p_{i},p_{j})}{\p p_{j\tau}} \Bigg)\non\\
&&\ +\ \varepsilon_{1,\mu\nu}p_{i}^{\mu}p_{i}^{\nu}\ \varepsilon_{2,b\rho}p_{j}^{\rho}k_{2a}\ (J^{ab})^{T}\ \Gamma^{(ij)}(p_{i},p_{j}) \  +\ \varepsilon_{2,\mu\nu}p_{j}^{\mu}p_{j}^{\nu}\ \varepsilon_{1,b\rho}p_{i}^{\rho}k_{1a}\ (J^{ab})^{T}\ \Gamma^{(ij)}(p_{i},p_{j})\Bigg]\ , \label{B2}
\ee
where, $\Gamma^{(ij)}(p_{i},p_{j})$ is the matrix form of:
\be
\Gamma^{(ij),\alpha_{i}\alpha_{j}}(p_{i},p_{j})\ =\ \Big{\lbrace} \prod_{\substack{l=1\\ l\neq i,j}}^{N}\epsilon_{l,\alpha_{l}}\Big{\rbrace}\ \Gamma^{\alpha_{1}\cdots\alpha_{N}}(p_{1},\cdots,p_{N})\ .
\ee

\begin{figure}[H]
\begin{center}
\includegraphics[scale=0.7]{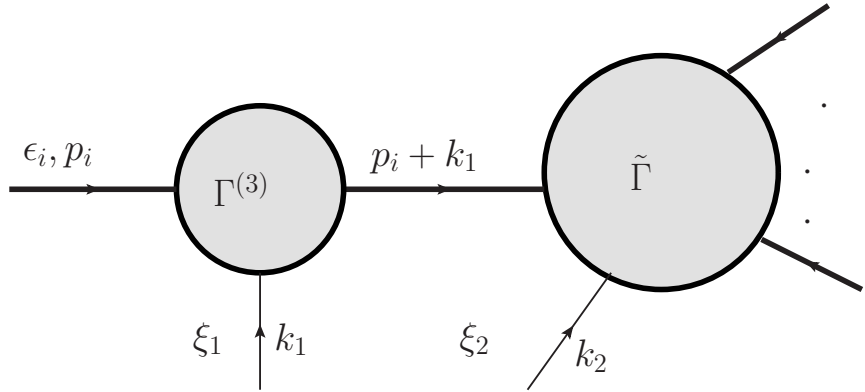}
\caption{ Diagram with one soft particle attached to external leg and another one attached to internal leg, starts contributing at subleading order.}\label{soft2d}
\end{center}
\end{figure}

To evaluate the contribution of Fig.\eqref{soft2d} for $i$'th leg we can directly follow the derivation of $A_{1}$, but we only need to keep the leading contribution here since the diagram starts contributing at subleading order. For soft insertions with polarisation $\xi_{2}$, we can use the result of \eqref{tildeGamma}. Then the contribution of Fig.\eqref{soft2d} turns out to be

\be
B_{3}\ &=& - \lbrace p_{i}\cdot k_{1}\rbrace^{-1} \epsilon_{i}^{T}\Big{\lbrace} e_{1,\mu}p_{i}^{\mu}Q_{i}^{T}\ +\ \varepsilon_{1,\mu\nu}p_{i}^{\mu}p_{i}^{\nu}\Big{\rbrace}\ \sum_{j=1}^{N}\epsilon_{j}^{T}\ 
\Bigg[ e_{2,\mu}\ Q_{j}^{T}\ \f{\p \Gamma^{(ij)}(p_{i},p_{j})}{\p p_{j\mu}}\ +\ \varepsilon_{2,\mu\nu}p_{j}^{\nu} \f{\p \Gamma^{(ij)}(p_{i},p_{j})}{\p p_{j\mu}}\Bigg]\ . \non\\ \label{B3}
\ee
To this we have to add another diagram contribution with $(\xi_{1},k_{1})\leftrightarrow (\xi_{2},k_{2})$ exchange of Fig.\eqref{soft2d}
\be
B_{3}^{\prime}\ &=& - \lbrace p_{i}\cdot k_{2}\rbrace^{-1}\ \epsilon_{i}^{T}\Big{\lbrace}  e_{2,\mu}p_{i}^{\mu}Q_{i}^{T}\ +\ \varepsilon_{2,\mu\nu}p_{i}^{\mu}p_{i}^{\nu}\Big{\rbrace}\ \sum_{j=1}^{N} 
\epsilon_{j}^{T}\Bigg[ e_{1,\mu}Q_{j}^{T}\  \f{\p \Gamma^{(ij)}(p_{i},p_{j})}{\p p_{j\mu}}\ +\ \varepsilon_{1,\mu\nu}p_{j}^{\nu}\f{\p \Gamma^{(ij)}(p_{i},p_{j})}{\p p_{j\mu}}\Bigg]\ . \non\\ \label{B3'}
\ee

\begin{figure}[H]
\begin{center}
\includegraphics[scale=0.7]{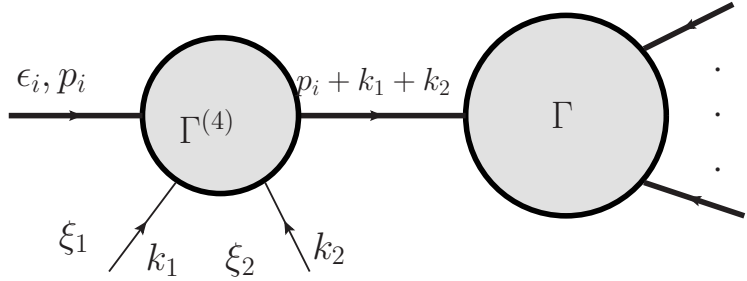}
\caption{ Diagram containing two soft particles attached to hard leg via four point 1PI vertex starts contributing at subleading oredr.}\label{soft2f}
\end{center}
\end{figure}

Fig.\eqref{soft2f} starts contributing at subleading order. So for the vertex and propagator expression we only need to keep leading order pieces in soft momenta expansion. The contribution takes the form

\be
B_{4}\ &=&\ \lbrace 2p_{i}\cdot (k_{1}+k_{2})\rbrace^{-1}\ \epsilon_{i}^{T}\ \Gamma^{(4)}(\xi_{1},k_{1};\ \xi_{2},k_{2};\ p_{i},-p_{i}-k_{1}-k_{2})\ \Xi_{i}(-p_{i}-k_{1}-k_{2})\ \Gamma^{(i)}(p_{i}+k_{1}+k_{2}).\non\\
\ee
Using the vertex expression from \eqref{Gamma4} and following the strategy given in sec.\eqref{strategy} we get

\be
B_{4}\ &=&\ \lbrace p_{i}\cdot (k_{1}+k_{2})\rbrace^{-1}\ \epsilon_{i}^{T}\  \Bigg[  -\f{i}{2} e_{1,\mu}e_{2,\rho}\Bigg{\lbrace} \f{\p \mathcal{K}_{i}(-p_{i})}{\p p_{i\mu}}\f{\p \Xi_{i}(-p_{i})}{\p p_{i\rho}}\ +\ \f{\p \mathcal{K}_{i}(-p_{i})}{\p p_{i\rho}}\f{\p \Xi_{i}(-p_{i})}{\p p_{i\mu}} \Bigg{\rbrace}\ Q_{i}^{T}Q_{i}^{T}\non\\
&&\  -\f{i}{2} \big(e_{1,\mu}\varepsilon_{2,\rho\sigma}p_{i}^{\sigma}+e_{2,\mu}\varepsilon_{1,\rho\sigma}p_{i}^{\sigma}\big)\Bigg{\lbrace} \f{\p \mathcal{K}_{i}(-p_{i})}{\p p_{i\mu}}\f{\p \Xi_{i}(-p_{i})}{\p p_{i\rho}}\ +\ \f{\p \mathcal{K}_{i}(-p_{i})}{\p p_{i\rho}}\f{\p \Xi_{i}(-p_{i})}{\p p_{i\mu}} \Bigg{\rbrace}\ Q_{i}^{T}\non\\
&&\  -\f{i}{2} \varepsilon_{1,\mu\nu}p_{i}^{\nu}\ \varepsilon_{2,\rho\sigma}p_{i}^{\sigma}\Bigg{\lbrace} \f{\p \mathcal{K}_{i}(-p_{i})}{\p p_{i\mu}}\f{\p \Xi_{i}(-p_{i})}{\p p_{i\rho}}\ +\ \f{\p \mathcal{K}_{i}(-p_{i})}{\p p_{i\rho}}\f{\p \Xi_{i}(-p_{i})}{\p p_{i\mu}} \Bigg{\rbrace}\non\\
&&\ -\ (e_{1}\cdot e_{2})Q_{i}^{T}Q_{i}^{T}\ -\ 2\ (p_{i}.\varepsilon_{1}\cdot e_{2})Q_{i}^{T}\ -2\ (p_{i}\cdot \varepsilon_{2}\cdot e_{1})Q_{i}^{T}\ -\ 2(p_{i}\cdot \varepsilon_{1}\cdot \varepsilon_{2}\cdot p_{i}) \Bigg]\ \Gamma^{(i)}(p_{i})\ .\non\\ \label{B4}
\ee

\begin{figure}[H]
\begin{center}
\includegraphics[scale=0.6]{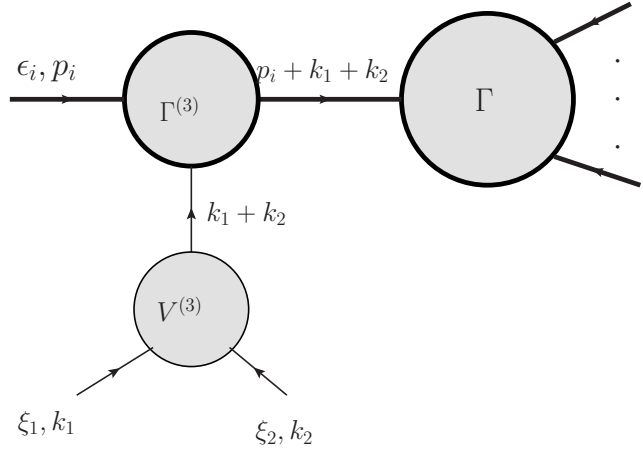}
\caption{ Diagram containing vartex having three soft particle interection starts contributing at subleading order.}\label{soft2g}
\end{center}
\end{figure}

Fig.\eqref{soft2g} also starts contributing from subleading order. Using the three soft particle vertex defined in \eqref{V3} after stripping off the arbitrary off-shell polarisation tensor and contracting with proper internal soft propagator, the contribution turns out to be :

\be
B_{5} &=& \lbrace p_{i}\cdot (k_{1}+k_{2})\rbrace^{-1}\ \lbrace k_{1}\cdot k_{2}\rbrace^{-1} \epsilon_{i}^{T}\non\\
&&\ \Bigg[\ -\ \Big{\lbrace}\ -\f{1}{D-2}\ p_{i}^{2}\ (k_{1}\cdot k_{2})\ (e_{1}\cdot e_{2})\ +\ \f{1}{D-2}\ p_{i}^{2}\ (e_{1}\cdot k_{2})(e_{2}\cdot k_{1})\non\\
&&\ +\ (e_{1}\cdot e_{2})\ (p_{i}\cdot k_{1})(p_{i}\cdot k_{2})\ +\ (k_{1}\cdot k_{2})\ (e_{1}\cdot p_{i})(e_{2}\cdot p_{i})\ -\ (e_{1}\cdot k_{2})\ (e_{2}\cdot p_{i})\ (p_{i}\cdot k_{1})\non\\
&&\ -\ (e_{1}\cdot p_{i})\ (e_{2}\cdot k_{1})\ (p_{i}\cdot k_{2})\Big{\rbrace}\non\\
&& +\Big{\lbrace}(e_{1}\cdot p_{i})\ (k_{1}\cdot \varepsilon_{2}\cdot k_{1})\ +\ (k_{2}\cdot k_{1})\ (p_{i}\cdot \varepsilon_{2}\cdot e_{1})- (p_{i}\cdot k_{1})\ (e_{1}\cdot \varepsilon_{2}\cdot k_{1}) - (e_{1}\cdot k_{2})\ (p_{i}\cdot \varepsilon_{2}\cdot k_{1})\Big{\rbrace}Q_{i}^{T} \non\\
&& + \Big{\lbrace}(e_{2}\cdot p_{i})\ (k_{2}\cdot \varepsilon_{1}\cdot k_{2})\ +\ (k_{1}\cdot k_{2})\ (p_{i}\cdot \varepsilon_{1}\cdot e_{2})- (p_{i}\cdot k_{2})\ (e_{2}\cdot \varepsilon_{1}\cdot k_{2})-(e_{2}\cdot k_{1})\ (p_{i}\cdot \varepsilon_{1}\cdot k_{2}) \Big{\rbrace}Q_{i}^{T}\non\\
&&\ +\ \Big{\lbrace} -(p_{i}\cdot k_{2})\ (k_{2}\cdot \varepsilon_{1}\cdot \varepsilon_{2}\cdot p_{i})- (p_{i}\cdot k_{1})\ (k_{1}\cdot \varepsilon_{2}\cdot \varepsilon_{1}\cdot p_{i})\ +\ (p_{i}\cdot k_{1})\ (k_{2}\cdot \varepsilon_{1}\cdot \varepsilon_{2}\cdot p_{i})\non\\
&&\ +\ (p_{i}\cdot k_{2})\ (k_{1}\cdot \varepsilon_{2}\cdot \varepsilon_{1}\cdot p_{i})\ -\ \varepsilon_{1,\mu\nu}\ \varepsilon_{2}^{\mu\nu}\ (p_{i}\cdot k_{1})\ (p_{i}\cdot k_{2})\ -\ 2(p_{i}\cdot \varepsilon_{1}\cdot k_{2})\ (p_{i}\cdot \varepsilon_{2}\cdot k_{1})\non\\
&&\ +\ (p_{i}\cdot \varepsilon_{2}\cdot p_{i})\ (k_{2}\cdot \varepsilon_{1}\cdot k_{2})\ +\ (p_{i}\cdot \varepsilon_{1}\cdot p_{i})\ (k_{1}\cdot \varepsilon_{2}\cdot k_{1}) \Big{\rbrace}\ \Bigg]\ \Gamma^{(i)}(p_{i})\ . \label{B5}
\ee

Now we have to perform sum over all external legs $i=1,2,\cdots ,N$ for the contributions  $B_{1},\ B_{1}^{\prime},\ B_{3},\ B_{3}^{\prime},\ B_{4}$ and $B_{5}$ and sum over pair $i,j=1,2,\cdots,N$ for $i\neq j$ for the contribution $B_{2}$. Then adding all the contributions and organising in the standard form we get the expression of double soft theorem when both the particles become soft at the same rate,

\be
&&\ \Gamma^{(N+2)}\big(\xi_{1},k_{1};\ \xi_{2},k_{2};\ \lbrace \epsilon_{i},p_{i}\rbrace\big)\non\\
&&=\ \Big[\sum_{i=1}^{N} \lbrace p_{i}\cdot k_{1}\rbrace^{-1} \epsilon_{i}^{T} \Big{\lbrace} Q_{i}^{T} e_{1,\mu}p_{i}^{\mu}\ +\ \varepsilon_{1,\mu\nu}p_{i}^{\mu}p_{i}^{\nu}\Big{\rbrace}\Big]\ \Big[\sum_{j=1}^{N} \lbrace p_{j}\cdot k_{2}\rbrace^{-1}\epsilon_{j}^{T} \Big{\lbrace} Q_{j}^{T}\ e_{2,\mu}p_{j}^{\mu}\ +\ \varepsilon_{2,\mu\nu}p_{j}^{\mu}p_{j}^{\nu}\Big{\rbrace}\Big]\Gamma^{(ij)}(p_{i},p_{j}) \non\\
&&\ +\ \Big[\sum_{i=1}^{N} \lbrace p_{i}\cdot k_{1}\rbrace^{-1}\epsilon_{i}^{T} \Big{\lbrace} Q_{i}^{T}\ e_{1,\mu}p_{i}^{\mu}\ +\ \varepsilon_{1,\mu\nu}p_{i}^{\mu}p_{i}^{\nu}\Big{\rbrace}\Big]\ \sum_{j=1}^{N} \lbrace p_{j}\cdot k_{2}\rbrace^{-1}\non\\
&&\Bigg{\lbrace} e_{2,\mu}k_{2\nu} \epsilon_{j}^{T}Q_{j}^{T}\Bigg(p_{j}^{\mu} \f{\p \Gamma^{(ij)}(p_{i},p_{j})}{\p p_{j\nu}}\ -p_{j}^{\nu}\f{\p \Gamma^{(ij)}(p_{i},p_{j})}{\p p_{j\mu}}\Bigg) + \big(e_{2,\mu}k_{2\nu}-e_{2,\nu}k_{2\mu}\big) \epsilon_{j}^{T} \mathcal{N}_{(j)}^{\mu\nu}(-p_{j}) \Gamma^{(ij)}(p_{i},p_{j})\ \non\\
&&+\ \varepsilon_{2,b\nu}p_{j}^{\nu}\ k_{2a}\ \epsilon_{j}^{T}\ \Bigg( p_{j}^{b}\f{\p \Gamma^{(ij)}(p_{i},p_{j})}{\p p_{ja}}\ -\ p_{j}^{a}\f{\p \Gamma^{(ij)}(p_{i},p_{j})}{\p p_{jb}}\ +\ (J^{ab})^{T}\ \Gamma^{(ij)}(p_{i},p_{j})\Bigg)\ \Bigg{\rbrace}\non\\
&& +\ \Big[\sum_{i=1}^{N} \lbrace p_{i}\cdot k_{2}\rbrace^{-1}\ \epsilon_{i}^{T} \Big{\lbrace} Q_{i}^{T}\ e_{2,\mu}p_{i}^{\mu}\ +\ \varepsilon_{2,\mu\nu}p_{i}^{\mu}p_{i}^{\nu}\Big{\rbrace}\Big]\  \sum_{j=1}^{N} \lbrace p_{j}\cdot k_{1}\rbrace^{-1}\non\\
&&\ \Bigg{\lbrace} e_{1,\mu}k_{1\nu} \epsilon_{j}^{T}Q_{j}^{T}\Bigg(p_{j}^{\mu} \f{\p \Gamma^{(ij)}(p_{i},p_{j})}{\p p_{j\nu}} -p_{j}^{\nu}\f{\p \Gamma^{(ij)}(p_{i},p_{j})}{\p p_{j\mu}}\Bigg) + \big(e_{1,\mu}k_{1\nu}-e_{1,\nu}k_{1\mu}\big) \epsilon_{j}^{T} \mathcal{N}_{(j)}^{\mu\nu}(-p_{j}) \Gamma^{(ij)}(p_{i},p_{j}) \non\\
&&+\ \varepsilon_{1,b\nu}p_{j}^{\nu}\ k_{1a}\ \epsilon_{j}^{T}\ \Bigg( p_{j}^{b}\f{\p \Gamma^{(ij)}(p_{i},p_{j})}{\p p_{ja}}-\ p_{j}^{a}\f{\p \Gamma^{(ij)}(p_{i},p_{j})}{\p p_{jb}}\ +\ (J^{ab})^{T}\ \Gamma^{(ij)}(p_{i},p_{j})\Bigg)\ \Bigg{\rbrace}\non\\
&& +\ \sum_{i=1}^{N}\ \lbrace p_{i}\cdot (k_{1}+k_{2}) \rbrace^{-1}\ \epsilon_{i}^{T} \Big{\lbrace} \mathcal{M}_{pp}\big(p_{i};\ e_{1},k_{1};\ e_{2},k_{2}\big)\ +\ \mathcal{M}_{pg}\big(p_{i};\ e_{1},k_{1}; \varepsilon_{2},k_{2}\big)\non\\
&& +\ \mathcal{M}_{pg}\big(p_{i};\ e_{2},k_{2}; \varepsilon_{1},k_{1}\big)\ +\ \mathcal{M}_{gg}\big(p_{i}; \ \varepsilon_{1},k_{1};\ \varepsilon_{2},k_{2}\ \big) \Big{\rbrace}\ \Gamma^{(i)}(p_{i}) \ . \label{doublesoft}
\ee
where
\be
&&\ \mathcal{M}_{pp}\big(p_{i};\ e_{1},k_{1};\ e_{2},k_{2}\big)\non\\ [15pt]
&&=\  Q_{i}^{T}Q_{i}^{T}\ (p_{i}\cdot k_{1})^{-1}\ (p_{i}\cdot k_{2})^{-1}\  \Big[-\ (k_{1}\cdot k_{2})\ (e_{1}\cdot p_{i})\ (e_{2}\cdot p_{i})\ +\ (p_{i}\cdot k_{2})\ (e_{1}\cdot p_{i})\ (e_{2}\cdot k_{1})\non\\
&&\ +\ (p_{i}\cdot k_{1})\ (e_{2}\cdot p_{i})\ (e_{1}\cdot k_{2})\ -\ (e_{1}\cdot e_{2})\ (p_{i}\cdot k_{1})\ (p_{i}\cdot k_{2})\Big]\non\\
&&\ -\ (k_{1}\cdot k_{2})^{-1} \Bigg[\ -\f{1}{D-2}\ p_{i}^{2}\ (k_{1}\cdot k_{2})\ (e_{1}\cdot e_{2})\ +\ \f{1}{D-2}\ p_{i}^{2}\ (e_{1}\cdot k_{2})\ (e_{2}\cdot k_{1})
\ +\ (e_{1}\cdot e_{2})\ (p_{i}\cdot k_{1})\non\\
&&\ (p_{i}\cdot k_{2})\ +\ (k_{1}\cdot k_{2})\ (e_{1}\cdot p_{i})(e_{2}\cdot p_{i})\ -\ (e_{1}\cdot k_{2})\ (e_{2}\cdot p_{i})\ (p_{i}\cdot k_{1})\ -\ (e_{1}\cdot p_{i})\ (e_{2}\cdot k_{1})\ (p_{i}\cdot k_{2})\Bigg],\non\\\label{Mpp}
\ee

\be
&&\mathcal{M}_{pg}\big(p_{i};\ e_{1},k_{1}; \varepsilon_{2},k_{2}\big)\ \non\\[15pt]
&&=\  Q_{i}^{T}\ \Bigg[(p_{i}\cdot k_{1})^{-1}\ (p_{i}\cdot k_{2})^{-1}\  \Bigg(2\ (e_{1}\cdot p_{i})\ (p_{i}\cdot k_{2})\ (p_{i}\cdot \varepsilon_{2}\cdot k_{1})\ +\ (e_{1}\cdot k_{2})\ (p_{i}\cdot k_{1})\ (p_{i}\cdot \varepsilon_{2}\cdot p_{i})\non\\
&&\  -\ (e_{1}\cdot p_{i})\ (p_{i}\cdot \varepsilon_{2}\cdot p_{i})\ (k_{1}\cdot k_{2})\ -\ 2(p_{i}\cdot \varepsilon_{2}\cdot e_{1})\ (p_{i}\cdot k_{1})\ (p_{i}\cdot k_{2})\Bigg)\non\\
&&\  +\ (k_{1}\cdot k_{2})^{-1}\ \Bigg((e_{1}\cdot p_{i})\ (k_{1}\cdot \varepsilon_{2}\cdot k_{1})\ +\ (k_{1}\cdot k_{2})\ (p_{i}\cdot \varepsilon_{2}\cdot e_{1})\ -\ (p_{i}\cdot k_{1})(e_{1}\cdot \varepsilon_{2}\cdot k_{1})\non\\
&&\ -\ (e_{1}\cdot k_{2})\ (p_{i}\cdot \varepsilon_{2}\cdot k_{1})\Bigg)\Bigg]\label{Mpg}
\ee
and
\be
&&\mathcal{M}_{gg}\big(p_{i}; \ \varepsilon_{1},k_{1};\ \varepsilon_{2},k_{2}\ \big)\non\\
&&=\  (p_{i}\cdot k_{1})^{-1}\ (p_{i}\cdot k_{2})^{-1}\ \Big[ -(k_{1}\cdot k_{2})\ (p_{i}\cdot \varepsilon_{1}\cdot p_{i})\ (p_{i}\cdot \varepsilon_{2}\cdot p_{i}) \ +\ 2(p_{i}\cdot k_{2})\ (p_{i}\cdot \varepsilon_{1}\cdot p_{i})\ (p_{i}\cdot \varepsilon_{2}\cdot k_{1})\non\\
&&\ \ +\ 2(p_{i}\cdot k_{1})\ (p_{i}\cdot \varepsilon_{2}\cdot p_{i})\ (p_{i}\cdot \varepsilon_{1}\cdot k_{2})\ -\ 2(p_{i}\cdot k_{1})\ (p_{i}\cdot k_{2})\ (p_{i}\cdot \varepsilon_{1}\cdot \varepsilon_{2}\cdot p_{i})\Big]\non\\[10pt]
&&\ +\ (k_{1}\cdot k_{2})^{-1}\ \Big[ -(k_{2}\cdot \varepsilon_{1}\cdot \varepsilon_{2}\cdot p_{i})\ (k_{2}\cdot p_{i})\ -\ (k_{1}\cdot \varepsilon_{2}\cdot \varepsilon_{1}\cdot p_{i})\ (k_{1}\cdot p_{i})\non\\
&&\ +\ (k_{2}\cdot \varepsilon_{1}\cdot \varepsilon_{2}\cdot p_{i})\ (k_{1}\cdot p_{i})+(k_{1}\cdot \varepsilon_{2}\cdot \varepsilon_{1}\cdot p_{i})\ (k_{2}\cdot p_{i})\ -\ (\varepsilon_{1,\rho\sigma}\ \varepsilon_{2}^{\rho\sigma})(k_{1}\cdot p_{i})(k_{2}\cdot p_{i})\non\\
&&\ -\ 2(p_{i}\cdot \varepsilon_{1}\cdot k_{2})(p_{i}\cdot \varepsilon_{2}\cdot k_{1})\ +\ (p_{i}\cdot \varepsilon_{2}\cdot p_{i})(k_{2}\cdot \varepsilon_{1}\cdot k_{2})\ +\ (p_{i}\cdot \varepsilon_{1}\cdot p_{i})\ (k_{1}\cdot \varepsilon_{2}\cdot k_{1})\Big]\ . \label{Mgg}
\ee

One can obtain the double soft photon theorem by setting both the graviton polarisations to zero. The contact terms that survive have been clubbed within $\mathcal{M}_{pp}$. Similarly, $\mathcal{M}_{pg}$ contains the contact terms that are present when one external graviton and one external photon are taken to be soft. This soft theorem is obtained by setting for example $e_{1,\mu}=0 = \varepsilon_{2,\mu\nu} $. Similarly, the double soft graviton theorem can be obtained by setting all the photon polarisations to be zero, which agrees with \cite{1707.06803}.

\begin{subsection}{Gauge invariance of the double soft theorem}\label{gaugeinv}
Using momentum, angular momentum and charge conservation of the $N$ particle amplitude, one can easily check that the double soft theorem \eqref{doublesoft} is invariant under  the following gauge transformations:
\be
\delta^{p}_{r}:\hspace{10mm} e_{r,\mu}(k_r)\ \rightarrow e_{r,\mu}(k_r)\ +\ \lambda_r(k_r)\ k_{r\mu}\ ,\hspace{10mm}\hbox{for $r=1,2$.}
\ee

\be
\delta^{g}_{r}:\hspace{8mm} \varepsilon_{r,\mu\nu}(k_r)\ \rightarrow \varepsilon_{r,\mu\nu}(k_r)\ +\ \zeta_{r\mu}(k_{r})k_{r\nu}\ +\ \zeta_{r\nu}(k_{r})k_{r\mu}\ , \hspace{8mm} \ \hbox{with\ $\ \zeta_{r\mu}(k_r)k_r^{\mu}=0$\ for $r=1,2$.}\non\\
\ee
The statements of momentum, angular momentum and charge conservation which needs to be used in the intermediate stages to prove gauge invariance are:
\be
&&\sum_{i=1}^{N}\ p_{i}^{\mu}\ \Gamma^{\alpha_{1}\alpha_{2}\cdots \alpha_{N}}\ =\ 0\ ,  \label{momentum}\\
&&\sum_{i=1}^{N}\ \Bigg[p_{i}^{b}\f{\p \Gamma^{\alpha_{1}\cdots\alpha_{N}}}{\p p_{ia}}\ -\ p_{i}^{a}\f{\p \Gamma^{\alpha_{1}\cdots\alpha_{N}}}{\p p_{ib}}\ +\ (J^{ab})_{\beta_{i}}^{\ \alpha_{i}}\ \Gamma^{\alpha_{1}\cdots\alpha_{i-1}\beta_{i}\alpha_{i+1}\cdots\alpha_{N}
}\Bigg]\ =\ 0\ ,  \label{angularmomentum}\\
&&\sum_{i=1}^{N}\ Q _{\beta_{i}} ^{\ \alpha_{i}}\ \Gamma^{\alpha_{1}\cdots\alpha_{i-1}\beta_{i}\alpha_{i+1}\cdots\alpha_{N}}\ =\ 0\ .\label{charge}
\ee
Under the gauge transformation for photon polarisation it is trivial to check that \eqref{doublesoft} is gauge invariant using charge conservation \eqref{charge}. Checking the gauge invariance under the gauge transformation of graviton is a little non-trivial. There in order to make use of \eqref{momentum}, one has to pass $p_{i}^{\mu}$ through $\p / \p p_{j\nu}$, picking extra contribution $\delta_{ij}\eta^{\mu\nu}$ . The terms which do not vanish by themselves have the following variation under gauge transformation
\be
&&\delta_{1}^{g}\bigg[\frac{1}{p_i.(k_1+k_2)}\epsilon_{i}^{T}\mathcal{M}_{pg}\big(p_{i};\ e_{2},k_{2}; \varepsilon_{1},k_{1}\big)\Gamma^{(i)}(p_{i})\bigg]\ =\ \frac{2}{p_i.k_2}\ e_2.p_i \ \zeta_1\cdot k_2\ \epsilon_{i}^{T}Q_{i}^{T}\ \Gamma^{(i)}(p_{i}) \non\\
&&\delta_{1}^{g}\bigg[\sum_{j=1}^N \frac{\varepsilon_{1}^{\mu\nu} p_{i\mu} p_{i\nu}}{p_i.k_1}\frac{1}{p_j.k_2}   e_{2\sigma} k_{2\lambda}\ \epsilon_{j}^{T}Q_{j}^{T}\Bigg(p_{j}^{\sigma}\frac{\partial\  \Gamma^{(j)}(p_{j})}{\partial p_{j\lambda}}-p_{j}^{\lambda}\frac{\partial\  \Gamma^{(j)}(p_{j})}{\partial p_{j\sigma}}\Bigg)\bigg]=-\frac{2}{p_i.k_2}\ e_2.p_i \ \zeta_1\cdot k_2\ \epsilon_{i}^{T}Q_{i}^{T}\ \Gamma^{(i)}(p_{i})  .\non
\ee
These terms add up to zero and other terms are invariant individually after using \eqref{momentum},\ \eqref{angularmomentum} and \eqref{charge} at different stages. Similarly we have following two variations adding up to zero, 
\be
&&\delta_{1}^{g}\bigg[\frac{1}{p_i.(k_1+k_2)}\mathcal{M}_{gg}\big(p_{i};  \varepsilon_{1},k_{1};\varepsilon_{2},k_{2} \big) \Gamma\bigg]\ = \frac{1}{p_i.(k_1+k_2)}\Big{\lbrace}\frac{2k_1.p_i}{p_i.k_2}\ p_i.\varepsilon_2.p_i\ \zeta_1.k_2\ +\ 2\zeta_1\cdot k_2\ p_i.\varepsilon_2.p_i\Big{\rbrace} \Gamma, \non\\
&&\delta_{1}^{g}\ \bigg[\sum_{j=1}^N \frac{\varepsilon_{1}^{\mu\nu} p_{i\mu} p_{i\nu}}{p_i.k_1}\frac{\varepsilon_{2,\rho\sigma}p_{j}^{\rho} k_{2\lambda}}{p_j.k_2}\ \epsilon_{j}^{T}\Bigg(p_{j}^{\sigma}\frac{\partial\  \Gamma^{(j)}(p_{j})}{\partial p_{j\lambda}}-p_{j}^{\lambda}\frac{\partial\  \Gamma^{(j)}(p_{j})}{\partial p_{j\sigma}}\Bigg)\bigg]\ =\ -\frac{2}{p_i.k_2}\ p_i.\varepsilon_2.p_i \ \zeta_1\cdot k_2\ \Gamma .\non
\ee
Analogous analysis goes through for gauge transformation of $\varepsilon_{2,\mu\nu}$. Also it is clear from above calculation that double soft photon, double soft graviton and soft gravtion-photon theorems are gauge invariant individually.

\end{subsection}
\end{section}

\begin{section}{Subleading soft theorem for multiple soft particles}\label{Msoft}

We will begin by writing down the multiple soft photon-graviton result for $M$ number of soft particles which we will eventually prove:

\be \label{semi-result}
&&\ \Gamma^{(N+M)}\ \big(\lbrace\epsilon_{i},p_{i}\rbrace ;\ \lbrace e_{r},k_{r}\rbrace ;\ \lbrace \varepsilon_{s},k_{s}\rbrace\big)\non\\[15pt]
&=&\ \Bigg{\lbrace}\prod_{j=1}^{N} \epsilon_{j,\alpha_{j}}(p_{j})\Bigg{\rbrace}\ \Bigg[\ \Big[\Bigg{\lbrace}\prod_{r=1}^{M}\ \Bigg(  S^{(0)}_{r}(\gamma) +\  S^{(0)}_{r}(g)\Bigg)\Bigg{\rbrace}\ \Gamma\Big]^{\alpha_{1}\cdots \alpha_{N}}\non\\
&&\ +\ \sum_{s=1}^{M}\ \Big[\ \Bigg{\lbrace}\prod_{\substack{r=1\\r\neq s}}^{M}\   \Bigg(S^{(0)}_{r}(\gamma)\ +\ S^{(0)}_{r}(g)\Bigg)\Bigg{\rbrace}\  \Big(S^{(1)}_{s}(\gamma)\ +\ \mathcal{N}_{s}(\gamma)\ +\ S^{(1)}_{s}(g)\  \Big)\ \Gamma\ \Big]^{\alpha_{1}\cdots\alpha_{N}} \non\\
&&\ +\ \sum_{\substack{r,u=1\\r< u}}^{M}\Big[ \ \Bigg{\lbrace}\prod_{\substack{s=1\\s\neq r,u}}^{M}\ \Bigg( S^{(0)}_{s}(\gamma)\ +\ S^{(0)}_{s}(g)\Bigg)\Bigg{\rbrace}\sum_{i=1}^{N}\ \lbrace p_{i}\cdot(k_{r}+k_{u})\rbrace ^{-1}\non\\
&&\ \Bigg( \mathcal{M}_{pp}\big(p_{i};\ e_{r},k_{r};\ e_{u},k_{u}\big)\ +\ \mathcal{M}_{gg}\big(p_{i};\ \varepsilon_{r},k_{r};\ \varepsilon_{u},k_{u}\big)\ +\ \mathcal{M}_{pg}\big(p_{i};\ e_{r},k_{r};\ \varepsilon_{u},k_{u}\big)\ \non\\
&&\ +\ \mathcal{M}_{pg}\big(p_{i};\ e_{u},k_{u};\ \varepsilon_{r},k_{r}\big)\Bigg)\ \Gamma \Big]^{\alpha_{1}\cdots \alpha_{N}}\ \Bigg] \ .
\ee

where

\be
\Big[S^{(0)}_{r}(\gamma)\Gamma\Big]^{\alpha_{1}\alpha_{2}\cdots\alpha_{N}}\ &=&\ \sum_{i=1}^{N}\ \f{Q_{\beta_{i}}^{\ \alpha_{i}}\ e_{r,\mu} p_{i}^{\mu}}{p_{i}\cdot k_{r}}\ \Gamma^{\alpha_{1}\cdots\alpha_{i-1}\beta_{i}\alpha_{i+1}\cdots\alpha_{N}} , \\[12pt]
\Big[S^{(0)}_{r}(g)\Gamma\Big]^{\alpha_{1}\cdots\alpha_{N}}\ &=&\ \sum_{i=1}^{N}\ \f{\varepsilon_{r,\mu\nu}\  p_{i}^{\mu}p_{i}^{\nu}}{p_{i}\cdot k_{r}} \ \Gamma^{\alpha_{1}\cdots\alpha_{N}}\ ,
\ee
\be
\Big[ S^{(1)}_{r}(\gamma)\ \Gamma\Big]^{\alpha_{1}\cdots\alpha_{N}}\ &=&\ \sum_{i=1}^{N}\ \f{Q_{\beta_{i}}^{\ \alpha_{i}}\ e_{r,\mu}\ k_{r,\nu}}{p_{i}\cdot k_{r}}\ \Bigg(p_{i}^{\mu}\f{\p \Gamma^{\alpha_{1}\cdots\alpha_{i-1}\beta_{i}\alpha_{i+1}\cdots\alpha_{N}}}{\p p_{i\nu}}\ -\ p_{i}^{\nu}\f{\p \Gamma^{\alpha_{1}\cdots\alpha_{i-1}\beta_{i}\alpha_{i+1}\cdots\alpha_{N}}}{\p p_{i\mu}}\Bigg)\ , \\[12pt]
\Big[S^{(1)}_{r}(g)\ \Gamma\Big]^{\alpha_{1}\cdots\alpha_{N}}\ &=&\ \sum_{i=1}^{N}\ \f{\varepsilon_{r,b\mu}\ p_{i}^{\mu}\ k_{ra}}{p_{i}\cdot k_{r}}\ \ \Bigg(p_{i}^{b}\f{\p \Gamma^{\alpha_{1}\cdots \alpha_{N}}}{\p p_{ia}} - p_{i}^{a}\f{\p \Gamma^{\alpha_{1}\cdots\alpha_{N}}}{\p p_{ib}}\ +(J^{ab})_{\beta_{i}}\ ^{\alpha_{i}}\ \Gamma^{\alpha_{1}\cdots \alpha_{i-1}\beta_{i}\alpha_{i+1}\cdots\alpha_{N}}\Bigg),\non\\
\ee
\be
\Big[\mathcal{N}_{s}(\gamma)\ \Gamma\Big]^{\alpha_{1}\cdots\alpha_{N}} &=&\ \sum_{i=1}^{N}\ (p_{i}\cdot k_{s})^{-1}\ \big(e_{s,\mu}k_{s\nu}-e_{s,\nu}k_{s\mu}\big)\ \big[\mathcal{N}_{(i)}^{\mu\nu}(-p_{i})\big]^{\alpha_{i}}\ _{\beta_{i}}\ \Gamma^{\alpha_{1}\cdots\alpha_{i-1}\beta_{i}\alpha_{i+1}
\cdots\alpha_{N}}\non\\
\ee
and the expressions of $\mathcal{M}_{pp}$\ ,\ $\mathcal{M}_{pg}$ \ ,\ $\mathcal{M}_{gg}$ and $\mathcal{N}_{(i)}^{\mu\nu}$ are given in eqs.\eqref{Mpp}, \eqref{Mpg},\eqref{Mgg} and \eqref{N} respectively.

Let us start proving the leading part of \eqref{semi-result}, which can be written as:
\be
\sum_{\substack{A_{1},A_{2},...,A_{N};\ A_{i}\subset \lbrace 1,2,...,M\rbrace \\
A_{i}\cap A_{j}=\phi,\ A_{1}\cup A_{2}\cup ...\cup A_{N}=\lbrace 1,2,...,M\rbrace}}\ \prod_{i=1}^{N}\ \Bigg{\lbrace}\ \epsilon_{i}^{T}\ \prod_{r\in A_{i}}(p_{i}.k_{r})^{-1}\Bigg{\rbrace}\Bigg{\lbrace}\prod_{r\in A_{i}}\Big(Q_{i}^{T}e_{r,\mu}p_{i}^{\mu}\ +\ \varepsilon_{r,\mu\nu}p_{i}^{\mu}p_{i}^{\nu}\Big) \Bigg{\rbrace}\Gamma \ . \label{leading}
\ee

\begin{figure}[H]
\includegraphics[scale=0.7]{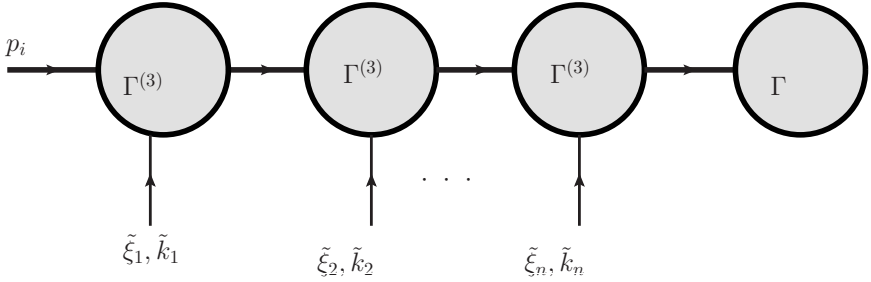}
\caption{Diagram with $n$ number of soft particles attached to $i$'th leg via three point vertices.}\label{multi_1}
\end{figure}
The i'th term in the product represents the multiplication of the leading soft factors for the soft particles attached to i'th hard particle line. To prove the leading order theorem, let us consider $A_{i}$ to be a configuration where n-number of soft particles are attached to i'th leg from outermost to innermost end with momenta $\lbrace \tilde{k}_{1},\tilde{k}_{2},...,\tilde{k}_{n}\rbrace$ and polarisations $\lbrace\tilde{\xi}_{1},\tilde{\xi}_{2},...,\tilde{\xi}_{n}\rbrace$. Now the leading contribution from Fig\eqref{multi_1} comes from manipulations of the leading part of the three point vertices and propagators. Using \eqref{Gamma3Xi}, we get :
\be
\epsilon_{i}^{T}\Bigg{\lbrace}\prod_{r=1}^{n}\Big(Q_{i}^{T}e_{r,\mu}p_{i}^{\mu}\ +\ \varepsilon_{r,\mu\nu}p_{i}^{\mu}p_{i}^{\nu}\Big) \Bigg{\rbrace}\ \lbrace p_{i}\cdot \tilde{k}_{1}\rbrace^{-1}\ \lbrace p_{i}\cdot(\tilde{k}_{1}+\tilde{k}_{2})\rbrace^{-1}\ ...\ \lbrace p_{i}\cdot (\tilde{k}_{1}+\tilde{k}_{2}+...+\tilde{k}_{n})\rbrace^{-1} \ . \non\\
\ee
Now to get the total contribution for the i'th leg one have to sum over all permutations of $\lbrace \tilde{k}_{1},\tilde{k}_{2},...,\tilde{k}_{n}\rbrace$. Using \eqref{I1} we get
\be
&&\epsilon_{i}^{T}\Bigg{\lbrace}\prod_{r=1}^{n}\Big(Q_{i}^{T}\tilde{e}_{r,\mu}p_{i}^{\mu} + \tilde{\varepsilon}_{r,\mu\nu}p_{i}^{\mu}p_{i}^{\nu}\Big) \Bigg{\rbrace} \sum_{perm\lbrace \tilde{k}_{1},\tilde{k}_{2},\cdots,\tilde{k}_{n} \rbrace}\lbrace p_{i}\cdot \tilde{k}_{1}\rbrace^{-1} \lbrace p_{i}\cdot(\tilde{k}_{1}+\tilde{k}_{2})\rbrace^{-1}\cdots \lbrace p_{i}\cdot (\tilde{k}_{1}+\tilde{k}_{2}+...+\tilde{k}_{n})\rbrace^{-1}\non\\
&&=\ \epsilon_{i}^{T}\Bigg{\lbrace}\prod_{r=1}^{n}\Big(Q_{i}^{T}\tilde{e}_{r,\mu}p_{i}^{\mu} + \tilde{\varepsilon}_{r,\mu\nu}p_{i}^{\mu}p_{i}^{\nu}\Big) \Bigg{\rbrace}\Bigg{\lbrace}\prod_{s=1}^{n}(p_{i}\cdot\tilde{k}_{s})^{-1}\Bigg{\rbrace}\non\\
&&=\ \epsilon_{i}^{T}\Bigg{\lbrace}\prod_{r\in A_{i}}\Big(Q_{i}^{T}\ e_{r,\mu}p_{i}^{\mu} + \varepsilon_{r,\mu\nu}p_{i}^{\mu}p_{i}^{\nu}\Big) \Bigg{\rbrace}\Bigg{\lbrace}\prod_{s\in A_{i}}(p_{i}\cdot k_{s})^{-1}\Bigg{\rbrace} \ .
\ee
This proves \eqref{leading}.\\

Now to prove the subleading soft theorem part let us start with the subleading part of Fig.\eqref{multi_1}. Following the structures of \eqref{Gamma3Xi} we can write the expression of Fig.\eqref{multi_1} as:
\be
&& \lbrace 2 p_{i}\cdot \tilde{k}_{1}\rbrace^{-1}\ \lbrace 2p_{i}\cdot(\tilde{k}_{1}+\tilde{k}_{2})+2\tilde{k}_{1}\cdot \tilde{k}_{2}\rbrace^{-1}\ ...\ \Big{\lbrace} 2p_{i}\cdot (\tilde{k}_{1}+\tilde{k}_{2}+...+\tilde{k}_{n})+2\sum_{\substack{r,u=1\\r< u}}^{n}\tilde{k}_{r}\cdot\tilde{k}_{u}\Big{\rbrace}^{-1}\non\\
&& \epsilon_{i}^{T}\ \Bigg[2\ \tilde{e}_{1,\mu}p_{i}^{\mu}Q_{i}^{T}\ +\ i\ \tilde{e}_{1,\mu}\ \mathcal{K}_{i}(-p_{i})\f{\p \Xi_{i}(-p_{i})}{\p p_{i\mu}}Q_{i}^{T}\ +\ 2\big(\tilde{e}_{1,\mu}\tilde{k}_{1\nu}-\tilde{e}_{1,\nu}\tilde{k}_{1\mu}\big)\ \mathcal{N}_{(i)}^{\mu\nu}(-p_{i})\ +\ \mathcal{K}_{i}(-p_{i})\mathcal{Q}^{P}_{(i)}(p_{i},\tilde{k}_{1})\non\\
&&+\ 2\tilde{\varepsilon}_{1,\mu\nu}p_{i}^{\mu}p_{i}^{\nu}\ +\ i\tilde{\varepsilon}_{1,\mu\nu}p_{i}^{\nu}\ \mathcal{K}_{i}(-p_{i})\ \f{\p \Xi_{i}(-p_{i})}{\p p_{i\mu}}\  +\ 2\ \tilde{\varepsilon}_{1,b\mu}\ p_{i}^{\mu}\tilde{k}_{1a}\ (J^{ab})^{T}\ +\ \mathcal{K}_{i}(-p_{i})\ \mathcal{Q}_{(i)}^{G}(p_{i},\tilde{k}_{1})\Bigg]\ \times\non\\
&&\ \Bigg[2\  \tilde{e}_{2,\mu}(p_{i}+\tilde{k}_{1})^{\mu}Q_{i}^{T}\ +\ i\ \tilde{e}_{2,\mu}\ \mathcal{K}_{i}(-p_{i}-\tilde{k}_{1})\f{\p \Xi_{i}(-p_{i}-\tilde{k}_{1})}{\p p_{i\mu}}Q_{i}^{T}\ +\ 2\big(\tilde{e}_{2,\mu}\tilde{k}_{2\nu}-\tilde{e}_{2,\nu}\tilde{k}_{2\mu}\big)\ \mathcal{N}_{(i)}^{\mu\nu}(-p_{i})\non\\
&&\ +\ \mathcal{K}_{i}(-p_{i})\mathcal{Q}^{P}_{(i)}(p_{i},\tilde{k}_{2})
+\  2\tilde{\varepsilon}_{2,\mu\nu}(p_{i}+\tilde{k}_{1})^{\mu}(p_{i}+\tilde{k}_{1})^{\nu}\ +\ i\tilde{\varepsilon}_{2,\mu\nu}(p_{i}+\tilde{k}_{1})^{\nu}\ \mathcal{K}_{i}(-p_{i}-\tilde{k}_{1})\ \f{\p \Xi_{i}(-p_{i}-\tilde{k}_{1})}{\p p_{i\mu}}\non\\
&&\ +\ 2\ \tilde{\varepsilon}_{2,b\mu}\ p_{i}^{\mu}\tilde{k}_{2a}\ (J^{ab})^{T}\ +\ \mathcal{K}_{i}(-p_{i})\ \mathcal{Q}_{(i)}^{G}(p_{i},\tilde{k}_{2})\Bigg]\ \times\cdot\cdot\cdot \times\non\\
&&\ \Bigg[2\  \tilde{e}_{n,\mu}(p_{i}+\tilde{k}_{1}+\tilde{k}_{2}+...+\tilde{k}_{n-1})^{\mu}Q_{i}^{T}\ +\ i\ \tilde{e}_{n,\mu}\ \mathcal{K}_{i}(-p_{i}-\tilde{k}_{1}-\tilde{k}_{2}-...-\tilde{k}_{n-1})\non\\
&&\ \f{\p \Xi_{i}(-p_{i}-\tilde{k}_{1}-\tilde{k}_{2}-...-\tilde{k}_{n-1})}{\p p_{i\mu}}Q_{i}^{T}\ +\ 2\big(\tilde{e}_{n,\mu}\tilde{k}_{n\nu}-\tilde{e}_{n,\nu}\tilde{k}_{n\mu}\big)\ \mathcal{N}_{(i)}^{\mu\nu}(-p_{i})\non\\
&&\ +\ \mathcal{K}_{i}(-p_{i})\mathcal{Q}^{P}_{(i)}(p_{i},\tilde{k}_{n})
+\  2\tilde{\varepsilon}_{n,\mu\nu}(p_{i}+\tilde{k}_{1}+\tilde{k}_{2}+...+\tilde{k}_{n-1})^{\mu}(p_{i}+\tilde{k}_{1}+\tilde{k}_{2}+...+\tilde{k}_{n-1})^{\nu}\non\\
&&\ +\ i\tilde{\varepsilon}_{n,\mu\nu}(p_{i}+\tilde{k}_{1}+\tilde{k}_{2}+...+\tilde{k}_{n-1})^{\nu}\ \mathcal{K}_{i}(-p_{i}-\tilde{k}_{1}-\tilde{k}_{2}-...-\tilde{k}_{n-1})\ \f{\p \Xi_{i}(-p_{i}-\tilde{k}_{1}-\tilde{k}_{2}-...-\tilde{k}_{n-1})}{\p p_{i\mu}}\non\\
&&\ +\ 2\ \tilde{\varepsilon}_{n,b\mu}\ p_{i}^{\mu}\tilde{k}_{na}\ (J^{ab})^{T}\ +\ \mathcal{K}_{i}(-p_{i})\ \mathcal{Q}_{(i)}^{G}(p_{i},\tilde{k}_{n})\Bigg]\ \Gamma^{(i)}(p_{i}+\tilde{k}_{1}+\tilde{k}_{2}+...+\tilde{k}_{n}) \ . \label{3-point_multi}
\ee

Now we will analyse the subleading terms one by one. First consider the contribution from $\tilde{k}_{r}\cdot \tilde{k}_{u}$ terms in the denominator of propagators in the first line. When we expand the denominator of propagators in powers of $\tilde{k}_{r}\cdot \tilde{k}_{u}$ we need to pick up order $\tilde{k}_{r}\cdot \tilde{k}_{u}$ term from one of the propagator.  In the rest of the propagators we can put $\tilde{k}_{r}\cdot \tilde{k}_{u}=0$ and for the rest of the coefficients we can only take the leading contribution. This term turns out to be
\be
&&-\Bigg{\lbrace}\sum_{m=2}^{n}\sum_{\substack{r,u=1\\r<u}}^{m}\f{\tilde{k}_{r}.\tilde{k}_{u}}{p_{i}.(\tilde{k}_{1}+\tilde{k}_{2}+...+\tilde{k}_{m})}\Bigg{\rbrace} \Bigg{\lbrace}\prod_{\ell=1}^{n}\big[p_{i}.(\tilde{k}_{1}+\tilde{k}_{2}+...+\tilde{k}_{\ell})\big]^{-1}\Bigg{\rbrace}\non\\
&&\epsilon_{i}^{T} \Bigg{\lbrace}\prod_{s=1}^{n}\Big(\tilde{e}_{s,\mu}p_{i}^{\mu}Q_{i}^{T}\ +\ \tilde{\varepsilon}_{s,\mu\nu}p_{i}^{\mu}p_{i}^{\nu}\Big) \Bigg{\rbrace}\ \Gamma^{(i)}(p_{i}) \ . \non\\
\ee
Summing over all the permutations of $\tilde{k}_{1},\tilde{k}_{2},...,\tilde{k}_{n} $ and using identity \eqref{I2} we get
\be
-\epsilon_{i}^{T} \Bigg{\lbrace}\prod_{s=1}^{n}\Big(p_{i}\cdot \tilde{k}_{s}\Big)^{-1}\ \Big(\tilde{e}_{s,\mu}p_{i}^{\mu}Q_{i}^{T}\ +\ \tilde{\varepsilon}_{s,\mu\nu}p_{i}^{\mu}p_{i}^{\nu}\Big) \Bigg{\rbrace}\ \sum_{\substack{r,u=1\\ r<u}}^{n}\ \tilde{k}_{r}\cdot\tilde{k}_{u}\ \big{\lbrace}p_{i}\cdot (\tilde{k}_{r}+\tilde{k}_{u})\big{\rbrace}^{-1}\ \Gamma^{(i)}(p_{i})\ . \non\\ \label{sl1}
\ee

Next consider the terms involving $\tilde{e}_{u}$ and $\tilde{\varepsilon}_{u}$ contracted with $\tilde{k}_{r}$ for $r<u$, appearing in the first and fifth terms within the square bracket of \eqref{3-point_multi}. Being subleading if we choose this kind of terms from $u$'th square bracket, from the rest of the square brackets of \eqref{3-point_multi} we only need to pick up the leading terms. This turns out to be
\be
&&\epsilon_{i}^{T}\sum_{\substack{r,u=1\\r<u}}^{n}\ \Bigg{\lbrace}\prod_{\substack{s=1\\s\neq u}}^{n}\Big(\tilde{e}_{s,\mu}p_{i}^{\mu}Q_{i}^{T}\ +\tilde{\varepsilon}_{s,\mu\nu}p_{i}^{\mu}p_{i}^{\nu}\Big)\Bigg{\rbrace}\ \Big(\tilde{e}_{u,\mu}\tilde{k}_{r}^{\mu}Q_{i}^{T}\ +\ 2 \tilde{\varepsilon}_{u,\mu\nu}p_{i}^{\mu}\tilde{k}_{r}^{\nu}\Big)\non\\
&& \Bigg{\lbrace}\prod_{\ell=1}^{n}\big[p_{i}.(\tilde{k}_{1}+\tilde{k}_{2}+...+\tilde{k}_{\ell})\big]^{-1}\Bigg{\rbrace} \  \Gamma^{(i)}(p_{i})\ . 
\ee
Now summing over all possible permutations of $\lbrace\tilde{\xi}_{1},\tilde{k}_{1}\rbrace$,$\lbrace\tilde{\xi}_{2},\tilde{k}_{2}\rbrace$,...,$\lbrace\tilde{\xi}_{n},\tilde{k}_{n}\rbrace$ and using identity \eqref{I3} we get
\be
&&\Bigg{\lbrace}\prod_{\ell=1}^{n}\big(p_{i}\cdot \tilde{k}_{\ell}\big)^{-1}\Bigg{\rbrace}\ \epsilon_{i}^{T} \sum_{\substack{r,u=1\\r<u}}^{n}\ \Big{\lbrace}p_{i}\cdot(\tilde{k}_{r}+\tilde{k}_{u})\Big{\rbrace}^{-1}\ \prod_{\substack{s=1\\s\neq r,u}}^{n}\Big( \tilde{e}_{s,\mu}p_{i}^{\mu}Q_{i}^{T}\ +\  \tilde{\varepsilon}_{s,\mu\nu}\ p_{i}^{\mu}p_{i}^{\nu}\Big) \non\\
&& \Bigg[(p_{i}\cdot \tilde{k}_{u})\ \Big(\ \tilde{e}_{r,\mu}p_{i}^{\mu}Q_{i}^{T}\ +\  \tilde{\varepsilon}_{r,\mu\nu}\ p_{i}^{\mu}p_{i}^{\nu}\Big)\ \Big(\tilde{e}_{u,\rho}\tilde{k}_{r}^{\rho}Q_{i}^{T}\ +\ 2\tilde{\varepsilon}_{u,\rho\sigma}\ p_{i}^{\rho}\ \tilde{k}_{r}^{\sigma}\Big)\non\\
&&+\ (p_{i}\cdot \tilde{k}_{r})\ \Big( \tilde{e}_{u,\mu}p_{i}^{\mu}Q_{i}^{T}\ +\  \tilde{\varepsilon}_{u,\mu\nu}\ p_{i}^{\mu}p_{i}^{\nu}\Big)\ \Big(\tilde{e}_{r,\rho}\tilde{k}_{u}^{\rho}Q_{i}^{T}\ +\ 2\tilde{\varepsilon}_{r,\rho\sigma}\ p_{i}^{\rho}\ \tilde{k}_{u}^{\sigma}\Big)\Bigg]\  \Gamma^{(i)}(p_{i})\ . \non\\ \label{sl2}
\ee

To analyse the rest of the subleading terms in \eqref{3-point_multi} we need to first substitute $\tilde{k}_{r}\cdot \tilde{k}_{u}=0$ in the first line and remove the contractions of $\tilde{e}_{u,\mu}$ and $\tilde{\varepsilon}_{u,\mu\nu}$ with $\tilde{k}_{r,\mu}$ from the first and fifth terms of each square bracket. Then we need to expand all the $\mathcal{K}_{i}$ and $\Xi_{i}$ up to subleading order in soft momenta. After Taylor expansion it is clear that from the second and sixth terms, the terms involving expansion of $\Xi_{i}$ vanish by successive application of $\epsilon_{i}^{T}\mathcal{K}_{i}(-p_{i})=0$. The terms having contraction of $\tilde{\varepsilon}_{u,\mu\nu}$ with $\tilde{k}_{r}^{\nu}$ in the sixth terms of each square brackets vanish by the same logic. Similarly $\mathcal{Q}_{(i)}^{P}$ and $\mathcal{Q}_{(i)}^{G}$ being linear in soft momenta and $\mathcal{K}_{i}(-p_{i})$ sitting in front of it in the fourth and eighth terms of square bracket, these terms vanish by successive use of $\epsilon_{i}^{T}\mathcal{K}_{i}(-p_{i})=0$. So, from \eqref{3-point_multi} we are left with

\be
&& \lbrace 2 p_{i}\cdot \tilde{k}_{1}\rbrace^{-1}\ \lbrace 2p_{i}\cdot(\tilde{k}_{1}+\tilde{k}_{2})\rbrace^{-1}\ ...\ \Big{\lbrace} 2p_{i}\cdot (\tilde{k}_{1}+\tilde{k}_{2}+...+\tilde{k}_{n})\Big{\rbrace}^{-1}\non\\
&& \epsilon_{i}^{T}\ \Bigg[2 \mathcal{E}_{1}(\gamma)\ +\ 2\mathcal{L}_{1}(\gamma) +\ 2\big(\tilde{e}_{1,\mu}\tilde{k}_{1\nu}-\tilde{e}_{1,\nu}\tilde{k}_{1\mu}\big)\ \mathcal{N}_{(i)}^{\mu\nu}(-p_{i})\non\\
&&\ +\ 2 \mathcal{E}_{1}(g)\ +\ 2\mathcal{L}_{1}(g) +\ 2\ \tilde{\varepsilon}_{1,b\mu}\ p_{i}^{\mu}\tilde{k}_{1a}\ (J^{ab})^{T}\ \Bigg]\ \times\non\\
&&\ \Bigg[2\ \mathcal{E}_{2}(\gamma)\ +\ 2\mathcal{L}_{2}(\gamma) +\ i\ \tilde{e}_{2,\mu}\tilde{k}_{1\rho}\ \f{\p \mathcal{K}_{i}(-p_{i})}{\p p_{i\rho}}\f{\p \Xi_{i}(-p_{i})}{\p p_{i\mu}}Q_{i}^{T}\ +\ 2\big(\tilde{e}_{2,\mu}\tilde{k}_{2\nu}-\tilde{e}_{2,\nu}\tilde{k}_{2\mu}\big)\ \mathcal{N}_{(i)}^{\mu\nu}(-p_{i})\non\\
&&\ +\  2 \mathcal{E}_{2}(g)\ +\ 2\mathcal{L}_{2}(g)\ +\ i\tilde{\varepsilon}_{2,\mu\nu}p_{i}^{\nu}\ \tilde{k}_{1}^{\rho}\ \f{\p \mathcal{K}_{i}(-p_{i})}{\p p_{i\rho}}\f{\p \Xi_{i}(-p_{i})}{\p p_{i\mu}}\ +\ 2\ \tilde{\varepsilon}_{2,b\mu}\ p_{i}^{\mu}\tilde{k}_{2a}\ (J^{ab})^{T}\ \Bigg]\ \times\cdot\cdot\cdot \times\non\\[13pt]
&&\ \Bigg[2\ \mathcal{E}_{n}(\gamma)\ +\ 2\mathcal{L}_{n}(\gamma) +\ i \tilde{e}_{n,\mu}(\tilde{k}_{1}+\tilde{k}_{2}+...+\tilde{k}_{n-1})_{\rho}\ \f{\p \mathcal{K}_{i}(-p_{i})}{\p p_{i\rho}}\f{\p \Xi_{i}(-p_{i})}{\p p_{i\mu}}Q_{i}^{T}\non\\
&&\ +\ 2\big(\tilde{e}_{n,\mu}\tilde{k}_{n\nu}-\tilde{e}_{n,\nu}\tilde{k}_{n\mu}\big)\ \mathcal{N}_{(i)}^{\mu\nu}(-p_{i})\ 
+\  2 \mathcal{E}_{n}(g)\ +\ 2\mathcal{L}_{n}(g)\non\\
&&\ +\ i\tilde{\varepsilon}_{n,\mu\nu}p_{i}^{\nu}\ (\tilde{k}_{1}+\tilde{k}_{2}+...+\tilde{k}_{n-1})_{\rho}\ \f{\p \mathcal{K}_{i}(-p_{i})}{\p p_{i\rho}}\ \f{\p \Xi_{i}(-p_{i})}{\p p_{i\mu}}\ +\ 2\ \tilde{\varepsilon}_{n,b\mu}\ p_{i}^{\mu}\tilde{k}_{na}\ (J^{ab})^{T}\ \Bigg]\ \times\non\\
&&\ \Gamma^{(i)}(p_{i}+\tilde{k}_{1}+\tilde{k}_{2}+...+\tilde{k}_{n}) \ . \label{inter}
\ee

where,
\be
\mathcal{E}_{s}(\gamma)\ =\ \tilde{e}_{s,\mu}p_{i}^{\mu}Q_{i}^{T}\ ,\hspace{10mm}\ \mathcal{L}_{s}(\gamma)\ =\ \f{i}{2}\  \tilde{e}_{s,\mu}\ \mathcal{K}_{i}(-p_{i})\f{\p \Xi_{i}(-p_{i})}{\p p_{i\mu}}Q_{i}^{T}\ , \label{def1}
\ee

\be
\mathcal{E}_{s}(g)\ =\ \tilde{\varepsilon}_{s,\mu\nu}p_{i}^{\mu}p_{i}^{\nu}\ ,\hspace{10mm}\ \mathcal{L}_{s}(g)\ =\ \f{i}{2}\ \tilde{\varepsilon}_{s,\mu\nu}p_{i}^{\nu}\ \mathcal{K}_{i}(-p_{i})\f{\p \Xi_{i}(-p_{i})}{\p p_{i\mu}}\ . \label{def2}
\ee

Now when we multiply all the terms and expand keeping terms only upto subleading order, $\epsilon_{i}^{T}$ operating on $\mathcal{L}_{s}(\gamma)$ and $\mathcal{L}_{s}(g)$ vanishes by \eqref{st1} unless there is some other matrix between them. The matrices that can appear in between $\epsilon_{i}^{T}$ and $\mathcal{L}_{s}(\gamma)$ and/or $\mathcal{L}_{s}(g)$ are proportional to $\p \mathcal{K}_{i}(-p_{i})/\p p_{i\mu}\ \p \Xi_{i}(-p_{i})/\p p_{i\nu}$\ ,\  $\mathcal{N}_{(i)}^{\mu\nu}(-p_{i})$\  or\  $(J^{ab})^{T}$\ . If we choose these kind of matrices from the $r$'th square bracket, then in the expression \eqref{inter} the $\mathcal{L}_{s}(\gamma)$ and $\mathcal{L}_{s}(g)$ vanishes for $s<r$ using \eqref{st1}. For $s>r$, \ $\mathcal{L}_{s}(\gamma)$ and $\mathcal{L}_{s}(g)$ terms are kept untouched. We need to also Taylor expand $\Gamma^{(i)}$ up to linear order in soft momenta and for the subleading order we have to keep only $\mathcal{E}_{s}(\gamma)$ and $\mathcal{E}_{s}(g)$ terms from the square brackets. This leads to

\be
&& \lbrace  p_{i}\cdot \tilde{k}_{1}\rbrace^{-1}\ \lbrace p_{i}\cdot(\tilde{k}_{1}+\tilde{k}_{2})\rbrace^{-1}\ ...\ \Big{\lbrace} p_{i}\cdot (\tilde{k}_{1}+\tilde{k}_{2}+...+\tilde{k}_{n})\Big{\rbrace}^{-1}\non\\
&&\epsilon_{i}^{T}\Bigg[\sum_{r=1}^{n}\ \Bigg{\lbrace}\prod_{s=1}^{r-1}\Big(\mathcal{E}_{s}(\gamma)\ +\ \mathcal{E}_{s}(g)\Big)\Bigg{\rbrace}\ \Bigg{\lbrace}\f{i}{2}\ \tilde{e}_{r,\mu}(\tilde{k}_{1}+\tilde{k}_{2}+...+\tilde{k}_{r-1})_{\rho}\ \f{\p \mathcal{K}_{i}(-p_{i})}{\p p_{i\rho}}\f{\p \Xi_{i}(-p_{i})}{\p p_{i\mu}}Q_{i}^{T}\non\\
&&\ +\ \big(\tilde{e}_{r,\mu}\tilde{k}_{r\nu}-\tilde{e}_{r,\nu}\tilde{k}_{r\mu}\big)\ \mathcal{N}_{(i)}^{\mu\nu}(-p_{i})\  +\ \f{i}{2}\tilde{\varepsilon}_{r,\mu\nu}p_{i}^{\nu}\ (\tilde{k}_{1}+\tilde{k}_{2}+...+\tilde{k}_{r-1})_{\rho}\ \f{\p \mathcal{K}_{i}(-p_{i})}{\p p_{i\rho}}\ \f{\p \Xi_{i}(-p_{i})}{\p p_{i\mu}}\non\\
&&\ +\  \tilde{\varepsilon}_{r,b\mu}\ p_{i}^{\mu}\tilde{k}_{ra}\ (J^{ab})^{T} \Bigg{\rbrace}\ \Bigg]\ \Bigg{\lbrace}\prod_{s=r+1}^{n}\Big(\mathcal{E}_{s}(\gamma)+\mathcal{L}_{s}(\gamma)+\mathcal{E}_{s}(g)+\mathcal{L}_{s}(g)\Big)\Bigg{\rbrace}\ \Gamma^{(i)}(p_{i})\non\\[13pt]
&&\ +\ \lbrace  p_{i}\cdot \tilde{k}_{1}\rbrace^{-1}\ \lbrace p_{i}\cdot(\tilde{k}_{1}+\tilde{k}_{2})\rbrace^{-1}\ ...\ \Big{\lbrace} p_{i}\cdot (\tilde{k}_{1}+\tilde{k}_{2}+\cdots +\tilde{k}_{n})\Big{\rbrace}^{-1}\ \Bigg{\lbrace}\prod_{s=1}^{n}\Big(\mathcal{E}_{s}(\gamma)\ +\ \mathcal{E}_{s}(g)\Big)\Bigg{\rbrace}\non\\
&&\ \ \sum_{r=1}^{n}\ \tilde{k}_{r\rho}\ \epsilon_{i}^{T}\ \f{\p \Gamma^{(i)}(p_{i})}{\p p_{i\rho}}\ . \label{step1}
\ee

Let us first focus on the non-universal term $\mathcal{N}_{(i)}^{\mu\nu}(-p_{i})$ defined in \eqref{N}. This contains  a piece  proportional to $\mathcal{B}^{\mu\nu}_{(i)}(-p_{i})\Xi_{i}(-p_{i})$. This part can be  moved through $\mathcal{L}_{s}(\gamma)$ and $\mathcal{L}_{s}(g)$  in \eqref{step1} using $\Xi_{i}(-p_{i})\mathcal{K}_{i}(-p_{i})=\ i(p_{i}^{2}+M_{i}^{2})=0$ on-shell. The terms proportional to $\f{\p \mathcal{K}_{i}(-p_{i})}{\p p_{i\nu}}\f{\p \Xi_{i}(-p_{i})}{\p p_{i\mu}}$ within $\mathcal{N}^{\mu\nu}_{(i)}(-p_{i})$ can also be moved to the right by first using \eqref{st6} and then \eqref{st2}. For the other terms having $\f{\p \mathcal{K}_{i}(-p_{i})}{\p p_{i\mu}}\f{\p \Xi_{i}(-p_{i})}{\p p_{i\nu}}$ we will not do anything. On the other hand we will move $(J^{ab})^{T}$ part through the $\mathcal{L}_{s}(\gamma)$ and $\mathcal{L}_{s}(g)$ terms\footnote{Here we are using the property that $U(1)$ charge generator $Q$ commutes through spin angular momentum generator $J^{ab}$.} to the extreme right using \eqref{st4} after expanding $\prod\limits_{s=r+1}^{n}\Big(\mathcal{E}_{s}(\gamma)+\mathcal{L}_{s}(\gamma)+\mathcal{E}_{s}(g)+\mathcal{L}_{s}(g)\Big)$ in the following way:
\be
&&\Big(\mathcal{E}_{r+1}(\gamma)+\mathcal{L}_{r+1}(\gamma)+\mathcal{E}_{r+1}(g)+\mathcal{L}_{r+1}(g)\Big)\ \cdots \Big(\mathcal{E}_{n}(\gamma)+\mathcal{L}_{n}(\gamma)+\mathcal{E}_{n}(g)+\mathcal{L}_{n}(g)\Big)\ \non\\
&&=\  \Big(\mathcal{E}_{r
+1}(\gamma)+\mathcal{E}_{r+1}(g)\Big)\cdots \Big(\mathcal{E}_{n}(\gamma)+\mathcal{E}_{n}(g)\Big)\ +\ \sum_{u=r+1}^{n}\Big(\mathcal{E}_{r+1}(\gamma)+\mathcal{E}_{r+1}(g)\Big)\cdots \Big(\mathcal{E}_{u-1}(\gamma)+\mathcal{E}_{u-1}(g)\Big)\  \non\\
&&\ \Big(\mathcal{L}_{u}(\gamma)+\mathcal{L}_{u}(g)\Big)\ \prod\limits_{s=u+1}^{n}\Big(\mathcal{E}_{s}(\gamma)+\mathcal{L}_{s}(\gamma)+\mathcal{E}_{s}(g)+\mathcal{L}_{s}(g)\Big)\ . \label{expansion}
\ee
Now we combine the first term in the expansion \eqref{expansion} with $(J^{ab})^{T}$ coefficient and with $\mathcal{N}_{i}^{\mu\nu}(-p_{i})$ coefficient with the last two lines of \eqref{step1}. For the rest of the expansion we use \eqref{st4} to move $(J^{ab})^{T}$ to the right and use \eqref{st6} and \eqref{st2} to move the $\f{\p \mathcal{K}_{i}(-p_{i})}{\p p_{i\nu}}\f{\p \Xi_{i}(-p_{i})}{\p p_{i\mu}}$ piece within $\mathcal{N}_{i}^{\mu\nu}(-p_{i})$ to the right. After combining similar terms we get:
\be
&&\  \lbrace  p_{i}\cdot \tilde{k}_{1}\rbrace^{-1}\ \lbrace p_{i}\cdot(\tilde{k}_{1}+\tilde{k}_{2})\rbrace^{-1}\ ...\ \Big{\lbrace} p_{i}\cdot (\tilde{k}_{1}+\tilde{k}_{2}+...+\tilde{k}_{n})\Big{\rbrace}^{-1}\ \epsilon_{i}^{T}\ \sum_{r=1}^{n}\Bigg{\lbrace}\prod_{\substack{s=1\\s\neq r}}^{n}\Big(\mathcal{E}_{s}(\gamma)\ +\ \mathcal{E}_{s}(g)\Big)\Bigg{\rbrace} \non\\
&&\ \Bigg[\Big( \tilde{e}_{r,\mu}p_{i}^{\mu}Q_{i}^{T}\ +\ \tilde{\varepsilon}_{r,\mu\nu}p_{i}^{\mu}p_{i}^{\nu}\Big)\  \tilde{k}_{r\rho}\  \f{\p \Gamma^{(i)}(p_{i})}{\p p_{i\rho}}\ +\ \tilde{\varepsilon}_{r,b\mu}\ p_{i}^{\mu}\tilde{k}_{ra}(J^{ab})^{T}\ \Gamma^{(i)}(p_{i})\non\\
&&\ +\  \big(\tilde{e}_{r,\mu}\tilde{k}_{r\nu}-\tilde{e}_{r,\nu}\tilde{k}_{r\mu}\big)\ \mathcal{N}_{(i)}^{\mu\nu}(-p_{i}) \Gamma^{(i)}(p_{i})\Bigg]\non\\[14pt]
&&\ +\ \lbrace  p_{i}\cdot \tilde{k}_{1}\rbrace^{-1}\ \lbrace p_{i}\cdot(\tilde{k}_{1}+\tilde{k}_{2})\rbrace^{-1}\ ...\ \Big{\lbrace} p_{i}\cdot (\tilde{k}_{1}+\tilde{k}_{2}+...+\tilde{k}_{n})\Big{\rbrace}^{-1} \times \non\\
&&\ \f{i}{2}\ \epsilon_{i}^{T}\Bigg[\sum_{r=1}^{n}\ \Bigg{\lbrace}\prod_{s=1}^{r-1}\Big(\mathcal{E}_{s}(\gamma)\ +\ \mathcal{E}_{s}(g)\Big)\Bigg{\rbrace}\ \Bigg{\lbrace}\  \tilde{e}_{r,\mu}(\tilde{k}_{1}+\tilde{k}_{2}+...+\tilde{k}_{r-1})_{\rho}\ \f{\p \mathcal{K}_{i}(-p_{i})}{\p p_{i\rho}}\f{\p \Xi_{i}(-p_{i})}{\p p_{i\mu}}Q_{i}^{T}\non\\
&&\   +\ \tilde{\varepsilon}_{r,\mu\nu}p_{i}^{\nu}\ (\tilde{k}_{1}+\tilde{k}_{2}+...+\tilde{k}_{r-1})_{\rho}\ \f{\p \mathcal{K}_{i}(-p_{i})}{\p p_{i\rho}}\ \f{\p \Xi_{i}(-p_{i})}{\p p_{i\mu}}\Bigg{\rbrace}\Bigg]\non\\
&&\ \times \Bigg{\lbrace}\prod_{s=r+1}^{n}\Big(\mathcal{E}_{s}(\gamma)+\mathcal{L}_{s}(\gamma)+\mathcal{E}_{s}(g)+\mathcal{L}_{s}(g)\Big)\Bigg{\rbrace}\ \Gamma^{(i)}(p_{i})\non\\[13pt]
&&\ +\ \lbrace  p_{i}\cdot \tilde{k}_{1}\rbrace^{-1}\ \lbrace p_{i}\cdot(\tilde{k}_{1}+\tilde{k}_{2})\rbrace^{-1}\ ...\ \Big{\lbrace} p_{i}\cdot (\tilde{k}_{1}+\tilde{k}_{2}+...+\tilde{k}_{n})\Big{\rbrace}^{-1} \times \non\\
&&\ \f{i}{2}\ \epsilon_{i}^{T}\sum_{\substack{r,u=1\\r<u}}^{n}\Bigg{\lbrace}
\prod_{\substack{s=1\\s\neq r}}^{u-1}
\Big(\mathcal{E}_{s}(\gamma)+\mathcal{E}_{s}(g)\Big)\ \Bigg{\rbrace}\ \ \Bigg[\tilde{\varepsilon}_{r,b\mu}p_{i}^{\mu}\tilde{k}_{ra}\ \Big(Q_{i}^{T}\tilde{e}_{u,\rho}\ +\ \tilde{\varepsilon}_{u,\rho\nu}p_{i}^{\nu}\Big)\Bigg{\lbrace}p_{i}^{a}\f{\p \mathcal{K}_{i}(-p_{i})}{\p p_{ib}}\ -\ p_{i}^{b}\f{\p \mathcal{K}_{i}(-p_{i})}{\p p_{ia}}\Bigg{\rbrace}\non\\
&&\ \times \f{\p \Xi_{i}(-p_{i})}{\p p_{i\rho}}\ +\ \Big(\f{i}{4}\ \Big)\ \big(\tilde{e}_{r,\mu}\tilde{k}_{r\nu}-\tilde{e}_{r,\nu}\tilde{k}_{r\mu}\big)\ \Big(Q_{i}^{T}\tilde{e}_{u,\rho}\ +\ \tilde{\varepsilon}_{u,\rho\sigma}p_{i}^{\sigma}\Big)\ \Bigg{\lbrace}2ip_{i}^{\mu}\f{\p \mathcal{K}_{i}(-p_{i})}{\p p_{i\nu}}\ -\ 2ip_{i}^{\nu}\f{\p \mathcal{K}_{i}(-p_{i})}{\p p_{i\mu}}\Bigg{\rbrace}\non\\
&&\ \times \f{\p \Xi_{i}(-p_{i})}{\p p_{i\rho}} \Bigg]Q_{i}^{T}\  \prod_{s=u+1}^{n}\ \Big(\mathcal{E}_{s}(\gamma)+\mathcal{L}_{s}(\gamma)+\mathcal{E}_{s}(g)+\mathcal{L}_{s}(g)\Big)\ \Gamma^{(i)}(p_{i})\ . \label{step2}
\ee
In expression \eqref{step2} we have combined terms in three parts. In the first part we combined the terms where we could move the non-trivial matrices\footnote{ The non-trivial matrices are $\f{\p \mathcal{K}_{i}(-p_{i})}{\p p_{i\mu}} \f{\p \Xi_{i}(-p_{i})}{\p p_{i\nu}}$, $(J^{ab})^{T},$ and $\mathcal{B}_{(i)}^{\mu\nu}(-p_{i})$.} to the extreme right. In the second part we have combined the terms where the non-trivial matrices are not moved at all. In the third part we have combined the terms where we have moved non-trivial matrices to one step right. Now let us write down the third part of \eqref{step2} after possible Lorentz index contractions:
\be
&&\  \lbrace  p_{i}\cdot \tilde{k}_{1}\rbrace^{-1}\ \lbrace p_{i}\cdot(\tilde{k}_{1}+\tilde{k}_{2})\rbrace^{-1}\ ...\ \Big{\lbrace} p_{i}\cdot (\tilde{k}_{1}+\tilde{k}_{2}+...+\tilde{k}_{n})\Big{\rbrace}^{-1} \times \non\\
&&\ \f{i}{2}\ \epsilon_{i}^{T}\sum_{\substack{r,u=1\\r<u}}^{n}\Bigg{\lbrace}
\prod_{\substack{s=1\\s\neq r}}^{u-1}
\Big(\mathcal{E}_{s}(\gamma)+\mathcal{E}_{s}(g)\Big)\ \Bigg{\rbrace}\ \ \Bigg[ (p_{i}\cdot \tilde{k}_{r})\ \tilde{\varepsilon}_{r,b\mu}p_{i}^{\mu}\ \big(Q_{i}^{T}\tilde{e}_{u,\rho}\ +\ \tilde{\varepsilon}_{u,\rho\nu}p_{i}^{\nu}\big)\f{\p \mathcal{K}_{i}(-p_{i})}{\p p_{ib}}\f{\p \Xi_{i}(-p_{i})}{\p p_{i\rho}}\non\\
&&\ -(\tilde{\varepsilon}_{r,b\mu}p_{i}^{\mu}p_{i}^{b})\ \big(Q_{i}^{T}\tilde{e}_{u,\rho}+\tilde{\varepsilon}_{u,\rho\nu}p_{i}^{\nu}\big)\ \tilde{k}_{ra}\ \f{\p \mathcal{K}_{i}(-p_{i})}{\p p_{ia}}\f{\p \Xi_{i}(-p_{i})}{\p p_{i\rho}}\ +\ (p_{i}\cdot \tilde{k}_{r})\  \tilde{e}_{r,\mu}\ \big(Q_{i}^{T}\tilde{e}_{u,\rho}+\tilde{\varepsilon}_{u,\rho\sigma}p_{i}^{\sigma}\big)\non\\
&&\ \f{\p \mathcal{K}_{i}(-p_{i})}{\p p_{i\mu}}\f{\p \Xi_{i}(-p_{i})}{\p p_{i\rho}}Q_{i}^{T}\ -\ ( \tilde{e}_{r,\mu}p_{i}^{\mu})\ \big(Q_{i}^{T}\tilde{e}_{u,\rho}+\tilde{\varepsilon}_{u,\rho\sigma}p_{i}^{\sigma}\big)\ \tilde{k}_{r\nu}\ \f{\p \mathcal{K}_{i}(-p_{i})}{\p p_{i\nu}}\f{\p \Xi_{i}(-p_{i})}{\p p_{i\rho}}Q_{i}^{T}\Bigg]\non\\
&&\  \prod_{s=u+1}^{n}\ \Big(\mathcal{E}_{s}(\gamma)+\mathcal{L}_{s}(\gamma)+\mathcal{E}_{s}(g)+\mathcal{L}_{s}(g)\Big)\Gamma^{(i)}(p_{i})\ . \label{3-part}
\ee 
Now it is quite obvious that second and fourth terms within the square bracket of above expression \eqref{3-part} completely cancels the second part of \eqref{step2}. So after cancellation we are left with the following terms in eq.\eqref{step2}
\be
&&\ \lbrace  p_{i}\cdot \tilde{k}_{1}\rbrace^{-1}\ \lbrace p_{i}\cdot(\tilde{k}_{1}+\tilde{k}_{2})\rbrace^{-1}\ ...\ \Big{\lbrace} p_{i}\cdot (\tilde{k}_{1}+\tilde{k}_{2}+...+\tilde{k}_{n})\Big{\rbrace}^{-1}\ \epsilon_{i}^{T}\ \sum_{r=1}^{n}\Bigg{\lbrace}\prod_{\substack{s=1\\s\neq r}}^{n}\Big(\mathcal{E}_{s}(\gamma)\ +\ \mathcal{E}_{s}(g)\Big)\Bigg{\rbrace} \non\\
&&\  \Bigg[\Big( \tilde{e}_{r,\mu}p_{i}^{\mu}Q_{i}^{T}\ +\ \tilde{\varepsilon}_{r,\mu\nu}p_{i}^{\mu}p_{i}^{\nu}\Big)\  \tilde{k}_{r\rho}\  \f{\p \Gamma^{(i)}(p_{i})}{\p p_{i\rho}}\ +\ \tilde{\varepsilon}_{r,b\mu}\ p_{i}^{\mu}\tilde{k}_{ra}(J^{ab})^{T}\ \Gamma^{(i)}(p_{i})\non\\
&&\ +\  \big(\tilde{e}_{r,\mu}\tilde{k}_{r\nu}-\tilde{e}_{r,\nu}\tilde{k}_{r\mu}\big)\ \mathcal{N}_{(i)}^{\mu\nu}(-p_{i}) \Gamma^{(i)}(p_{i})\Bigg]\non\\[14pt]
&&\ + \lbrace  p_{i}\cdot \tilde{k}_{1}\rbrace^{-1}\ \lbrace p_{i}\cdot(\tilde{k}_{1}+\tilde{k}_{2})\rbrace^{-1}\ ...\ \Big{\lbrace} p_{i}\cdot (\tilde{k}_{1}+\tilde{k}_{2}+...+\tilde{k}_{n})\Big{\rbrace}^{-1} \times \non\\
&&\ \f{i}{2}\ \epsilon_{i}^{T}\sum_{\substack{r,u=1\\r<u}}^{n}\Bigg{\lbrace}
\prod_{\substack{s=1\\s\neq r}}^{u-1}
\Big(\mathcal{E}_{s}(\gamma)+\mathcal{E}_{s}(g)\Big)\ \Bigg{\rbrace}\ \ \Bigg[ (p_{i}\cdot \tilde{k}_{r})\ \Big(\ Q_{i}^{T}\ \tilde{e}_{r,\mu}\ +\ \tilde{\varepsilon}_{r,\mu\nu}p_{i}^{\nu}\ \Big)\ \big(Q_{i}^{T}\ \tilde{e}_{u,\rho}\ +\ \tilde{\varepsilon}_{u,\rho\sigma}p_{i}^{\sigma}\big)\non\\
&&\ \f{\p \mathcal{K}_{i}(-p_{i})}{\p p_{i\mu}}\f{\p \Xi_{i}(-p_{i})}{\p p_{i\rho}} \Bigg] \prod_{s=u+1}^{n}\ \Big(\mathcal{E}_{s}(\gamma)+\mathcal{L}_{s}(\gamma)+\mathcal{E}_{s}(g)+\mathcal{L}_{s}(g)\Big)\Gamma^{(i)}(p_{i})\ . \label{step3}
\ee

Let us first consider the first part of the above expression \eqref{step3} and sum over all permutations of $\lbrace\tilde{\xi}_{1},\tilde{k}_{1}\rbrace\ ,\ \lbrace \tilde{\xi}_{2},\tilde{k}_{2}\rbrace\ \cdots \lbrace \tilde{\xi}_{n},\tilde{k}_{n}\rbrace$.\ Since the summation over $r$ index part is already invariant under this permutation, sum over permutations effectively only acts on the propagator denominator factors and then using \eqref{I1} the first part of \eqref{step3} becomes
\be
&&\ \Bigg{\lbrace} \prod_{\ell=1}^{n}\ (p_{i}\cdot \tilde{k}_{\ell})^{-1}\ \Bigg{\rbrace} \epsilon_{i}^{T}\sum_{r=1}^{n}\Bigg{\lbrace}\prod_{\substack{s=1\\s\neq r}}^{n}\Big(\mathcal{E}_{s}(\gamma)\ +\ \mathcal{E}_{s}(g)\Big)\Bigg{\rbrace} \non\\
&&\  \Bigg[\Big(\ \tilde{e}_{r,\mu}p_{i}^{\mu}Q_{i}^{T}\ +\ \tilde{\varepsilon}_{r,\mu\nu}p_{i}^{\mu}p_{i}^{\nu}\Big)\  \tilde{k}_{r\rho}\  \f{\p \Gamma^{(i)}(p_{i})}{\p p_{i\rho}}\ +\ \tilde{\varepsilon}_{r,b\mu}\ p_{i}^{\mu}\tilde{k}_{ra}(J^{ab})^{T}\ \Gamma^{(i)}(p_{i})\non\\
&&\ +\  \big(\tilde{e}_{r,\mu}\tilde{k}_{r\nu}-\tilde{e}_{r,\nu}\tilde{k}_{r\mu}\big)\ \mathcal{N}_{(i)}^{\mu\nu}(-p_{i}) \ \Gamma^{(i)}(p_{i})\Bigg] \ . 
\ee
Since this expression is already subleading, from the other hard particle legs we only have to keep leading contribution having form $\prod\limits_{r\in A_{j} }\ (p_{j}.k_{r})^{-1}\ \epsilon_{j}^{T}\Big(e_{r,\mu}p_{j}^{\mu}Q_{j}^{T}+\varepsilon_{r,\mu\nu}p_{j}^{\mu}
p_{j}^{\nu}\Big)$. Now we have to sum over all possible distribution of soft particles among the hard legs. Then total contribution of this part turns out to be

\be
&&\Big{\lbrace}\prod_{\ell=1}^{N}\epsilon_{\ell,\alpha_{\ell}}\Big{\rbrace} \sum_{r=1}^{M}\sum_{i=1}^{N}\ (p_{i}.k_{r})^{-1}\Bigg( \prod_{\substack{s=1\\s\neq r}}^{M}\Bigg{\lbrace}\ \sum_{\substack{j=1\\j\neq i}}^{N}(p_{j}\cdot k_{s})^{-1}\ \Big(\ e_{s,\mu}p_{j}^{\mu}Q_{j}^{T}\ +\ \varepsilon_{s,\mu\nu}p_{j}^{\mu}p_{j}^{\nu}\Big)\Bigg{\rbrace}\non\\
&& \Bigg[\Big( e_{r,\mu}p_{i}^{\mu}Q_{i}^{T}\ +\ \varepsilon_{r,\mu\nu}p_{i}^{\mu}p_{i}^{\nu}\Big)\  k_{r\rho}\  \f{\p \Gamma}{\p p_{i\rho}}\ +\ \varepsilon_{r,b\mu}\ p_{i}^{\mu}k_{ra}(J^{ab})^{T}\ \Gamma\non\\
&&\ +\  \big(e_{r,\mu}k_{r\nu}-e_{r,\nu}k_{r\mu}\big)\ \mathcal{N}_{(i)}^{\mu\nu}(-p_{i}) \ \Gamma\Bigg]\ \Bigg)^{\alpha_{1}\alpha_{2}\cdots\alpha_{N}}\  . \non\\ \label{subleading1}
\ee
With this contribution we will add the contribution coming from  sum of the diagrams having one soft particle attached to n-hard particle amplitude via $\widetilde{\Gamma}$ and other $(M-1)$ soft particles attached to external hard particle legs via three point couplings $\Gamma^{(3)}$, analogous to Fig.\eqref{soft2d}. Following the result of \eqref{B3}, the sum of the contributions from these kind of diagrams become
\be
&&\ -\Big{\lbrace}\prod_{\ell=1}^{N}\epsilon_{\ell,\alpha_{\ell}}\Big{\rbrace}\ \sum_{r=1}^{M}\Bigg[ \Bigg{\lbrace}\prod_{\substack{s=1\\s\neq r}}^{M}\sum_{j=1}^{N}(p_{j}.k_{s})^{-1}\ \Big(e_{s,\mu}p_{j}^{\mu}Q_{j}^{T}+\varepsilon_{s,\mu\nu}p_{j}^{\mu}p_{j}^{\nu}\Big)\Bigg{\rbrace}\non\\
&&\  \sum_{i=1}^{N}\ \Bigg(Q_{i}^{T}e_{r,\mu}\ +\ \varepsilon_{r,\mu\nu}p_{i}^{\nu}\Bigg)\  \f{\p \Gamma}{\p p_{i\mu}}\Bigg]^{\alpha_{1}\alpha_{2}\cdots\alpha_{N}} . \non\\ \label{subleading2}
\ee

Now let us analyse the second part of \eqref{step3}, where the $(p_{i}\cdot \tilde{k}_{r})$ factor within the square bracket can be written as:
\be
p_{i}\cdot \tilde{k}_{r}\ =\ p_{i}\cdot (\tilde{k}_{1}+\tilde{k}_{2}+\cdots +\tilde{k}_{r})\ -\ p_{i}\cdot (\tilde{k}_{1}+\tilde{k}_{2}+\cdots +\tilde{k}_{r-1}) \ . 
\ee
Substituting this expression in the second part of \eqref{step3} it is easy to see that first term above cancels the propagator right to $(\tilde{\xi}_{r},\tilde{k}_{r})$ insertion and second term above cancels the propagator left to $(\tilde{\xi}_{r},\tilde{k}_{r})$ insertion. Now we have to sum over all permutations of $\lbrace\tilde{\xi}_{r},\tilde{k}_{r}\rbrace$ but we will achieve this in the following steps. First we will fix all the point of insertions of soft particles in fig.\eqref{multi_1} except the one carrying momentum $\tilde{k}_{r}$ and then sum over all possible insertion of  $(\tilde{\xi}_{r},\tilde{k}_{r})$ left to $(\tilde{\xi}_{u},\tilde{k}_{u})$. Due to pairwise cancellation between terms (similar cancellation happens while proving Ward Identity for gauge theories) we will be left with only the term having insertion of $(\tilde{\xi}_{r},\tilde{k}_{r})$ just left to $(\tilde{\xi}_{u},\tilde{k}_{u})$. To write the second part of \eqref{step3} in a convenient form we will relabel the polarisation and momenta of soft particles except the one having momenta $\tilde{k}_{r}$, from left to right as,
\be
(\hat{\xi}_{1},\hat{k}_{1}) ,\ (\hat{\xi}_{2},\hat{k}_{2}),\cdots ,(\hat{\xi}_{u-2},\hat{k}_{u-2}),\ (\tilde{\xi}_{u},\tilde{k}_{u}),(\hat{\xi}_{u+1},\hat{k}_{u+1}),\cdots , (\hat{\xi}_{n},\hat{k}_{n}) \ . 
\ee

Hence after this relabelling and summing over all  $(\tilde{\xi}_{r},\tilde{k}_{r})$ upto the position left of $(\tilde{\xi}_{u},\tilde{k}_{u})$ for fixed $r$ and $u$ the second part of \eqref{step3} reduces to
\be
&&\ \lbrace  p_{i}\cdot \hat{k}_{1}\rbrace^{-1}\ \lbrace p_{i}\cdot(\hat{k}_{1}+\hat{k}_{2})\rbrace^{-1}\ ...\Big{\lbrace}p_{i}\cdot(\hat{k}_{1}+\cdots \hat{k}_{u-2})\Big{\rbrace}^{-1}\ \Big{\lbrace}p_{i}\cdot (\hat{k}_{1}+\cdots +\hat{k}_{u-2}+\tilde{k}_{r}+\tilde{k}_{u})\Big{\rbrace}^{-1}\non\\
&&\cdots \Big{\lbrace} p_{i}\cdot (\hat{k}_{1}+\cdots +\hat{k}_{u-2}+\tilde{k}_{r}+\tilde{k}_{u}+\hat{k}_{u+1}+\cdots +\hat{k}_{n})\Big{\rbrace}^{-1} \times \non\\
&&\  \epsilon_{i}^{T}\Bigg{\lbrace}
\prod_{\substack{s=1}}^{u-2}
\Big(\widehat{\mathcal{E}}_{s}(\gamma)+\widehat{\mathcal{E}}_{s}(g)\Big)\ \Bigg{\rbrace}\ \ \Bigg[ \f{i}{2}\ \Big(\ Q_{i}^{T}\ \tilde{e}_{r,\mu}\ +\ \tilde{\varepsilon}_{r,\mu\nu}p_{i}^{\nu}\ \Big)\ \big(Q_{i}^{T}\ \tilde{e}_{u,\rho}\ +\ \tilde{\varepsilon}_{u,\rho\sigma}p_{i}^{\sigma}\big)\non\\
&&\ \f{\p \mathcal{K}_{i}(-p_{i})}{\p p_{i\mu}}\f{\p \Xi_{i}(-p_{i})}{\p p_{i\rho}} \Bigg] \Bigg{\lbrace}\prod_{s=u+1}^{n}\ \Big(\widehat{\mathcal{E}}_{s}(\gamma)+\widehat{\mathcal{L}}_{s}(\gamma)+\widehat{\mathcal{E}}_{s}(g)+\widehat{\mathcal{L}}_{s}(g)\Big)\Bigg{\rbrace}\ \Gamma^{(i)}(p_{i})\ , \label{step4}
\ee
where $\widehat{\mathcal{E}}_{s}(\gamma),\ \widehat{\mathcal{L}}_{s}(\gamma),\ \widehat{\mathcal{E}}_{s}(g),\ \widehat{\mathcal{L}}_{s}(g)$
are the same as given in \eqref{def1} and \eqref{def2} with the tilde replaced by hat for the polarisations and momenta. Now with the above expression we also need to add the term with $r$ and $u$ exchange, which 
we can get just by interchanging $\mu$ and $\rho$ indices in the term containing $\f{\p \mathcal{K}_{i}(-p_{i})}{\p p_{i\mu}}\f{\p \Xi_{i}(-p_{i})}{\p p_{i\rho}}$ in \eqref{step4}. Then we have to sum over all permutations considering $r$-th and $u$-th soft particles as single unit. Before doing this let us analyse the remaining two kind of diagrams \eqref{multi_2} and \eqref{multi_3} which also contributes in the subleading order of soft momenta.

\begin{figure}[H]
\includegraphics[scale=0.6]{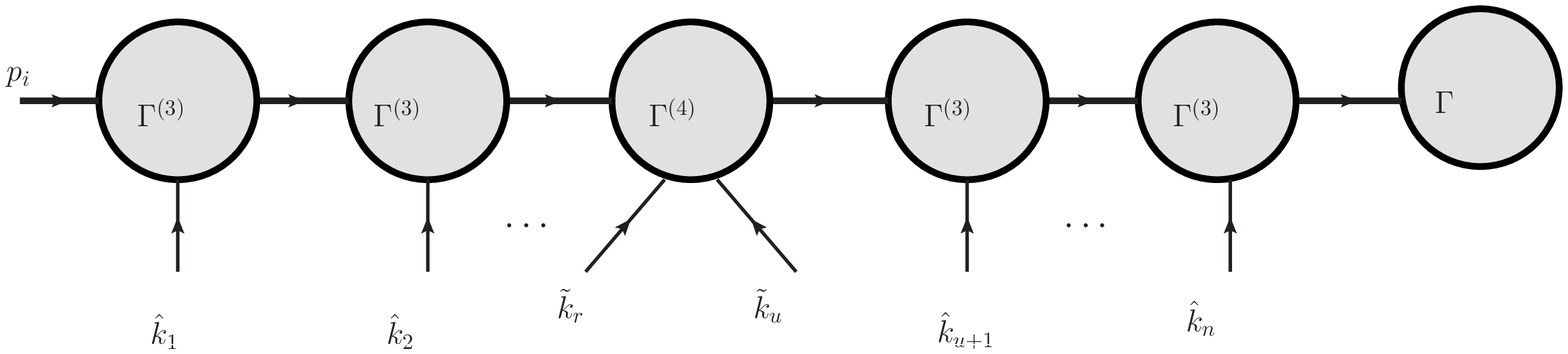}
\caption{diagram contributes in subleading order with two of the soft particles attached via $\Gamma^{(4)}$ \ vertex.}\label{multi_2}
\end{figure}

Insertion of two soft particles via $\Gamma^{(4)}$ vertex in Fig.\eqref{multi_2} implies it has one less propagator than Fig.\eqref{multi_1} and makes the contribution subleading. So from the rest of diagram we only need to pick leading contribution from the vertices and propagators. Rest of the diagram contribution contains $(n-2)$ number of $\Gamma^{(3)}\Xi_{i}$ factors  which one can write down using \eqref{Gamma3Xi} and each of them will contribute $\Big(\widehat{\mathcal{E}}_{s}(\gamma)+\widehat{\mathcal{L}}_{s}(\gamma)+\widehat{\mathcal{E}}_{s}(g)+\widehat{\mathcal{L}}_{s}(g)\Big)$ for $s=1,...,n$ except for $s=r,u$. Then using \eqref{st1} one can drop the $\widehat{\mathcal{L}}_{s}(\gamma)+\widehat{\mathcal{L}}_{s}(g)$ factors appearing left to the vertex, where the soft particle with momentum $\tilde{k}_{r}+\tilde{k}_{u}$ attached to the hard leg. In this way the hard particle polarisation $\epsilon_{i}^{T}$ moved just left to $\Gamma^{(4)}$ vertex and that can be evaluated similarly like \eqref{B4}. Hence the contribution of diagram\eqref{multi_2} turns out to be
\be
&&\ \lbrace  p_{i}\cdot \hat{k}_{1}\rbrace^{-1}\ \lbrace p_{i}\cdot(\hat{k}_{1}+\hat{k}_{2})\rbrace^{-1}\ ...\Big{\lbrace}p_{i}\cdot(\hat{k}_{1}+\cdots \hat{k}_{u-2})\Big{\rbrace}^{-1}\ \Big{\lbrace}p_{i}\cdot (\hat{k}_{1}+\cdots +\hat{k}_{u-2}+\tilde{k}_{r}+\tilde{k}_{u})\Big{\rbrace}^{-1}\non\\
&&\cdots \Big{\lbrace} p_{i}\cdot (\hat{k}_{1}+\cdots +\hat{k}_{u-2}+\tilde{k}_{r}+\tilde{k}_{u}+\hat{k}_{u+1}+\cdots +\hat{k}_{n})\Big{\rbrace}^{-1} \times \non\\
&&\  \epsilon_{i}^{T}\Bigg{\lbrace}
\prod_{\substack{s=1}}^{u-2}
\Big(\widehat{\mathcal{E}}_{s}(\gamma)+\widehat{\mathcal{E}}_{s}(g)\Big)\ \Bigg{\rbrace}\ \ \Bigg[  -\f{i}{2} \Big(\ Q_{i}^{T}\ \tilde{e}_{r,\mu}\ +\ \tilde{\varepsilon}_{r,\mu\nu}p_{i}^{\nu}\ \Big)\ \big(Q_{i}^{T}\ \tilde{e}_{u,\rho}\ +\ \tilde{\varepsilon}_{u,\rho\sigma}p_{i}^{\sigma}\big)\non\\
&&\times \ \Bigg{\lbrace} \f{\p \mathcal{K}_{i}(-p_{i})}{\p p_{i\mu}}\f{\p \Xi_{i}(-p_{i})}{\p p_{i\rho}}\ +\ \f{\p \mathcal{K}_{i}(-p_{i})}{\p p_{i\rho}}\f{\p \Xi_{i}(-p_{i})}{\p p_{i\mu}}\ \Bigg{\rbrace} -\ Q_{i}^{T}Q_{i}^{T}\ (\tilde{e}_{r}\cdot \tilde{e}_{u})\ -\ 2\ Q_{i}^{T}(p_{i}.\tilde{\varepsilon}_{r}\cdot \tilde{e}_{u})\non\\
&&\ -2\ Q_{i}^{T}(p_{i}\cdot \tilde{\varepsilon}_{u}\cdot \tilde{e}_{r})\ -\ 2(p_{i}\cdot \tilde{\varepsilon}_{r}\cdot \tilde{\varepsilon}_{u}\cdot p_{i}) \Bigg] \ \Bigg{\lbrace}\prod_{s=u+1}^{n}\ \Big(\widehat{\mathcal{E}}_{s}(\gamma)+\widehat{\mathcal{L}}_{s}(\gamma)+\widehat{\mathcal{E}}_{s}(g)+\widehat{\mathcal{L}}_{s}(g)\Big)\Bigg{\rbrace}\ \Gamma^{(i)}(p_{i})\ . \non\\\label{step5}
\ee

\begin{figure}[H]
\includegraphics[scale=0.6]{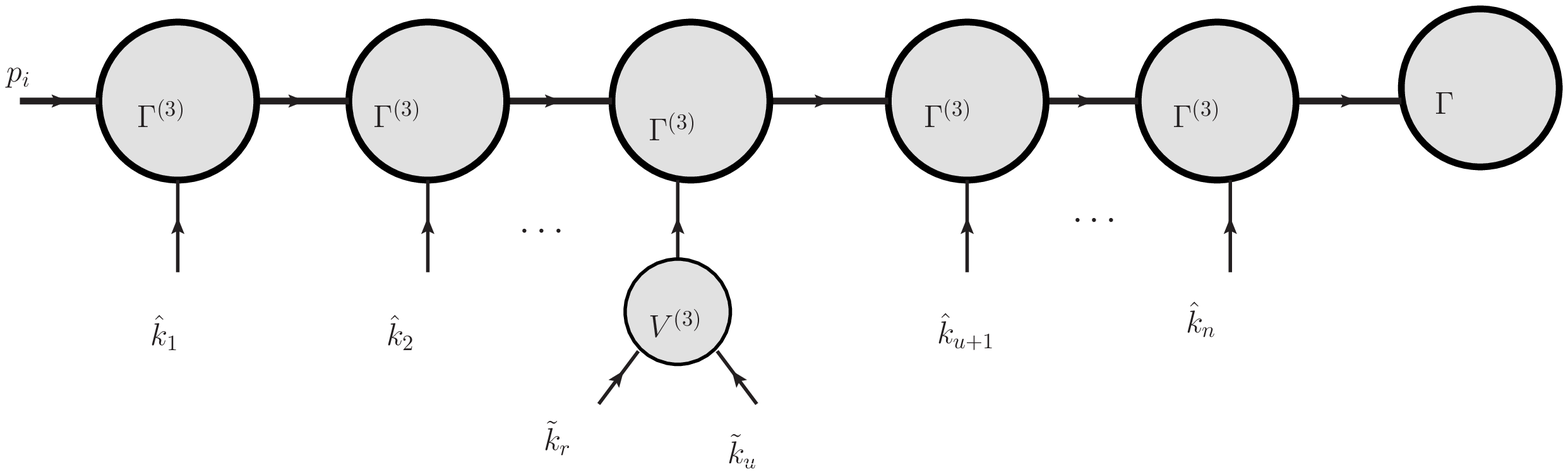}
\caption{diagram contributes in subleading order with two of the soft particles attached via $V^{(3)}$ vertex.}\label{multi_3}
\end{figure}

For the evaluation of Fig.\eqref{multi_3} contribution from the $(n-2)$ number of $\Gamma^{(3)}\Xi_{i}$ will be same as the contribution of Fig.\eqref{multi_2} except here two of the soft particles are attached to the hard leg via $V^{(3)}$ vertex followed by a soft particle propagator with momentum $(\tilde{k}_{r}+\tilde{k}_{u})$ and $\Gamma^{(3)}$ vertex. This can be worked out following \eqref{B5} . The full contribution of Fig.\eqref{multi_3} turns out to be
\be
&&\ \lbrace  p_{i}\cdot \hat{k}_{1}\rbrace^{-1}\ \lbrace p_{i}\cdot(\hat{k}_{1}+\hat{k}_{2})\rbrace^{-1}\ ...\Big{\lbrace}p_{i}\cdot(\hat{k}_{1}+\cdots \hat{k}_{u-2})\Big{\rbrace}^{-1}\ \Big{\lbrace}p_{i}\cdot (\hat{k}_{1}+\cdots +\hat{k}_{u-2}+\tilde{k}_{r}+\tilde{k}_{u})\Big{\rbrace}^{-1}\non\\
&&\cdots \Big{\lbrace} p_{i}\cdot (\hat{k}_{1}+\cdots +\hat{k}_{u-2}+\tilde{k}_{r}+\tilde{k}_{u}+\hat{k}_{u+1}+\cdots +\hat{k}_{n})\Big{\rbrace}^{-1} \ \lbrace \tilde{k}_{r}\cdot \tilde{k}_{u}\rbrace^{-1}\times \non\\
&&\ \epsilon_{i}^{T}\Bigg{\lbrace}
\prod_{\substack{s=1}}^{u-2}
\Big(\widehat{\mathcal{E}}_{s}(\gamma)+\widehat{\mathcal{E}}_{s}(g)\Big)\ \Bigg{\rbrace}\ \Bigg[\ -\ \Big{\lbrace}\ -\f{1}{D-2}\ p_{i}^{2}\ (\tilde{k}_{r}\cdot \tilde{k}_{u})\ (\tilde{e}_{r}\cdot \tilde{e}_{u})\ +\ \f{1}{D-2}\ p_{i}^{2}\ (\tilde{e}_{r}\cdot \tilde{k}_{u})(\tilde{e}_{u}\cdot \tilde{k}_{r})\non\\
&&\ +\ (\tilde{e}_{r}\cdot \tilde{e}_{u})\ (p_{i}\cdot \tilde{k}_{r})(p_{i}\cdot \tilde{k}_{u})\ +\ (\tilde{k}_{r}\cdot \tilde{k}_{u})\ (\tilde{e}_{r}\cdot p_{i})(\tilde{e}_{u}\cdot p_{i})\ -\ (\tilde{e}_{r}\cdot \tilde{k}_{u})\ (\tilde{e}_{u}\cdot p_{i})\ (p_{i}\cdot \tilde{k}_{r})\non\\
&&\ -\ (\tilde{e}_{r}\cdot p_{i})\ (\tilde{e}_{u}\cdot \tilde{k}_{r})\ (p_{i}\cdot \tilde{k}_{u})\Big{\rbrace}\non\\
&& +\ \Big{\lbrace}(\tilde{e}_{r}\cdot p_{i})\ (\tilde{k}_{r}\cdot \tilde{\varepsilon}_{u}\cdot \tilde{k}_{r})\ +\ (\tilde{k}_{u}\cdot \tilde{k}_{r})\ (p_{i}\cdot \tilde{\varepsilon}_{u}\cdot \tilde{e}_{r})\ - (p_{i}\cdot \tilde{k}_{r})\ (\tilde{e}_{r}\cdot \tilde{\varepsilon}_{u}\cdot \tilde{k}_{r})\ -(\tilde{e}_{r}\cdot \tilde{k}_{u})\ (p_{i}\cdot \tilde{\varepsilon}_{u}\cdot \tilde{k}_{r})\ \Big{\rbrace}Q_{i}^{T} \non\\
&& +\  \Big{\lbrace}(\tilde{e}_{u}\cdot p_{i})\ (\tilde{k}_{u}\cdot \tilde{\varepsilon}_{r}\cdot \tilde{k}_{u})\ +\ (\tilde{k}_{r}\cdot \tilde{k}_{u})\ (p_{i}\cdot \tilde{\varepsilon}_{r}\cdot \tilde{e}_{u})\ - (p_{i}\cdot \tilde{k}_{u})\ (\tilde{e}_{u}\cdot \tilde{\varepsilon}_{r}\cdot \tilde{k}_{u})\ - (\tilde{e}_{u}\cdot \tilde{k}_{r})\ (p_{i}\cdot \tilde{\varepsilon}_{r}\cdot \tilde{k}_{u})\ \Big{\rbrace}Q_{i}^{T}\non\\
&&\ +\ \Big{\lbrace} -(p_{i}\cdot \tilde{k}_{u})\ (\tilde{k}_{u}\cdot \tilde{\varepsilon}_{r}\cdot \tilde{\varepsilon}_{u}\cdot p_{i})\ -\ (p_{i}\cdot \tilde{k}_{r})\ (\tilde{k}_{r}\cdot \tilde{\varepsilon}_{u}\cdot \tilde{\varepsilon}_{r}\cdot p_{i})\ +\ (p_{i}\cdot \tilde{k}_{r})\ (\tilde{k}_{u}\cdot \tilde{\varepsilon}_{r}\cdot \tilde{\varepsilon}_{u}\cdot p_{i})\non\\
&&\ +\ (p_{i}\cdot \tilde{k}_{u})\ (\tilde{k}_{r}\cdot \tilde{\varepsilon}_{u}\cdot \tilde{\varepsilon}_{r}\cdot p_{i})\ -\ \tilde{\varepsilon}_{r,\mu\nu}\ \tilde{\varepsilon}_{u}^{\mu\nu}\ (p_{i}\cdot \tilde{k}_{r})\ (p_{i}\cdot \tilde{k}_{u})\ -\ 2(p_{i}\cdot \tilde{\varepsilon}_{r}\cdot \tilde{k}_{u})\ (p_{i}\cdot \tilde{\varepsilon}_{u}\cdot \tilde{k}_{r})\non\\
&&\ +\ (p_{i}\cdot \tilde{\varepsilon}_{u}\cdot p_{i})\ (\tilde{k}_{u}\cdot \tilde{\varepsilon}_{r}\cdot \tilde{k}_{u})\ +\ (p_{i}\cdot \tilde{\varepsilon}_{r}\cdot p_{i})\ (\tilde{k}_{r}\cdot \tilde{\varepsilon}_{u}\cdot \tilde{k}_{r}) \Big{\rbrace}\ \Bigg]\non\\
&&\ \times \Bigg{\lbrace}\prod_{s=u+1}^{n}\ \Big(\widehat{\mathcal{E}}_{s}(\gamma)+\widehat{\mathcal{L}}_{s}(\gamma)+\widehat{\mathcal{E}}_{s}(g)+\widehat{\mathcal{L}}_{s}(g)\Big)\Bigg{\rbrace}\ \Gamma^{(i)}(p_{i})\ . \label{step6}
\ee

Now in the sum of \eqref{step4}, and its $r,u$ exchanged part  and \eqref{step5} the terms having derivative of $\mathcal{K}_{i}$ and $\Xi_{i}$ cancel. Then for the rest of the terms we can drop $\widehat{\mathcal{L}}_{s}(\gamma)+\widehat{\mathcal{L}}_{s}(g)$ for $s=u+1$ to $n$ using \eqref{st1}. Hence the sum over \eqref{step4}, $r,u$ exchanged part of \eqref{step4} , \eqref{step5} and \eqref{step6} turns out to be

\be
&&\ \lbrace  p_{i}\cdot \hat{k}_{1}\rbrace^{-1}\ \lbrace p_{i}\cdot(\hat{k}_{1}+\hat{k}_{2})\rbrace^{-1}\ ...\Big{\lbrace}p_{i}\cdot(\hat{k}_{1}+\cdots \hat{k}_{u-2})\Big{\rbrace}^{-1}\ \Big{\lbrace}p_{i}\cdot (\hat{k}_{1}+\cdots +\hat{k}_{u-2}+\tilde{k}_{r}+\tilde{k}_{u})\Big{\rbrace}^{-1}\non\\
&&\cdots \Big{\lbrace} p_{i}\cdot (\hat{k}_{1}+\cdots +\hat{k}_{u-2}+\tilde{k}_{r}+\tilde{k}_{u}+\hat{k}_{u+1}+\cdots +\hat{k}_{n})\Big{\rbrace}^{-1} \ \lbrace \tilde{k}_{r}\cdot \tilde{k}_{u}\rbrace^{-1}\times \non\\
&&\ \epsilon_{i}^{T}\Bigg{\lbrace}
\prod_{\substack{s=1}}^{u-2}
\Big(\widehat{\mathcal{E}}_{s}(\gamma)+\widehat{\mathcal{E}}_{s}(g)\Big)\ \Bigg{\rbrace}\ \Bigg{\lbrace}\prod_{s=u+1}^{n}\ \Big(\widehat{\mathcal{E}}_{s}(\gamma)+\widehat{\mathcal{E}}_{s}(g)\Big)\Bigg{\rbrace}\non\\
&&\ \Bigg[\ -\ \Big{\lbrace}\ -\f{1}{D-2}\ p_{i}^{2}\ (\tilde{k}_{r}\cdot \tilde{k}_{u})\ (\tilde{e}_{r}\cdot \tilde{e}_{u})\ +\ \f{1}{D-2}\ p_{i}^{2}\ (\tilde{e}_{r}\cdot \tilde{k}_{u})(\tilde{e}_{u}\cdot \tilde{k}_{r})\non\\
&&\ +\ (\tilde{e}_{r}\cdot \tilde{e}_{u})\ (p_{i}\cdot \tilde{k}_{r})(p_{i}\cdot \tilde{k}_{u})\ +\ (\tilde{k}_{r}\cdot \tilde{k}_{u})\ (\tilde{e}_{r}\cdot p_{i})(\tilde{e}_{u}\cdot p_{i})\ -\ (\tilde{e}_{r}\cdot \tilde{k}_{u})\ (\tilde{e}_{u}\cdot p_{i})\ (p_{i}\cdot \tilde{k}_{r})\non\\
&&\ -\ (\tilde{e}_{r}\cdot p_{i})\ (\tilde{e}_{u}\cdot \tilde{k}_{r})\ (p_{i}\cdot \tilde{k}_{u})\Big{\rbrace}\non\\
&&\ +\ \Big{\lbrace}(\tilde{e}_{r}\cdot p_{i})\ (\tilde{k}_{r}\cdot \tilde{\varepsilon}_{u}\cdot \tilde{k}_{r})\ +\ (\tilde{k}_{u}\cdot \tilde{k}_{r})\ (p_{i}\cdot \tilde{\varepsilon}_{u}\cdot \tilde{e}_{r})\ -\ (p_{i}\cdot \tilde{k}_{r})\ (\tilde{e}_{r}\cdot \tilde{\varepsilon}_{u}\cdot \tilde{k}_{r})\ -\ (\tilde{e}_{r}\cdot \tilde{k}_{u})\ (p_{i}\cdot \tilde{\varepsilon}_{u}\cdot \tilde{k}_{r})\ \Big{\rbrace}Q_{i}^{T} \non\\
&&\ +\ \Big{\lbrace}(\tilde{e}_{u}\cdot p_{i})\ (\tilde{k}_{u}\cdot \tilde{\varepsilon}_{r}\cdot \tilde{k}_{u})\ +\ (\tilde{k}_{r}\cdot \tilde{k}_{u})\ (p_{i}\cdot \tilde{\varepsilon}_{r}\cdot \tilde{e}_{u})\ -\ (p_{i}\cdot \tilde{k}_{u})\ (\tilde{e}_{u}\cdot \tilde{\varepsilon}_{r}\cdot \tilde{k}_{u})\ -\ (\tilde{e}_{u}\cdot \tilde{k}_{r})\ (p_{i}\cdot \tilde{\varepsilon}_{r}\cdot \tilde{k}_{u})\ \Big{\rbrace}Q_{i}^{T}\non\\
&&\ +\ \Big{\lbrace} -(p_{i}\cdot \tilde{k}_{u})\ (\tilde{k}_{u}\cdot \tilde{\varepsilon}_{r}\cdot \tilde{\varepsilon}_{u}\cdot p_{i})\ -\ (p_{i}\cdot \tilde{k}_{r})\ (\tilde{k}_{r}\cdot \tilde{\varepsilon}_{u}\cdot \tilde{\varepsilon}_{r}\cdot p_{i})\ +\ (p_{i}\cdot \tilde{k}_{r})\ (\tilde{k}_{u}\cdot \tilde{\varepsilon}_{r}\cdot \tilde{\varepsilon}_{u}\cdot p_{i})\non\\
&&\ +\ (p_{i}\cdot \tilde{k}_{u})\ (\tilde{k}_{r}\cdot \tilde{\varepsilon}_{u}\cdot \tilde{\varepsilon}_{r}\cdot p_{i})\ -\ \tilde{\varepsilon}_{r,\mu\nu}\ \tilde{\varepsilon}_{u}^{\mu\nu}\ (p_{i}\cdot \tilde{k}_{r})\ (p_{i}\cdot \tilde{k}_{u})\ -\ 2(p_{i}\cdot \tilde{\varepsilon}_{r}\cdot \tilde{k}_{u})\ (p_{i}\cdot \tilde{\varepsilon}_{u}\cdot \tilde{k}_{r})\non\\
&&\ +\ (p_{i}\cdot \tilde{\varepsilon}_{u}\cdot p_{i})\ (\tilde{k}_{u}\cdot \tilde{\varepsilon}_{r}\cdot \tilde{k}_{u})\ +\ (p_{i}\cdot \tilde{\varepsilon}_{r}\cdot p_{i})\ (\tilde{k}_{r}\cdot \tilde{\varepsilon}_{u}\cdot \tilde{k}_{r}) \Big{\rbrace}\ \non\\
&&\ -\ (\tilde{k}_{r}\cdot \tilde{k}_{u})\ \Big{\lbrace}\ Q_{i}^{T}Q_{i}^{T} (\tilde{e}_{r}\cdot \tilde{e}_{u})\ +\ 2 Q_{i}^{T}\ (p_{i}.\tilde{\varepsilon}_{r}\cdot \tilde{e}_{u})\ +\ 2 Q_{i}^{T}\ (p_{i}\cdot \tilde{\varepsilon}_{u}\cdot \tilde{e}_{r})\ +\ 2(p_{i}\cdot \tilde{\varepsilon}_{r}\cdot \tilde{\varepsilon}_{u}\cdot p_{i})\Big{\rbrace}\ \Bigg]\ \Gamma^{(i)}(p_{i})\ . \non\\ \label{step7}
\ee
Now we will sum over all permutations separately for $\hat{k}_{1},\hat{k}_{2},\cdots \hat{k}_{u-2}$ and $\hat{k}_{u+1},\hat{k}_{u+2},\cdots \hat{k}_{n}$ and the corresponding polarisations  keeping  $(r,u)$ fixed as a unit. Under this sum over permutations the only change happens in the first three lines of \eqref{step7}. Using \eqref{I1} and combining $\Big(\widehat{\mathcal{E}}_{s}(\gamma)+\widehat{\mathcal{E}}_{s}(g)\Big)$, the first three lines of \eqref{step7} turn out to be
\be
\Big{\lbrace}p_{i}\cdot (\tilde{k}_{r}+\tilde{k}_{u})\Big{\rbrace}^{-1}\ \Bigg{\lbrace}\prod_{\substack{s=1\\ s\neq r,u}}^{n}\ (p_{i}\cdot \tilde{k}_{s})^{-1}\Bigg{\rbrace}\ \lbrace \tilde{k}_{r}\cdot \tilde{k}_{u}\rbrace^{-1}\ \epsilon_{i}^{T}\ \Bigg{\lbrace}
\prod_{\substack{s=1\\s\neq r,u}}^{n}
\Big(\mathcal{E}_{s}(\gamma)+\mathcal{E}_{s}(g)\Big)\ \Bigg{\rbrace}\ .
\ee

Now we have to sum over all possible choice of $r,u$ from $\lbrace 1,2,\cdots n\rbrace$ and add the contributions of \eqref{sl1} and \eqref{sl2}. Consequently, we get

\be
&&\ \sum_{\substack{r,u=1\\r<u}}^{n}\ \Bigg{\lbrace}\prod_{\substack{s=1\\s\neq r,u}}^{n}\Big(p_{i}\cdot \tilde{k}_{s}\Big)^{-1}\epsilon_{i}^{T} \Big(\tilde{e}_{s,\mu}p_{i}^{\mu}Q_{i}^{T}\ +\ \tilde{\varepsilon}_{s,\mu\nu}p_{i}^{\mu}p_{i}^{\nu}\Big) \Bigg{\rbrace}\ \Big{\lbrace}p_{i}\cdot (\tilde{k}_{r}+\tilde{k}_{u})\Big{\rbrace}^{-1}\non\\
&&\ \Bigg( \mathcal{M}_{pp}\big(p_{i};\ \tilde{e}_{r},\tilde{k}_{r};\ \tilde{e}_{u},\tilde{k}_{u}\big)\ +\ \mathcal{M}_{gg}\big(p_{i};\ \tilde{\varepsilon}_{r},\tilde{k}_{r};\ \tilde{\varepsilon}_{u},\tilde{k}_{u}\big)\non\\
&&\ +\ \mathcal{M}_{pg}\big(p_{i};\ \tilde{e}_{r},\tilde{k}_{r};\ \tilde{\varepsilon}_{u},\tilde{k}_{u}\big)\  +\ \mathcal{M}_{pg}\big(p_{i};\ \tilde{e}_{u},\tilde{k}_{u};\ \tilde{\varepsilon}_{r},\tilde{k}_{r}\big) \Bigg)\  \Gamma^{(i)}(p_{i})\non
\ee

\be
&=&\ \sum_{\substack{r,u\in A_{i}\\r<u}}\ \Bigg{\lbrace}\prod_{\substack{s\in A_{i}\\s\neq r,u}}\ \Big(p_{i}\cdot k_{s}\Big)^{-1}\ \epsilon_{i}^{T}\ \Big( e_{s,\mu}p_{i}^{\mu}Q_{i}^{T}\ +\ \varepsilon_{s,\mu\nu}p_{i}^{\mu}p_{i}^{\nu}\Big) \Bigg{\rbrace}\ \Big{\lbrace}p_{i}\cdot (k_{r}+k_{u})\Big{\rbrace}^{-1}\non\\
&&\ \Bigg( \mathcal{M}_{pp}\big(p_{i};\ e_{r},k_{r};\ e_{u},k_{u}\big)\ +\ \mathcal{M}_{gg}\big(p_{i};\ \varepsilon_{r},k_{r};\ \varepsilon_{u},k_{u}\big)\non\\
&&\ +\ \mathcal{M}_{pg}\big(p_{i};\ e_{r},k_{r};\ \varepsilon_{u},k_{u}\big)\  +\ \mathcal{M}_{pg}\big(p_{i};\ e_{u},k_{u};\ \varepsilon_{r},k_{r}\big) \Bigg)\  \Gamma^{(i)}(p_{i})\ . 
\ee

Now since this $i$'th leg soft insertion already contributes in subleading order, for the other external legs we only have to consider soft insertion via $\Gamma^{(3)}$ vertices and pick up leading contributions. Summing over all external hard legs with possible soft insertions we get:

\be
&&\sum_{\substack{A_{1},A_{2},...,A_{N};\ A_{i}\subset \lbrace 1,2,...,M\rbrace \\
A_{i}\cap A_{j}=\phi,\ A_{1}\cup A_{2}\cup ...\cup A_{N}=\lbrace 1,2,...,M\rbrace}}\sum_{i=1}^{N}\ \Bigg[ \prod_{\substack{j=1\\j\neq i}}^{N}\ \Bigg{\lbrace}\prod_{\ell\in A_{j}}(p_{i}.k_{\ell})^{-1}\Bigg{\rbrace}\ \epsilon_{j}^{T}\Bigg{\lbrace}\prod_{\ell \in A_{j}}\Big(Q_{j}e_{\ell,\mu}p_{j}^{\mu}\ +\ \varepsilon_{\ell,\mu\nu}p_{j}^{\mu}p_{j}^{\nu}\Big) \Bigg{\rbrace}\Bigg]\non\\
&&\ \times \sum_{\substack{r,u\in A_{i}\\r<u}}\ \Bigg{\lbrace}\prod_{\substack{s\in A_{i}\\s\neq r,u}}\ \Big(p_{i}\cdot k_{s}\Big)^{-1}\ \epsilon_{i}^{T}\ \Big(Q_{i}\ e_{s,\mu}p_{i}^{\mu}\ +\ \varepsilon_{s,\mu\nu}p_{i}^{\mu}p_{i}^{\nu}\Big) \Bigg{\rbrace}\ \Big{\lbrace}p_{i}\cdot (k_{r}+k_{u})\Big{\rbrace}^{-1}\non\\
&&\ \Bigg( \mathcal{M}_{pp}\big(p_{i};\ e_{r},k_{r};\ e_{u},k_{u}\big)\ +\ \mathcal{M}_{gg}\big(p_{i};\ \varepsilon_{r},k_{r};\ \varepsilon_{u},k_{u}\big)\non\\
&&\ +\ \mathcal{M}_{pg}\big(p_{i};\ e_{r},k_{r};\ \varepsilon_{u},k_{u}\big)\  +\ \mathcal{M}_{pg}\big(p_{i};\ e_{u},k_{u};\ \varepsilon_{r},k_{r}\big) \Bigg)\ \Gamma(\epsilon_{1},p_{1};\ \cdots ;\ \epsilon_{N},p_{N})\ . \label{subleading3}
\ee

Now summing over \eqref{leading}, \eqref{subleading1}, \eqref{subleading2} and \eqref{subleading3} we get the contribution given is \eqref{semi-result}. This proves multiple soft photon-graviton theorem up to subleading order in soft momenta. We can analogously prove the gauge invariance of the multiple soft theorem result eq.\eqref{semi-result}, following section \ref{gaugeinv} .

\end{section}

\begin{section}{Special cases}\label{special}
 If we  choose first $M_{1}$ number of soft particles to be photons with polarisation and momenta $\lbrace e_{r},\ell_{r} \rbrace$, last $M_{2}=M-M_{1}$ number of soft particles to be gravitons with polarisation and momenta $\lbrace \varepsilon_{r},k_{r}\rbrace$ and hard particles are charge eigenstate of the $U(1)$ generator $Q$, then from eq.\eqref{semi-result} we get\\

\be
&&\ \Gamma^{(N+M_{1}+M_{2})}\ \big(\lbrace\epsilon_{i},p_{i}\rbrace ;\ \lbrace e_{r},\ell_{r}\rbrace ;\ \lbrace \varepsilon_{s},k_{s}\rbrace\big)\non\\[15pt]
&=&\ \Bigg{\lbrace}\prod_{j=1}^{N} \epsilon_{j,\alpha_{j}}(p_{j})\Bigg{\rbrace}\ \Bigg[\ \Bigg{\lbrace}\prod_{r=1}^{M_{1}}\  S^{(0)}_{r}(\gamma)\Bigg{\rbrace}\ \Bigg{\lbrace}\prod_{s=1}^{M_{2}}\  S^{(0)}_{s}(g)\Bigg{\rbrace}\ \Gamma^{\alpha_{1}\alpha_{2}\cdots \alpha_{N}}\non\\
&&\ +\ \sum_{m=1}^{M_{1}}\ \Bigg{\lbrace}\prod_{\substack{r=1\\r\neq m}}^{M_{1}}\  S^{(0)}_{r}(\gamma)\Bigg{\rbrace}\ \Bigg{\lbrace}\prod_{s=1}^{M_{2}}\  S^{(0)}_{s}(g)\Bigg{\rbrace}\ \Big[\ S^{(1)}_{m}(\gamma)\ \Gamma\Big]^{\alpha_{1}\alpha_{2}\cdots\alpha_{N}} \non\\
&&\ +\ \sum_{m=1}^{M_{2}}\ \Bigg{\lbrace}\prod_{r=1}^{M_{1}}\  S^{(0)}_{r}(\gamma)\Bigg{\rbrace}\ \Bigg{\lbrace}\prod_{\substack{s=1\\s\neq m}}^{M_{2}}\  S^{(0)}_{s}(g)\Bigg{\rbrace}\ \Big[S^{(1)}_{m}(g)\Gamma\Big]^{\alpha_{1}\alpha_{2}\cdots\alpha_{N}}\non\\
&&\ +\ \sum_{\substack{m,n=1\\m< n}}^{M_{1}}\ \Bigg{\lbrace}\prod_{\substack{r=1\\r\neq m,n}}^{M_{1}}\  S^{(0)}_{r}(\gamma)\Bigg{\rbrace}\ \Bigg{\lbrace}\prod_{s=1}^{M_{2}}\  S^{(0)}_{s}(g)\Bigg{\rbrace}\sum_{i=1}^{N}\ \f{1}{p_{i}.(\ell_{m}+\ell_{n})} \mathcal{M}_{pp}\big(p_{i};\ e_{m},\ell_{m};\ e_{n},\ell_{n}\big)\ \Gamma^{\alpha_{1}...\alpha_{N}}\non\\
&&\ +\ \sum_{\substack{m,n=1\\m< n}}^{M_{2}}\ \Bigg{\lbrace}\prod_{r=1}^{M_{1}}\  S^{(0)}_{r}(\gamma)\Bigg{\rbrace}\ \Bigg{\lbrace}\prod_{\substack{s=1\\s\neq p,q}}^{M_{2}}\  S^{(0)}_{s}(g)\Bigg{\rbrace}\sum_{i=1}^{N}\ \f{1}{p_{i}.(k_{m}+k_{n})} \mathcal{M}_{gg}\big(p_{i};\ \varepsilon_{m},k_{m};\ \varepsilon_{n},k_{n}\big)\ \Gamma^{\alpha_{1}...\alpha_{N}}\non\\
&&\ +\ \sum_{m=1}^{M_{1}}\sum_{n=1}^{M_{2}} \Bigg{\lbrace}\prod_{\substack{r=1\\r\neq m}}^{M_{1}}\  S^{(0)}_{r}(\gamma)\Bigg{\rbrace}\ \Bigg{\lbrace}\prod_{\substack{s=1\\s\neq n}}^{M_{2}}\  S^{(0)}_{s}(g)\Bigg{\rbrace}\sum_{i=1}^{N}\ \f{1}{p_{i}.(\ell_{m}+k_{n})} \mathcal{M}_{pg}\big(p_{i};\ e_{m},\ell_{m};\ \varepsilon_{n},k_{n}\big)\ \Gamma^{\alpha_{1}...\alpha_{N}}\non\\
&&\ +\ \sum_{m=1}^{M_{1}}\ \Bigg{\lbrace}\prod_{\substack{r=1\\r\neq m}}^{M_{1}}\  S^{(0)}_{r}(\gamma)\Bigg{\rbrace}\ \Bigg{\lbrace}\prod_{s=1}^{M_{2}}\  S^{(0)}_{s}(g)\Bigg{\rbrace}\ \sum_{i=1}^{N}\ \f{1}{p_{i}.\ell_{m}}\ \big(e_{m\mu}\ell_{m\nu} - e_{m\nu}\ell_{m\mu}\big) \ \Big{(}\mathcal{N}_{(i)}^{\mu\nu}(-p_{i})\Gamma\Big{)}^{\alpha_{1}...\alpha_{N}} \Bigg]\   \non\\ \label{final}
\ee
where in the expression of $S_{r}^{(0)}(\gamma)$, $S_{r}^{(1)}(\gamma)$, $\mathcal{M}_{pp} $ and $\mathcal{M}_{pg}$ we have taken the charge eigenvalue of the corresponding external hard particle instead of the $U(1)$ charge generator i.e. $\epsilon_{i,\alpha_{i}}Q_{\beta_{i}}^{\ \alpha_{i}}=q_{i}\epsilon_{i,\beta_{i}}$ . For example subleading soft photon factor takes form:
\be
\Big[ S^{(1)}_{r}(\gamma)\ \Gamma\Big]^{\alpha_{1}\cdots\alpha_{N}}\ &=&\ \sum_{i=1}^{N}\ \f{q_i\ e_{r,\mu}\ k_{r,\nu}}{p_{i}\cdot k_{r}}\ \Bigg(p_{i}^{\mu}\f{\p \Gamma^{\alpha_{1}\cdots\alpha_{N}}}{\p p_{i\nu}}\ -\ p_{i}^{\nu}\f{\p \Gamma^{\alpha_{1}\cdots\alpha_{N}}}{\p p_{i\mu}}\Bigg)
\ee

\subsection{Classical limit}
Now, let us use the above formula to write down the low frequency gravitational and electromagnetic radiation emitted in classical scattering.  We will first estimate relative strengths of various soft factors in the classical limit following the logic of \cite{1801.07719}.

As discussed in \cite{1801.07719},  in the classical limit $|S^{(0)}|$ is large. This can be seen as follows : for a given process the differential cross section for emitting $M_1$ soft photons and $M_2$ soft gravitons in frequency range $\omega$ to $\omega(1+\delta)$, within solid angle $\Delta\Omega$ around $\hat{k}$  is 
\be
\frac{A_1^{M_1}}{M_1!}\frac{A_2^{M_2}}{M_2!} \Gamma\ . \label{dsection}
\ee
where $\Gamma$ is the amplitude without soft particles and expressions for $A_{1}$ and $A_{2}$ to the leading order in soft particle energy $\omega$ are,
\be
A_1 = \frac{1}{2^D\pi^{D-1}}|S^{(0)}(\gamma)  |^2\ \omega^{D-2}\ \Delta\Omega\ \delta\ ,\hspace{10mm} 
A_2 = \frac{1}{2^D\pi^{D-1}} |S^{(0)}(g) |^2\ \omega^{D-2}\ \Delta\Omega\ \delta.
\ee
Here we have parametrised soft particle momenta by $k^{\mu}=\omega(1,\hat{k})$. Classical limit corresponds to large $M_1, M_2$.  In this limit, we need to find the maxima of soft particle number distribution :  
\be
\frac{\partial}{\partial M_r}[M_r\log A_r -M_r\log M_r+M_r]=0, \hspace{10mm} \hbox{ for r=1,2}.
\ee
Thus the distribution is maximised at  $M_1=A_1$ and $M_2=A_2$. Hence, $A_1$ and $A_2$ are very large implying $|S^{(0)}(\gamma)|$ and $|S^{(0)}(g)|$ are large as well. 

Next we will estimate the subleading soft factors. Let the external momenta $p_i$ be of order $\mu$ and external angular momenta be of order $\mu\lambda$. Here $\lambda$ is the length scale of the process and will be typically given by impact parameter for scattering processes. Since, we are interested in soft radiation in classical limit, $\mu, \lambda$ have to be very large. Hence the soft momentum is of order : $k, \ell \sim \lambda^{-1} \tau$, where $\tau$ is a small number as the soft particle wavelength has to be large compared to the  impact parameter. Let the order hard particle charges $q_i$ be given by $\beta$. As we require $|S^{(0)}(\gamma)|$ to be large, $\beta$ have to be large.  Thus, the order of magnitude of the following quantities in classical limit turns out to be :
\be
S^{(0)}_{r}(g) \sim \mu \lambda \tau^{-1},\  \ \  S^{(1)}_{r}(g) \sim \mu \lambda ,\  \  \ S^{(0)}_{r}(\gamma) \sim \beta\lambda \tau^{-1},\ \  \ S^{(1)}_{r}(\gamma) \sim  \beta\lambda\ ,   \non\\
 \mathcal{M}_{gg}\sim \mu^2,\ \ \ \mathcal{M}_{pg}\sim \beta\mu,\ \ \ \mathcal{M}_{pp}\sim \beta^2+\mu^2,\ \ \ \mathcal{N}_{r}(\gamma)\sim \beta\lambda\mu^{-1} .
\ee
Above, for determining the order of magnitude of theory dependent term $\mathcal{N}_{r}(\gamma)$, we use the fact that the full non-universal three point coupling $N_{(i)}^{\mu\nu}(-p_{i})$ turns out typically proportional to covariant electromagnetic moment in classical limit, being coupled to the electromagnetic field strength. So we used the order of magnitude for $N_{(i)}^{\mu\nu}(-p_{i})$ to be same as the order of magnitude of electromagnetic moment, which is $\beta\lambda$.
 
Now, if we consider a process where photon and graviton emission is comparable, we have to consider the limit $\beta\sim \mu$. The terms in \eqref{final} respectively have the following order of magnitude:
\be
& \lambda^{M_1+M_2} \beta^{M_1}\mu^{M_2} \tau^{-(M_1+M_2)}, & \lambda^{M_1+M_2} \beta^{M_1}\mu^{M_2} \tau^{-(M_1+M_2)+1},\non\\ 
& \lambda^{M_1+M_2} \beta^{M_1}\mu^{M_2} \tau^{-(M_1+M_2)+1},& \lambda^{M_1+M_2-1} \beta^{M_1-2}\mu^{M_2-1} \tau^{-(M_1+M_2)+1}(\beta^2+\mu^2),\non\\
 & \lambda^{M_1+M_2-1}  \beta^{M_1} \mu^{M_2-1} \tau^{-(M_1+M_2)+1},
& \lambda^{M_1+M_2-1} \beta^{M_1}\mu^{M_2-1} \tau^{-(M_1+M_2)+1},\non\\ 
 & \lambda^{M_1+M_2} \beta^{M_1}\mu^{M_2-1} \tau^{-(M_1+M_2)+1}.&
\ee
The contact terms containing $\mathcal{M}_{gg}$, $\mathcal{M}_{gp}$ and $\mathcal{N}_{r}(\gamma)$ are clearly sub-dominant with respect to the first three lines in \eqref{final}; one can check that $\mathcal{M}_{pp}$ is also sub-dominant for $\beta\sim \mu$. This is intuitive as classically we expect  interaction between soft particles to be a small correction to the radiation. Also, the commutator between $S^{(0)}$ and $S^{(1)}$ factors is suppressed by powers of $\mu$. So, the ordering between the two is irrelevant. Finally, an amplitude with $M_1$ soft photons and $M_2$ soft gravitons when $M_1, M_2$ are large(classical limit) to the subleading order in $\omega$ can be written as :
\be 
 \prod_{r=1}^{M_1}\Big(  S^{(0)}_{r}(\gamma) +\  S^{(1)}_{r}(\gamma)\Big) \prod_{s=1}^{M_2}\Big(  S^{(0)}_{s}(g) +\  S^{(1)}_{s}(g)\Big)\ \ \Gamma. \label{classicalsoft}
\ee
Hence, the differential cross section \eqref{dsection} will be corrected according to   :
$$ A_1 = \frac{1}{2^D\pi^{D-1}}|S^{(0)}(\gamma) +\  S^{(1)}(\gamma) |^2\  \omega^{D-1}\ \Delta\Omega\ \delta, $$
$$ A_2 = \frac{1}{2^D\pi^{D-1}} |S^{(0)}(g) +\  S^{(1)}(g)|^2\ \omega^{D-1}\ \Delta\Omega\ \delta. $$
The energy emitted is just number of soft particles times $\omega$. So, the radiation energy in frequency range $\omega$ to $\omega(1+\delta)$, within solid angle $\Delta\Omega$ around $\hat{k}$  is  $(A_{1}+A_2)\omega$. 

Eq.\eqref{classicalsoft} is the  classical version of soft theorem. The soft factors for graviton and photon have decoupled and as expected gravitational and electromagnetic radiations decoupled in classical limit.\\

\end{section}

\noindent{\bf Acknowledgments:}\ 
We are deeply thankful to Prof. Ashoke Sen for suggesting this problem, for insightful discussions and crucial ideas throughout the course of this work and for going through and correcting the draft with immense patience.  We also want to thank Nabamita Banerjee and Dileep Jatkar for encouragement in pursuing this project. SB is grateful to HRI, Allahabad  where this work has been completed. We express our gratitude to people of India for their continuous support to theoretical sciences.

\appendix
\begin{section}{Summation Identities}
To prove the multiple soft theorem in many intermediate stages we have used the following identities, which are proven in \cite{1707.06803}.

\be
\sum_{perm\lbrace 1,2,...,n\rbrace}\ \prod_{\ell=1}^{n}\ (a_{1}+a_{2}+...+a_{\ell})^{-1}\ =\ \prod_{\ell=1}^{n}\ (a_{\ell})^{-1}\ . \label{I1}
\ee

\be
&&\sum_{perm\lbrace 1,2,...,n\rbrace}\ \sum_{m=2}^{n}\sum_{\substack{r,u=1\\r<u}}^{m}\ b_{ru}\ (a_{1}+a_{2}+...+a_{m})^{-1}\ \prod_{\ell=1}^{n}\ (a_{1}+a_{2}+...+a_{\ell})^{-1}\non\\
&& =\ \prod_{\ell=1}^{n}\ (a_{\ell})^{-1}\ \sum_{\substack{r,u=1\\r<u}}^{n}\ b_{ru}\ (a_{r}+a_{u})^{-1},\hspace{6mm}\ for\ b_{rs}=b_{sr}\ for\ 1\leq r<s\leq n.\label{I2}
\ee

\be
\sum_{perm\lbrace 1,2,...,n\rbrace}\ \sum_{\substack{r,u=1\\r<u}}^{n}\ c_{ur}\ \prod_{\ell=1}^{n}\ (a_{1}+a_{2}+...+a_{\ell})^{-1}\ =\ \prod_{\ell=1}^{n}\ (a_{\ell})^{-1}\ \sum_{\substack{r,u=1\\r<u}}^{n}\ (a_{r}+a_{u})^{-1}\ (a_{u}c_{ur}+a_{r}c_{ru})\ . \non\\ \label{I3}
\ee
\end{section}

\begin{section}{Comparison of non-universal term with existing result in four dimension}\label{appB}

Here for some specific massless theories in 4D we compute the non-universal term appeared in the last line of single soft photon theorem result \eqref{singlesoft}:
\be
&& \Gamma^{(N+1)}(e,k;\ \lbrace \epsilon_{i},p_{i}\ \rbrace)\Big{|}_{non-uni}\non\\
&=&\  \sum_{i=1}^{N}\ \lbrace p_{i}\cdot k\rbrace^{-1}\  (e_{\mu}k_{\nu}-e_{\nu}k_{\mu})\ \epsilon_{i}^{T}\ \Big[ \f{iQ_{i}^{T}}{4}\ \f{\p \mathcal{K}_{i}(-p_{i})}{\p p_{i\nu}}\ \f{\p \Xi_{i}(-p_{i})}{\p p_{i\mu}}\ +\ \f{1}{2}\ \mathcal{B}^{\mu\nu}_{(i)}(-p_{i})\Xi_{i}(-p_{i})\Big]\ \Gamma^{(i)}(p_{i}).\non\\ \label{non-uni}
\ee
Let us first consider the theory massless spinor $\chi$ minimally coupled to photon. To analyse the first term within the square bracket of eq.\eqref{non-uni} we first need to extract the kinetic operator for Dirac fermion $\chi$ having property $\mathcal{K}^{\alpha\beta}(p)=-\mathcal{K}^{\beta\alpha}(-p)$. We work in Majorana representation by breaking each $\chi$ spinor component into real and imaginary part and considered them to be independent. In Majorana representation the gamma matrices satisfy the following properties
\be
\lbrace \gamma^{\mu},\gamma^{\nu}\rbrace=-2\eta^{\mu\nu}\ ,\ (\gamma^{\mu})^{*}=-\gamma^{\mu}\ ,\ (\gamma^{\mu})^{\dagger}=\gamma^{0}\gamma^{\mu}\gamma^{0} \ .\label{gamma}
\ee
In this representation,
\be
\f{\p \mathcal{K}^{\alpha\beta}(-p)}{\p p_{\mu}}=(\gamma^{0}\gamma^{\mu})^{\alpha\beta}\ ,\ \f{\p \Xi_{\alpha\beta}(-p)}{\p p_{\mu}}= -i(\gamma^{\mu}\gamma^{0})_{\alpha\beta}
\ee
where $\alpha$ and $\beta$ take values from 1 to 4. After substituting this in the first part of the non-universal term in eq\eqref{non-uni}, we simplify the result using the properties \eqref{gamma} and get
\be
\sum_{i=1}^{N}\f{e_{\mu}k_{\nu}}{p_{i}\cdot k}\ \ \epsilon_{i}^{T}(p_{i})Q_{i}^{T}\ (J^{\mu\nu}_{S})^{T}\Gamma^{(i)}(p_{i})
\ee
where $(J_{S}^{\mu\nu})_{\alpha}^{\ \beta}=-\f{1}{2}(\gamma^{\mu\nu})_{\alpha}^{\ \beta}$ is the spin angular momenta of the spinor with $\gamma^{\mu\nu}\ \equiv \ \f{1}{2}[\gamma^{\mu},\gamma^{\nu}] $ . Hence this contribution we can include within the universal part of subleading single soft photon theorem i.e. second line of the RHS of eq.\eqref{singlesoft} to write the subleading soft factor in terms of total angular momentum of finite energy particles.

Now consider the non-minimal coupling term $g_1\bar{\chi}\gamma^{\mu\nu}F_{\mu\nu}\chi $ in this theory, where $g_{1}$ is an arbitrary coupling of mass dimension $-1$. Comparing this non-minimal interaction term with eq.\eqref{non-minimal} we get $\mathcal{B}^{\mu\nu,\alpha\beta}=\ 2g_{1}(\gamma^{\mu\nu})^{\alpha\beta}$. Now if we consider the photon is outgoing with momentum $k^{\mu}$ and helicity $+1$ and all the external fermions being outgoing with helicity $+\f{1}{2}$ then the non-vanishing contribution of the second part of eq.\eqref{non-uni} becomes
 \be
 4\sqrt{2}g_{1}\ \sum_{i=1}^{N}\ \f{[ki]}{<ki>}\ \widetilde{\Gamma}^{(i)}(p_{i}) , 
\ee
where $p_{i,a\dot{b}}=-|i]_{a}<i|_{\dot{b}}$, $k_{a\dot{b}}=-|k]_{a}<k|_{\dot{b}}$, etc. following the convention of \cite{1308.1697} and $\widetilde{\Gamma}^{(i)}(p_{i})= \bar{u}_{-}(p_{i})\Gamma^{(i)}(p_{i})$. This agrees with the result of \cite{1611.07534}.

Similarly consider a theory with real scalar and photon fields. There can not be any minimal coupling between them. In this theory for both real scalar and photon $\p \mathcal{K}(-p)/\p p_{\nu}\p \Xi(-p)/\p p_{\mu}$ vanishes. So first part of eq.\eqref{non-uni} contributes to zero. Now consider a non-minimal interaction term $g_2\phi F_{\mu\nu}F^{\mu\nu}$ with $g_{2}$ be an arbitrary coupling with mass dimension $-1$. Comparing with eq.\eqref{non-minimal} we get 
\[ \mathcal{B}^{\mu\nu,\rho}(-p_{i})=\begin{bmatrix} 0& ig_{2}\ (\eta^{\mu\rho}p_{i}^{\nu}-\eta^{\nu\rho}p_{i}^{\mu})\\
ig_{2}\ (\eta^{\mu\rho}p_{i}^{\nu}-\eta^{\nu\rho}p_{i}^{\mu})&0\\
\end{bmatrix}\] where we have clubbed the fields into bigger multiplet $\Phi_\alpha$ with components : 
\[ \Phi = \begin{bmatrix}A_\rho \\ \phi \\ 
\end{bmatrix}\].
In this case for simplicity if we choose all the outgoing finite energy particles to be photon with helicity $+1$ with the outgoing soft photon having $+1$ helicity then from eq.\eqref{non-uni} we get the following non-vanishing contribution
\be
\Gamma^{(N+1)}(e^{+},k;\ \lbrace e^{+}_{i},p_{i}\ \rbrace)|_{non-uni}\ =\ ig_{2}\ \sum_{i=1}^{N}\ \f{[ki]}{<ki>}\ \Gamma^{(i)}(p_{i}).
\ee
Hence this result also agrees with \cite{1611.07534}.

\end{section}

\end{document}